\documentclass[a4paper,fleqn,usenatbib]{mnras}

\usepackage[T1]{fontenc}
\usepackage{ae,aecompl}

\usepackage{rotating}

\usepackage{graphicx}	% Including figure files
\usepackage{amsmath}	% Advanced maths commands
\usepackage{amssymb}	% Extra maths symbols

\newcommand{\Msun}{$M_{\odot}$}

\newcommand{\kms}{km\,s$^{-1}$}
\newcommand{\vs}{$v \sin i$}
\newcommand{\teff}{$T_{\rm eff}$}
\newcommand{\lgg}{$\log\,{g}$}

\title[Evolution of magnetic fields in solar-type stars]
{The evolution of surface magnetic fields in young solar-type stars I: the first 250 Myr\thanks{Based on observations obtained at the Canada-France-Hawaii Telescope (CFHT), which is operated by the National Research Council of Canada, 
the Institut National des Sciences de l'Univers of the Centre National de la Recherche Scientifique of France, and the University of Hawaii.  Also based on observations obtained at the Bernard Lyot Telescope (TBL, Pic du Midi, France) of the Midi-Pyr\'en\'ees Observatory, 
which is operated by the Institut National des Sciences de l'Univers of the Centre National de la Recherche Scientifique of France. }
}

\author[Folsom et al.]{C.P. Folsom$^{1,2}$\thanks{\tt folsomc@obs.ujf-grenoble.fr}, P. Petit$^{3,4}$, J. Bouvier$^{1,2}$, A. L\`ebre$^{5}$, L. Amard$^{5}$, A. Palacios$^{5}$, 
\newauthor
J. Morin$^{5}$, J.-F. Donati$^{3,4}$, S.V. Jeffers$^{6}$, S.C. Marsden$^{7}$, A.A. Vidotto$^{8}$\\
$^1$Universit\'e Grenoble Alpes, IPAG, F-38000 Grenoble, France\\
$^2$CNRS, IPAG, F-38000 Grenoble, France\\
$^3$Universit\'e de Toulouse, UPS-OMP, IRAP, Toulouse, France\\
$^4$CNRS, Institut de Recherche en Astrophysique et Planetologie, 14, avenue Edouard Belin, F-31400 Toulouse, France\\
$^5$LUPM, Universit\'e de Montpellier, CNRS, Place Eug\`ene Bataillon, 34095, France\\
$^6$Institut f\"ur Astrophysik, Georg-August-Universit\"at G\"ottingen, Friedrich-Hund-Platz 1, 37077, G\"ottingen, Germany \\
$^7$Computational Engineering and Science Research Centre, University of Southern Queensland, Toowoomba 4350, Australia \\
$^8$Observatoire de Gen\`eve, Universit\'e de Gen\`eve, Chemin des Maillettes 51, CH-1290 Versoix, Switzerland \\
}

\date{Accepted XXX. Received YYY; in original form ZZZ}

\pubyear{2015}

\begin{document}
\label{firstpage}
\pagerange{\pageref{firstpage}--\pageref{lastpage}}
\maketitle

\begin{abstract}
The surface rotation rates of young solar-type stars vary rapidly with age from the end of the pre-main sequence through the early main sequence.  Important changes in the dynamos operating in these stars may result from this evolution, which should be observable in their surface magnetic fields.  Here we present a study aimed at observing the evolution of these magnetic fields through this critical time period.  We observed stars in open clusters and stellar associations of known ages, and used Zeeman Doppler Imaging to characterize their complex magnetic large-scale fields.  Presented here are results for 15 stars, from 5 associations, with ages from 20 to 250 Myr, masses from 0.7 to 1.2 $M_\odot$, and rotation periods from 0.4 to 6 days.  We find complex large-scale magnetic field geometries, with global average strengths from 14 to 140 G.  There is a clear trend towards decreasing average large-scale magnetic field strength with age, and a tight correlation between magnetic field strength and Rossby number.  Comparing the magnetic properties of our zero-age main sequence sample to those of both younger and older stars, it appears that the magnetic evolution of solar-type stars during the pre-main sequence is primarily driven by structural changes, while it closely follows the stars' rotational evolution on the main sequence. 
\end{abstract}


\section{Introduction}

Solar-type stars undergo a dramatic evolution in their rotation rates as they leave the pre-main sequence and settle into the main sequence \citep[for a recent review see][]{Bouvier2013-rotation-evol-review}.  Early on the pre-main sequence stellar rotation rates are regulated, likely due to interactions between a star and its disk.  Eventually, after a few Myr, solar-type stars decouple from their disks and around this time the disk begins dissipating.  Since the stars are still contracting on the pre-main sequence, they spin up.  On a slower timescale, solar-type stars lose angular momentum through a magnetized wind.  Thus once a star has reached the main sequence it begins to spin down 
\citep[e.g.][]{Schatzman1962-rot-mag-evol,Skumanich1972-spindown-law,Mestel1987-mag-braking-late}. 
Since solar-type stars have dynamo driven magnetic fields, there is likely an important evolution in their magnetic properties over this time period.  Such changes in magnetic properties could be driven both by changes in rotation rate and by changes in the internal structure of PMS stars \citep[e.g.][]{Gregory2012-TTauri-B-structure}.  In turn, stellar magnetic fields play a key role in angular momentum loss.  Thus understanding these magnetic fields is critical for understanding the rotational evolution of stars \citep[e.g.][]{Vidotto2011-MHD-wind-V374Peg, Matt2012-magnetic-breaking-formulation, Reville2015-MHD-wind-torque}. 

Rotation rates deeper in a star may differ somewhat from this description of observed surface rotation rates.  
During the spin-down phase angular momentum is lost from the surface of a star, potentially creating enhanced radial differential rotation.  Recent rotational models by \citet{Gallet2013-Bouvier-ang-mom-evol, Gallet2015-Bouvier-ang-mom-evol2} use differences between the core and envelope rotation rates to explain the evolution of observed surface rotation rates.  These models predict a period of greatly enhanced radial differential rotation as a star reaches the main sequence. 
Other models of the rotational evolution of stars also have important impacts on the magnetic properties of these stars, such as the `Metastable Dynamo Model' of \citet{Brown2014-metastable-dynamo-model-rotation}. 

The large-scale magnetic fields of main sequence solar-type stars were first observed in detail many years ago \citep[e.g.][]{Donati1997-ABDor}, and more recently the magnetic strengths and geometries of these stars have been characterized for a significant sample of these stars \citep[e.g.][]{Petit2008-sunlike-mag-geom}. 
Some trends are apparent: there are clearly contrasting magnetic properties between solar-like stars and M-dwarfs \citep{Morin2008-Mdwarf-topo, Morin2010-Mdwarfs}.  There is some evidence for a correlation between more poloidal magnetic geometries and slow rotation rates \citep{Petit2008-sunlike-mag-geom}.  With a large sample of stars, there appears to be trends in large-scale magnetic field strength with age and rotation \citep{Vidotto2014-magnetism-age-rot}.  
Zeeman broadening measurements from \citet{Saar1996-maybe-saturation-zeeman-broad} and \citet{Reiners2009-ZeemanBroad-saturation-Mdwarfs} find trends in the small-scale magnetic field strength with rotation and Rossby number, particularly for M-dwarfs.  
\citet{Donati2009-ARAA-magnetic-fields} provide a detailed review of magnetic properties for a wide range of non-degenerate stars.  
Currently, the BCool collaboration is carrying out the largest systematic characterization of magnetic fields in main-sequence solar-type stars, with early results in \citet{Marsden2014-Bcool-survey1} and Petit et al.\ (in prep).  

On the pre-main sequence, large-scale magnetic fields have been observed and characterized for a large number of stars \citep[e.g.][]{Donati2008-BPTau-ZDI, Donati2010MNRAS-AATau-ZDI, Donati2011-TWHya-ZDI}.  These observations are principally from the `Magnetic Protostars and Planets' (MaPP) and `Magnetic Topologies of Young Stars and the Survival of massive close-in Exoplanets' (MaTYSSE) projects.  There are clear differences between the magnetic properties of T Tauri stars and older main sequence stars, which seem to be a consequence of the internal structure of the star \citep{Gregory2012-TTauri-B-structure}.  This appears to be similar to the difference between main sequence solar mass stars and M-dwarfs \citep{Morin2008-Mdwarf-topo}.

We aim to provide the first systematic study of the magnetic properties of stars from the late pre-main sequence through the zero age main sequence (ZAMS) up to $\sim$250 Myr.  This covers the most dramatic portion of the rotational evolution of solar mass stars.  Observations of a few individual stars in this age and mass range have been made 
(e.g. HD 171488, \citealt{Jeffers2008-HD171488-ZDI}, \citealt{Jeffers2011-nonSolar-HD171488-3epochs}; HD 141943, \citealt{Marsden2011-hd141943-sempol-hd101412}; HD 106506, \citealt{Waite2011-HD106506-young}; HN Peg, \citealt{BoroSaikia2015-HNPeg}; HD 35296 \& HD 29615, \citealt{Waite2015-2-young-solar-B}), 
but to date only a modest number of stars have been observed on an individual basis.  

In this paper we focus on young (not accreting) stars, in the age range 20 to 250 Myr, and in the restricted mass range from 0.7 to 1.2 $M_\odot$.  This fills the gap between the T Tauri star observations of MaPP \& MaTYSSE and the older main sequence observations of BCool.  The observations focus on stars in young clusters and associations, in order to constrain stellar ages.  Spectropolarimetric observations and Zeeman Doppler Imaging (ZDI) are used to determine the strength and geometry of the large-scale stellar magnetic fields.  
This work is being carried out as part of the `TOwards Understanding the sPIn Evolution of Stars' (TOUPIES) project\footnote{http://ipag.osug.fr/Anr\_Toupies/}.  Observations are ongoing in the large program `History of the Magnetic Sun' at the Canada-France-Hawaii Telescope.  Future papers in this series will expand the size of the sample, and extend the age range up to 600 Myr.

\section{Observations}
\label{observations}
We obtained time-series of spectropolarimetric observations using the ESPaDOnS instrument at the Canada France Hawaii Telescope (CFHT; \citealt{Donati2003-ESPaDOnS-descript}; see also \citealt{Silvester2012-data-paper}), and the Narval instrument \citep{Auriere2003-Narval-early} on the T\'elescope Bernard Lyot (TBL) at the Observatoire du Pic du Midi, France.  Narval is a direct copy of ESPaDOnS, and thus virtually identical observing and data reduction procedures were used for observations from the two instruments.  
ESPaDOnS and Narval are both high resolution \'echelle spectropolarimeters, with R$\sim$65000 and nearly continuous wavelength coverage from 3700 to 10500 \AA.  The instruments consist of a Cassegrain mounted polarimeter module, which is attached by optical fiber to a cross-dispersed bench mounted \'echelle spectrograph.  
Observations were obtained in spectropolarimetric mode, which obtains circularly polarized Stokes $V$ spectra, in addition to the total intensity Stokes $I$ spectra.  
Data reduction was performed with the Libre-ESpRIT package \citep{Donati1997-major}, which is optimized for ESPaDOnS and Narval, and performs calibration and optimal spectrum extraction in an automated fashion.  

Observations for a single star were usually obtained within a two week period, and always over as small a time period as practical, (in individual cases this ranged from 1 to 4 weeks, as detailed in Table \ref{observations-table}).  This was done in order to avoid any potential intrinsic evolution of the large-scale stellar magnetic field.  Observations were planned to obtain a minimum of 15 spectra, distributed as evenly as possible in rotational phase, over a few consecutive rotational cycles.  However, in some cases fewer observations were achieved due to imperfect weather during our two week time frame.  A minimum target S/N of 100 was used, although this was increased for earlier type stars and slower rotators, which were expected to have weaker magnetic fields.  A few observations fell below this target value, and a detailed consideration of the potential impact of low S/N on spurious signals, and how to avoid this spurious signal, are discussed in Appendix \ref{Spurious signal}.
A summary of the observations obtained can be found in Table \ref{observations-table}.

\begin{table*}
\centering
\caption{Summary of observations obtained. Exposure times are for a full sequence of 4 sub-exposures, 
and the S/N values are the peak for $V$ spectrum (per 1.8 \kms\ spectral pixel, typically near 730 nm). }
\begin{tabular}{lccccccc}
\hline\hline
Object          & Coordinates & Assoc.      & Dates of             &Telescope &Integration& Num.   & S/N   \\
                & RA, Dec     &             & Observations         &Semester   & Time (s)  &Obs.   & Range \\
\hline
HII 296         &03:44:11.20 +23:22:45.6 & Pleiades    & 13 Oct - 30 Oct 2009 & TBL 09B  &3600 &  18 & 70-110  \\
HII 739         &03:45:42.12 +24:54:21.7 & Pleiades    & 4  Oct - 1  Nov 2009 & TBL 09B  &3600 &  17 & 140-270 \\
HIP 12545       &02:41:25.89 +05:59:18.4 & $\beta$ Pic & 25 Sept - 29 Sept 2012 & CFHT 12B & 640 &  16 & 110-130 \\
BD-16351        &02:01:35.61 -16:10:00.7 & Columba     & 25 Sept - 1  Oct 2012 & CFHT 12B & 600 &  16 & 70-100  \\
HIP 76768       &15:40:28.39 -18:41:46.2 & AB Dor      & 18 May - 30 May 2013 & CFHT 13A & 800 &  24 & 110-150 \\
TYC 0486-4943-1 &19:33:03.76 +03:45:39.7 & AB Dor      & 24 Jun - 1  Jul 2013 & CFHT 13A &1400 &  15 & 95-120  \\
TYC 5164-567-1  &20:04:49.36 -02:39:20.3 & AB Dor      & 15 Jun - 1  Jul 2013 & CFHT 13A & 800 &  19 & 100-130 \\
TYC 6349-0200-1 &20:56:02.75 -17:10:53.9 & $\beta$ Pic & 15 Jun - 30 Jun 2013 & CFHT 13A & 800 &  16 & 120-130 \\
TYC 6878-0195-1 &19:11:44.67 -26:04:08.9 & $\beta$ Pic & 15 Jun - 1  Jul 2013 & CFHT 13A & 800 &  16 & 110-140 \\
PELS 031        &03:43:19.03 +22:26:57.3 & Pleiades    & 15 Nov - 23 Nov 2013 & CFHT 13B &3600 &  14 & 90-150  \\
DX Leo          &09:32:43.76 +26:59:18.7 & Her-Lyr     & 7  May - 18 May 2014 & TBL 14A  & 600 &  8  & 250-295 \\
V447 Lac        &22:15:54.14 +54:40:22.4 & Her-Lyr     & 7  Jun - 16 Jul 2014 & TBL 14A  & 600 &  7  & 187-227 \\
LO Peg          &21:31:01.71 +23:20:07.4 & AB Dor      & 16 Aug - 31 Aug 2014 & TBL 14A  & 600 &  47 & 70-124  \\
V439 And        &00:06:36.78 +29:01:17.4 & Her-Lyr     & 1  Sept - 27 Sept 2014 & TBL 14B  & 180 &  14 & 182-271 \\
PW And	        &00:18:20.89 +30:57:22.2 & AB Dor      & 3  Sept - 19 Sept 2014 & TBL 14B  &1000 &  11 & 161-194 \\
\hline\hline  
\end{tabular} 
\label{observations-table} 
\end{table*}

\subsection{Sample Selection}

Our observations focus on well established solar-type members of young stellar clusters and associations, in order to provide relatively accurate ages.  In this paper we focus on stars younger than 250 Myr old, but in future papers we will extend this to 600 Myr.  
In order to provide the high S/N necessary to reliably detect magnetic fields, the sample is restricted to relatively bright targets, $V < 12$, and hence nearby stellar associations and young open clusters.  

We attempt to focus on stars with well established rotation periods in the literature.  
The current sample includes stars with rotation periods between 0.42 and 6.2 days, however the majority of the stars have periods between 2 and 5 days.  

So far the study has focused on stars between G8 and K6 spectral types (\teff\ approximately from 5500 to 4500 K, with one hotter 6000 K star).  
This provides a sample of stars with qualitatively similar internal structure, consisting of large convective envelopes and radiative cores.  Focusing on stars slightly cooler than the Sun has the advantage of selecting stars with stronger magnetic fields, due to their larger convective zones, and increasing our sensitivity to large-scale magnetic fields, through the increased number of lines available to Least Squares Deconvolution (see Sect. \ref{Least squares deconvolution}).  Having some spread in \teff\ is valuable, as this allows us to consider variations in magnetic field as a function of the varying convective zone depths.  

Details on individual targets are included in Appendix \ref{Individual Targets}.  The physical parameters of individual targets are presented in Table \ref{fundimental-param-table} and Fig.~\ref{hr-diagram}. 

\subsection{Least squares deconvolution}
\label{Least squares deconvolution}
Least Squares Deconvolution \citep[LSD;][]{Donati1997-major,Kochukhov2010-LSD} was applied to our observations, in order to detect and characterize stellar magnetic fields.  
LSD is a cross-correlation technique which uses many lines in the observed spectrum to produce effectively a `mean' observed line profile, with much higher S/N than any individual line.
Line masks, needed as input for LSD, were constructed based on data extracted from the Vienna Atomic Line Database (VALD) \citep{Ryabchikova1997-VALD-early, Kupka1999-VALD}, using `extract stellar' requests.  The line masks were constructed assuming solar chemical abundances, and using the effective temperature and surface gravity for each star found in Sect.~\ref{spectrum-fitting} (and Table \ref{fundimental-param-table}), rounded to the nearest 500 K in \teff\ and 0.5 in \lgg.  The line masks used lines with a VALD depth parameter greater than 0.1, and lines from 500 nm to 900 nm excluding Balmer lines (see Appendix \ref{Spurious signal} for a discussion of the wavelength range used), and include $\sim$3500 lines.  

The normalization of the LSD profiles is intrinsically somewhat arbitrary \citep{Kochukhov2010-LSD}, as long as the normalization values are used self-consistently throughout an analysis.  We used the same normalization for all stars in the sample, with the values taken from the means from a typical line mask.  The normalizing values were a line depth of 0.39, Land\'e factor of 1.195, and a wavelength of 650~nm.  This normalization has no direct impact on our results, as long as the normalization values are consistent with the values used for measuring $B_{\ell}$ (Eq.~\ref{bz-equation}) and for modeling Stokes $V$ profiles in ZDI.  

The resulting LSD profiles were used to measure longitudinal magnetic fields and radial velocities, as well as input for ZDI.  Sample LSD profiles for all our stars are plotted in Fig.~\ref{fig-lsd-grid}.

\section{Fundamental physical parameters}
\label{fund-param}

\subsection{Spectroscopic analysis}
\label{spectrum-fitting}

\subsubsection{Primary analysis}

\begin{figure*}
\centering
\includegraphics[width=5.0in]{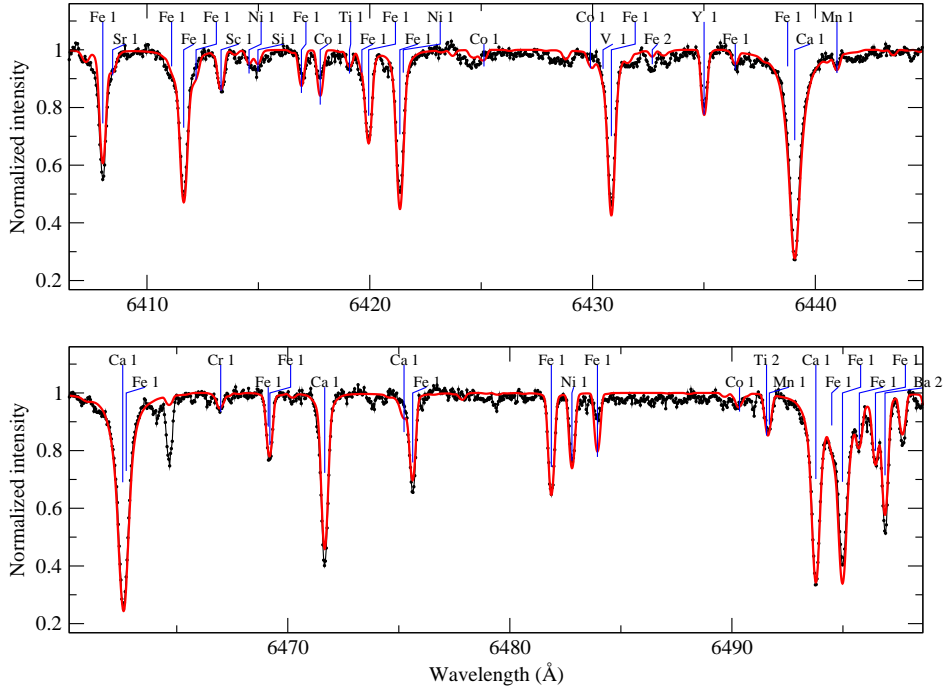}
\caption{Sample fit of the synthetic spectrum (red line) to the observation (black points) for HIP 12545.  }
\label{spec-fitting-ex}
\end{figure*}

Many of the stars in this study have poorly determined physical parameters in the literature, and in several cases no spectroscopic analysis.  Thus in order to provide precise, self-consistent physical parameters, we performed a detailed spectroscopic analysis of all the stars.  The same high resolution spectra with a wide wavelength range that are necessary to detect magnetic fields in Stokes $V$ are also ideal for spectroscopic analysis in Stokes $I$.  

The observations were first normalized to continuum level, by fitting a low order polynomial to carefully selected continuum regions, and then dividing the spectrum by the polynomial.
The quantitative analysis proceeded by fitting synthetic spectra to the observations, by $\chi^2$ minimization, and simultaneously fitting for \teff, \lgg, \vs, microturbulence, and radial velocity.  Synthetic spectra were calculated using the {\sc Zeeman} spectrum synthesis program \citep{Landstreet1988-Zeeman1,Wade2001-zeeman2_etc}, which solves the polarized radiative transfer equations assuming Local Thermodynamic Equilibrium (LTE).  Further optimizations for negligible magnetic fields were used \citep{Folsom2012-HAeBe-abundances, Folsom2013-PhD-thesis}, and a Levenberg-Marquardt $\chi^2$ minimization algorithm was used.  

Atomic data were extracted from VALD, with an `extract stellar' request,  with temperatures approximately matching those we find for the stars (within 250 K).  
Model atmospheres from {\sc atlas9} \citep{Kurucz1993-ATLAS9etc} were used, 
which have a plane-parallel structure, assume LTE, and include solar abundances.  
For fitting \teff\, and  \lgg, a grid of model atmospheres was used (with a spacing of 250 K in \teff\ and of 0.5 in \lgg), and interpolated between (logarithmically) to produce exact models for the fit.  
The fitting was done on five independent spectral windows, each $\sim$100 \AA\ long, from 6000 \AA\ to 6700 \AA\ (6000-6100, 6100-6276, 6314-6402, 6402-6500, and 6600-6700 \AA).  Regions contaminated by telluric lines were excluded from the fit, as was the region around the H$\alpha$ Balmer line due to its ambiguous normalization in \'echelle spectra.  The averages of the results from the independent windows were taken as the final best fit values, and the standard deviations of the results were used as the uncertainty estimates.  An example of such a fit is provided in Fig.~\ref{spec-fitting-ex}, and the final best parameters are reported in Table \ref{fundimental-param-table}.

In the computation of synthetic spectra we assumed solar abundances, from \citet{Asplund2009-solar-abun}.  We checked this assumption for a few stars  (HII~739, TYC~6349-0200-1, and TYC~6878-0195-1) by performing a full abundance analysis simultaneously with the determination of the other stellar parameters.  Solar abundances were consistently found, thus we conclude that this is a sufficiently good approximation for our analysis. 

\begin{figure*}
\centering
\includegraphics[width=6.9in]{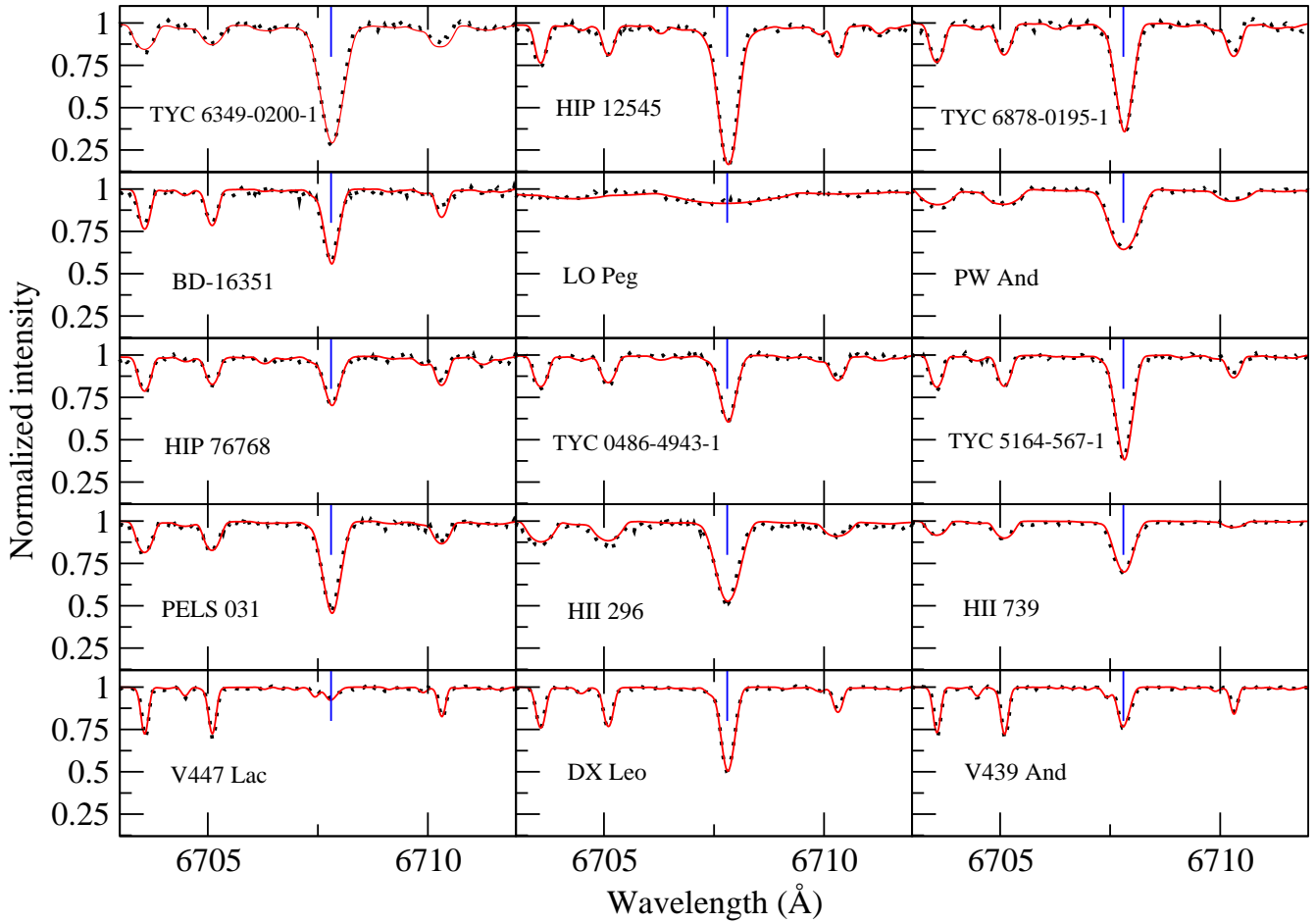}
\caption{Lithium line fits for the stars in this study, obtained with MARCS spectral synthesis.  Observations are dotted lines, the fits are solid lines and the Li line position is indicated by a thin vertical (blue) line.}
\label{fig-Li-fits}
\end{figure*}

\subsubsection{Secondary analysis}

For all stars in the sample, we performed a secondary spectral analysis, using spectral synthesis from the 1D hydrostatic MARCS models of stellar atmospheres \citep{Gustafsson2008-MARCS-grid}.  This analysis produced lithium abundances ($A_{\rm Li}$), in addition to \teff, \lgg, \vs, and microturbulence values.  
We used a grid of plane-parallel model atmospheres in LTE, with solar abundances. The grid has steps of 250 K in \teff\ and steps of 0.5 dex in \lgg\ (note that specific abundances as well as metallicity - [Fe/H] - can be adjusted precisely through the spectral synthesis). To produce the high-resolution synthetic spectra of the lithium line region (at 6707.8 \AA), we used the TurboSpectrum code \citep{Alvarez1998-TurboSpectrum-etc} and an interpolation routine for MARCS model structures kindly provided by Dr. T. Masseron (private communication).  Finally, we convolved the computed synthetic spectra with a Gaussian profile (in order to reproduce the instrumental profiles of ESPaDOnS and NARVAL), and by a rotational profile (to account for rotational velocity). Details of the complete method and of the detailed atomic and molecular line lists (initially extracted from the VALD database) can be found in \citet{CantoMartins2011-Li-abun-M67}.  Fits to the Li lines are presented in Fig.~\ref{fig-Li-fits}.  

This MARCS spectral synthesis analysis was done independently from the previously described analysis ({\sc Zeeman} spectral synthesis analysis). It then provided a crosscheck for all the stellar parameters produced through the {\sc Zeeman} spectral synthesis. Indeed \teff, \lgg, [Fe/H], \vs, and microturbulence velocity have also been determined using MARCS synthetic spectra (mainly from the lithium line region, with checks from regions around the Ca IR triplet and H$\beta$). The stellar parameters derived from MARCS spectral synthesis have been used for the $A_{\rm Li}$ determination.  A conservative accuracy of 0.15 dex has been adopted for the $A_{\rm Li}$ determinations, considering an accuracy of 50 K on \teff; 0.5 dex on \lgg; 0.15 dex on [Fe/H]; 0.5 \kms\ on \vs; 0.5 \kms\ on microturbulence velocity.

\subsubsection{Spectroscopic comparison}

In general, a good agreement for all the stellar parameters was found between the two approaches ({\sc Zeeman} vs MARCS spectral synthesis). Some specific cases, however, present discrepancies on some parameters. 
Our \vs\ measurements are consistent within $1\sigma$ for all stars except LO Peg, where we disagree at $1.9\sigma$ ($73.1 \pm 1.1$ and $67 \pm 3$ \kms).  The \vs\ value of \citet{Barnes2005-LOPeg-DI} ($65.84 \pm 0.06$ \kms) is smaller than both our values, but consistent with the MARCS value.  LO Peg has the broadest line profiles, and the line profiles most affected by spots.  Thus it is not surprising that our methods disagree slightly, due to the distorted line profiles.  Our measurements of microturbulence agree typically within $1\sigma$, and always within $1.5\sigma$.  Our \lgg\ measurements always agree within $1\sigma$. This suggests we may have overestimated the uncertainties on \lgg, however in the interest of caution we retain the current values.  Our \teff\ values generally agree within $1\sigma$, however there are a few significant disagreements.  Our results disagree at $3\sigma$ ($\sim$300 K: $6066 \pm 89$ and $5750 \pm 50$ K ) for HII 739.  This star is a spectroscopic binary, with a small radial velocity separation, and the lines of the two components largely superimposed.  Since the analyses use different selections of lines in different wavelength windows this produces different results.  We adopt the hotter \teff, obtained from the bluer part of the spectrum, as this is likely less influenced by contamination from the secondary.  There is a significant disagreement in our \teff\ for LO Peg, by $3.5\sigma$ (500 K: $4739 \pm 138$ and $5250 \pm 50$ K).  This is likely due to the large distortions to the line profiles by star spots, and blending of lines due to the large \vs.  We adopt the value based on the larger number of spectral lines, which should mitigate the impact of line profile distortions, and this value is consistent with the literature values \citep{Jeffries1994-LOPeg,BailerJones2011-stellar-params-bayseian, McCarthy2012-teff-l-size}. However, this difference may represent a real uncertainty on the \teff\ of LO Peg, due to its large spots.  For BD-16351 the \teff\ measurements differ by $1.8\sigma$ (210 K: $5211 \pm 109$ and $5000 \pm 50$ K), however there is no clear error with either value. \citet{daSilva2009-young-associations} find a \teff\ of 5083 K for BD-16351, approximately halfway between our values.  
Our \teff\ measurements for HII 296 differ by $2.1\sigma$ (236 K: $5236 \pm 101$ and $5000 \pm 50$ K), again with no clear errors in either value.  Several literature \teff\ measurements exist for HII 296, which fall in the range 5100 to 5200 \citep{Cayrel1990-teff-hii296, Cenarro2007-teff-hii296, Soubiran2010-teff-params, Prugniel2011-teff-param}, thus between our two values but favoring the higher value.  
The formal disagreements in \teff\ for BD-16351 and HII 296 are acceptable, and are likely a reflection of the real systematic uncertainties involved.  Ultimately we adopt the \teff\ values from the {\sc Zeeman} analysis, in order to provide a homogeneous set of values, since the values are largely consistent, and of comparable quality.

\begin{table*}
%\centering
\caption{Derived fundamental parameters for the stars in our sample.  $P_{\rm rot}$ are the adopted rotation periods, containing a mix of literature and our spectropolarimetric periods, as discussed in Appendix~\ref{Individual Targets}.  Radial velocities ($v_r$) are the averages and standard deviations of our observations.  Lithium abundances are in the form $\log(N_{\rm Li}/N_{\rm H})+12$.  }  
\begin{sideways}
\begin{tabular}{lccccccccc}
\hline\hline
Star           & Assoc.       & Age          & $P_{\rm rot}$             & \teff          & \lgg            & \vs            & $\xi$           &$v_r$              & $i$             \\ 
               &              & (Myr)        & (days)                  & (K)            &                 & (km/s)         & (km/s)          & (km/s)            & ($^{\circ}$)     \\ 
\hline
TYC 6349-0200-1&  $\beta$ Pic & $ 24 \pm  3$ & $    3.41 \pm     0.05$ & $4359 \pm 131$ & $4.19 \pm 0.31$ & $15.8 \pm 0.5$ & $1.4 \pm 0.3$ & $ -7.17 \pm 0.14$ & $52 ^{+20}_{-20}$ \\
HIP 12545      &  $\beta$ Pic & $ 24 \pm  3$ & $    4.83 \pm     0.01$ & $4447 \pm 130$ & $4.33 \pm 0.23$ & $10.2 \pm 0.4$ & $1.4 \pm 0.3$ & $  7.70 \pm 0.14$ & $39 ^{+20}_{-20}$ \\
TYC 6878-0195-1&  $\beta$ Pic & $ 24 \pm  3$ & $    5.70 \pm     0.06$ & $4667 \pm 120$ & $4.38 \pm 0.29$ & $11.2 \pm 0.4$ & $1.4 \pm 0.3$ & $ -8.62 \pm 0.09$ & $68 ^{+22}_{-20}$ \\
BD-16351       &      Columba & $ 42 \pm  6$ & $    3.21 \pm     0.01$ & $5211 \pm 109$ & $4.65 \pm 0.16$ & $10.2 \pm 0.3$ & $1.5 \pm 0.3$ & $ 11.05 \pm 0.11$ & $42 ^{+17}_{ -9}$ \\
LO Peg         &       AB Dor & $120 \pm 10$ & $0.423229 \pm 0.000048$ & $4739 \pm 138$ & $4.36 \pm 0.25$ & $73.1 \pm 1.1$ & $1.8 \pm 0.6$ & $-19.81 \pm 2.18$ & $45 ^{+ 3}_{ -3}$ \\
PW And         &       AB Dor & $120 \pm 10$ & $ 1.76159 \pm  0.00006$ & $5012 \pm 108$ & $4.42 \pm 0.18$ & $22.9 \pm 0.2$ & $1.7 \pm 0.4$ & $-11.11 \pm 0.48$ & $46 ^{+ 7}_{ -7}$ \\
HIP 76768      &       AB Dor & $120 \pm 10$ & $    3.70 \pm     0.02$ & $4506 \pm 153$ & $4.53 \pm 0.25$ & $10.1 \pm 0.6$ & $0.6 \pm 0.3$ & $ -6.87 \pm 0.38$ & $60 ^{+30}_{-13}$ \\
TYC 0486-4943-1&       AB Dor & $120 \pm 10$ & $    3.75 \pm     0.30$ & $4706 \pm 161$ & $4.45 \pm 0.27$ & $10.9 \pm 0.4$ & $1.1 \pm 0.5$ & $-19.95 \pm 0.04$ & $75 ^{+15}_{ -8}$ \\
TYC 5164-567-1 &       AB Dor & $120 \pm 10$ & $    4.68 \pm     0.06$ & $5130 \pm 161$ & $4.45 \pm 0.22$ & $ 9.6 \pm 0.3$ & $1.2 \pm 0.6$ & $-16.25 \pm 0.15$ & $65 ^{+25}_{-12}$ \\
HII 739        &     Pleiades & $125 \pm  8$ & $    1.58 \pm     0.01$ & $6066 \pm  89$ & $4.64 \pm 0.09$ & $14.8 \pm 0.3$ & $1.8 \pm 0.3$ & $  5.64 \pm 0.08$ & $51 ^{+20}_{-20}$ \\
PELS 031       &     Pleiades & $125 \pm  8$ & $     2.5 \pm      0.1$ & $5046 \pm 108$ & $4.59 \pm 0.17$ & $11.9 \pm 0.3$ & $1.6 \pm 0.3$ & $  6.17 \pm 0.31$ & $35 ^{+ 8}_{ -7}$ \\
HII 296        &     Pleiades & $125 \pm  8$ & $ 2.60863 \pm  0.00009$ & $5236 \pm 101$ & $4.33 \pm 0.16$ & $17.6 \pm 0.2$ & $1.6 \pm 0.3$ & $  6.37 \pm 0.20$ & $73 ^{+17}_{-20}$ \\
V447 Lac       &      Her-Lyr & $257 \pm 46$ & $  4.4266 \pm     0.05$ & $5274 \pm  74$ & $4.64 \pm 0.15$ & $ 4.6 \pm 0.3$ & $1.1 \pm 0.3$ & $ -7.37 \pm 0.04$ & $29 ^{+ 5}_{ -4}$ \\
DX Leo         &      Her-Lyr & $257 \pm 46$ & $   5.377 \pm    0.073$ & $5354 \pm  76$ & $4.71 \pm 0.12$ & $ 6.5 \pm 0.3$ & $1.3 \pm 0.3$ & $  8.53 \pm 0.08$ & $58 ^{+ 8}_{ -6}$ \\
V439 And       &      Her-Lyr & $257 \pm 46$ & $    6.23 \pm     0.01$ & $5393 \pm  71$ & $4.50 \pm 0.10$ & $ 4.6 \pm 0.3$ & $1.3 \pm 0.3$ & $ -6.38 \pm 0.03$ & $38 ^{+ 4}_{ -4}$ \\
\hline\hline
Star           & Assoc.       & $L$            & $R$             & $M$                & $\tau_{\rm conv}$        & Rossby            & $A_{\rm Li}$ \\ 
               &              & ($L_\odot$)     & ($R_\odot$)      & ($M_\odot$)         & (days)                 & number            &  (dex)    \\
\hline                          
TYC 6349-0200-1&  $\beta$ Pic & $0.30\pm 0.02$ & $0.96 \pm 0.07$ & $0.85^{+0.05}_{-0.05}$ & $50.5^{+14.4}_{- 6.6}$ & $ 0.07^{+0.01}_{-0.02}$ & $3.30 \pm 0.15$  \\
HIP 12545      &  $\beta$ Pic & $0.40\pm 0.06$ & $1.07 \pm 0.05$ & $0.95^{+0.05}_{-0.05}$ & $55.8^{+ 6.8}_{- 6.3}$ & $ 0.14^{+0.02}_{-0.02}$ & $3.40 \pm 0.15$  \\
TYC 6878-0195-1&  $\beta$ Pic & $0.80\pm 0.32$ & $1.37 \pm 0.28$ & $1.17^{+0.13}_{-0.21}$ & $59.9^{+23.5}_{-17.8}$ & $ 0.10^{+0.04}_{-0.03}$ & $2.45 \pm 0.15$  \\
BD-16351       &      Columba & $0.52\pm 0.21$ & $0.88 \pm 0.18$ & $0.90^{+0.07}_{-0.05}$ & $22.4^{+ 4.5}_{- 2.1}$ & $ 0.14^{+0.01}_{-0.02}$ & $2.25 \pm 0.15$  \\
LO Peg         &       AB Dor & $0.20\pm 0.01$ & $0.66 \pm 0.04$ & $0.75^{+0.05}_{-0.05}$ & $27.9^{+ 1.4}_{- 1.4}$ & $ 0.02^{+0.01}_{-0.01}$ & $2.55 \pm 0.15$  \\
PW And         &       AB Dor & $0.35\pm 0.14$ & $0.78 \pm 0.16$ & $0.85^{+0.05}_{-0.05}$ & $25.5^{+ 4.5}_{- 2.6}$ & $ 0.07^{+0.01}_{-0.01}$ & $2.85 \pm 0.15$  \\
HIP 76768      &       AB Dor & $0.27\pm 0.06$ & $0.85 \pm 0.11$ & $0.80^{+0.07}_{-0.05}$ & $39.7^{+11.6}_{- 9.4}$ & $ 0.09^{+0.03}_{-0.02}$ & $1.25 \pm 0.15$  \\
TYC 0486-4943-1&       AB Dor & $0.21\pm 0.08$ & $0.69 \pm 0.15$ & $0.75^{+0.05}_{-0.05}$ & $28.8^{+ 4.3}_{- 3.0}$ & $ 0.13^{+0.03}_{-0.03}$ & $1.77 \pm 0.15$  \\
TYC 5164-567-1 &       AB Dor & $0.50\pm 0.20$ & $0.89 \pm 0.19$ & $0.90^{+0.08}_{-0.05}$ & $24.8^{+ 7.5}_{- 4.1}$ & $ 0.19^{+0.04}_{-0.05}$ & $3.10 \pm 0.15$  \\
HII 739        &     Pleiades & $1.35\pm 0.20$ & $1.07 \pm 0.06$ & $1.15^{+0.06}_{-0.06}$ & $6.21^{+ 2.6}_{- 0.3}$ & $ 0.25^{+0.01}_{-0.08}$ & $3.00 \pm 0.15$  \\
PELS 031       &     Pleiades & $0.62\pm 0.06$ & $1.05 \pm 0.06$ & $0.95^{+0.05}_{-0.05}$ & $29.0^{+ 5.4}_{- 3.8}$ & $ 0.09^{+0.02}_{-0.02}$ & $2.80 \pm 0.15$  \\
HII 296        &     Pleiades & $0.49\pm 0.05$ & $0.93 \pm 0.05$ & $0.90^{+0.05}_{-0.05}$ & $20.1^{+ 1.0}_{- 1.0}$ & $ 0.13^{+0.01}_{-0.01}$ & $3.10 \pm 0.15$  \\
V447 Lac       &      Her-Lyr & $0.46\pm 0.02$ & $0.81 \pm 0.03$ & $0.90^{+0.05}_{-0.05}$ & $20.2^{+ 1.0}_{- 1.0}$ & $ 0.22^{+0.01}_{-0.01}$ & $1.95 \pm 0.15$  \\
DX Leo         &      Her-Lyr & $0.49\pm 0.02$ & $0.81 \pm 0.03$ & $0.90^{+0.05}_{-0.05}$ & $20.1^{+ 1.0}_{- 1.0}$ & $ 0.27^{+0.02}_{-0.02}$ & $2.65 \pm 0.15$  \\
V439 And       &      Her-Lyr & $0.64\pm 0.02$ & $0.92 \pm 0.03$ & $0.95^{+0.05}_{-0.05}$ & $17.9^{+ 0.9}_{- 0.9}$ & $ 0.35^{+0.02}_{-0.02}$ & $2.25 \pm 0.15$  \\
\hline\hline
\end{tabular} 
\end{sideways}
\label{fundimental-param-table} 
\end{table*}

%%%%%%%%%%%%%%%%%%%%%%%%%%%%%%%%%%%%%%%%%%%%%%%%%%%%%%%%%%%%%%%%%%%

\begin{table*}
%\centering
\caption{Literature fundamental parameters for the stars in our sample. 
Age references:  $^1$ \citet{Bell2015-young-assoc-ages}, $^2$ \citet{Luhman2005-ABDor-age} and \citet{Barenfeld2013-ABDor-memb-age}, $^3$ \citet{Stauffer1998-Pleiades-age-Li}, $^4$\citet{Lopez-Santiago2006-HerLyr-ABDor-assoc} and \citet{Eisenbeiss2013-HerLyr-age}.     
Distance references: $^5$ \citet{van_Leeuwen2007-Hipparcos_book}, $^6$ \citet{Torres2008-youngNearbyAssoc}, $^7$ \citet{vandenAncker2000-HD199143-assoc}, $^8$ \citet{Torres2006-search-youngAssocaitions}, $^{9}$ \citet{Montes2001-moving-groups}, $^{10}$ \citet{Melis2014-Pleiades-distance-VLBI}.  
}
\begin{tabular}{lcccc}
\hline\hline
Star           & Assoc.       & Age               & Distance                & Distance \\
               &              & (Myr)             & (pc)                    & Method   \\
\hline
TYC 6349-0200-1&  $\beta$ Pic & $ 24 \pm  3$ $^1$ & $45.7  \pm 1.6$ $^{5,7}$ & Assoc. Parallax\\ 
     HIP 12545 &  $\beta$ Pic & $ 24 \pm  3$ $^1$ & $42.0  \pm 2.7$ $^{5}$   & Parallax \\ 
TYC 6878-0195-1&  $\beta$ Pic & $ 24 \pm  3$ $^1$ & $79    \pm 16$  $^{8}$   & Dynamical \\ 
      BD-16351 &      Columba & $ 42 \pm  6$ $^1$ & $78    \pm 16$  $^{6}$   & Dynamical \\ 
        LO Peg &       AB Dor & $120 \pm 10$ $^2$ & $40.3  \pm 1.1$ $^{5}$   & Parallax \\ 
        PW And &       AB Dor & $120 \pm 10$ $^2$ & $30.6  \pm 6.1$ $^{9}$  & Dynamical \\ 
     HIP 76768 &       AB Dor & $120 \pm 10$ $^2$ & $40.2  \pm 4.4$ $^{5}$   & Parallax \\ 
TYC 0486-4943-1&       AB Dor & $120 \pm 10$ $^2$ & $71    \pm 14$  $^{6}$   & Dynamical \\ 
TYC 5164-567-1 &       AB Dor & $120 \pm 10$ $^2$ & $70    \pm 14$  $^{6}$   & Dynamical \\ 
       HII 739 &     Pleiades & $125 \pm  8$ $^3$ & $136.2 \pm 2.3$ $^{10}$& Assoc. Parallax \\ 
      PELS 031 &     Pleiades & $125 \pm  8$ $^3$ & $136.2 \pm 2.3$ $^{10}$& Assoc. Parallax\\ 
       HII 296 &     Pleiades & $125 \pm  8$ $^3$ & $136.2 \pm 2.3$ $^{10}$& Assoc. Parallax\\ 
      V447 Lac &      Her-Lyr & $257 \pm 46$ $^4$ & $46.4  \pm 0.5$ $^{5}$   & Parallax \\ 
        DX Leo &      Her-Lyr & $257 \pm 46$ $^4$ & $56.2  \pm 0.6$ $^{5}$   & Parallax \\ 
      V439 And &      Her-Lyr & $257 \pm 46$ $^4$ & $73.2  \pm 0.6$ $^{5}$   & Parallax \\ 
\hline\hline

\end{tabular} 
\label{litt-param-table} 
\end{table*}

\subsection{H-R diagram and evolutionary tracks}
\label{HRD_evolutionary_tracks}

\subsubsection{H-R diagram positions}
Absolute luminosities for the stars in this sample were derived from $J$-band photometry, from the 2MASS project \citep{Cutri2003-2MASS}.  Infrared photometry is preferable to optical photometry since it is likely impacted less by star spots.  This is because the brightness contrast between spots and the quiescent photosphere is less in the IR.  
The stars are all nearby, so interstellar extinction is likely negligible.  However, IR photometry further mitigates any possible impact of extinction.  
Finally, 2MASS provides a homogeneous catalog of data for the stars in our study.
The bolometric correction from \citet{Pecaut2013-PMS-BC-withJ} was used combined with our effective temperatures (from Sect.\ \ref{spectrum-fitting}).  \citet{Pecaut2013-PMS-BC-withJ} include results for the 2MASS $J$-band, and for pre-main sequence stars.  
We assume reddening is negligible, since the stars are all near the Sun ($<130$ pc).  From this we calculate the absolute luminosities, presented in Table~\ref{fundimental-param-table}.  

Six stars have precise, reliable Hipparcos parallax measurements. We use the re-reduction of the Hipparcos data by \citet{van_Leeuwen2007-Hipparcos_book}.  The star TYC 6349-0200-1 does not have a Hipparcos measurement, but it has a physical association with HD 199143 \citep{vandenAncker2000-HD199143-assoc}, which was observed by Hipparcos.  Therefore we use the parallax of HD 199143 for TYC 6349-0200-1 \citep[e.g.][]{Evans2012-youngMovingGroups}.  
For the Pleiades, there has been a longstanding disagreement regarding the distance, particularly between the Hipparcos parallax of \citet{vanLeeuwen2009-cluster-distances} and the HST trigonometric parallax of \citet{Soderblom2005-Pleiades-distance}.  The recent VLBI parallax of \citet{Melis2014-Pleiades-distance-VLBI} strongly supports the HST value, thus we adopt their value and take the uncertainties on stellar distances as their estimate of the dispersion in cluster depth.  The difference between these distances is $\sim$10\%, and this degree of uncertainty has no major impact on our results.  
For the other stars we use the dynamical distances, mostly from \citet{Torres2008-youngNearbyAssoc}, supplemented by \citet{Montes2001-moving-groups} and \citet{Torres2006-search-youngAssocaitions}.  These dynamical distances are based on the proper motion of a star, and are the distance to the star that makes its real space velocity closest to that of its association's velocity.  For these dynamical distances we adopt a 20\% uncertainty, which is a conservative assumption since the authors do not provide uncertainties.  The adopted distances are included in Table \ref{litt-param-table}.  

\begin{figure}
\centering
\includegraphics[width=3.3in]{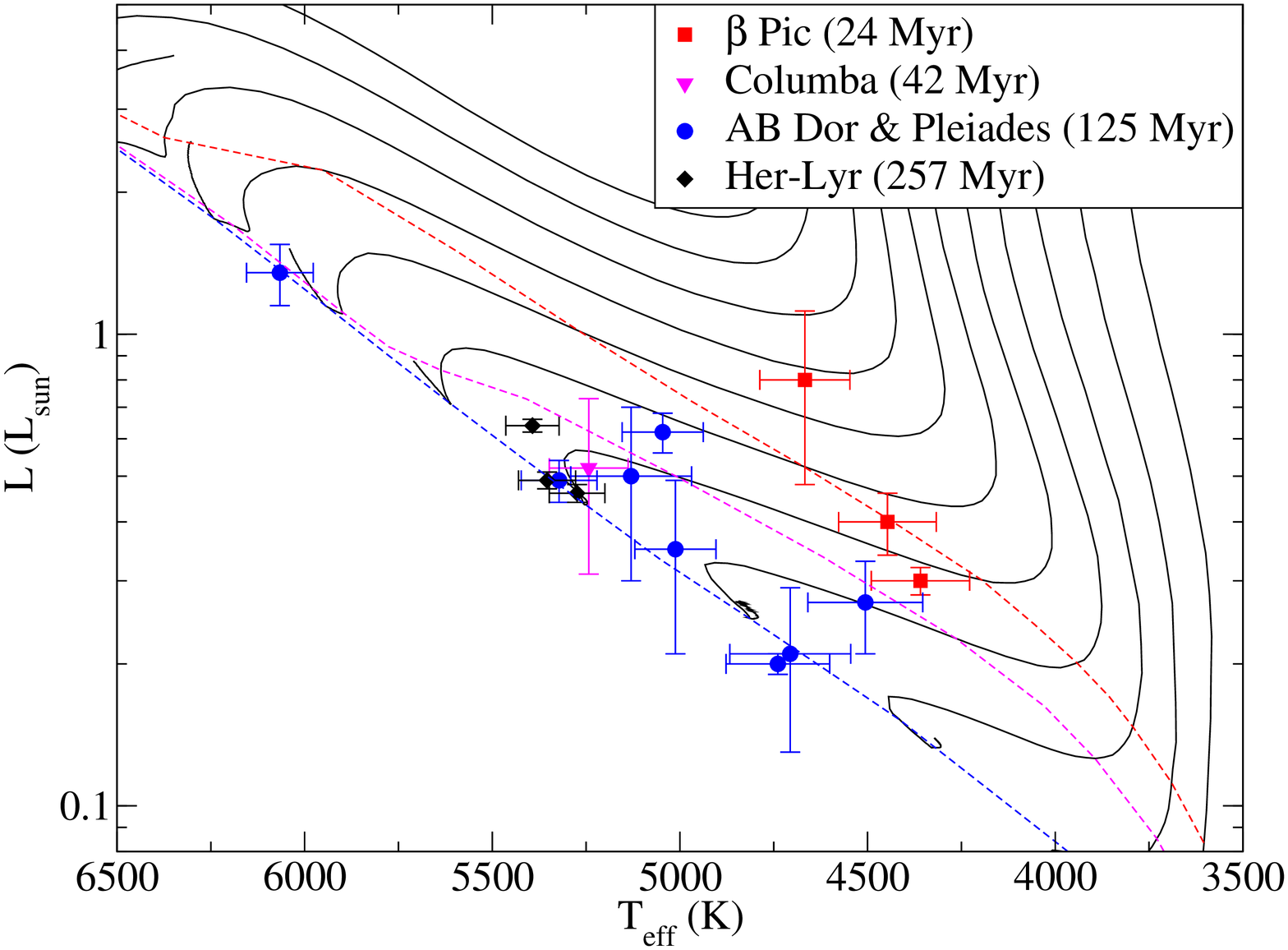}
\caption{H-R diagram for the stars in this study.  
Evolutionary tracks (solid lines) are from Amard et al.\ (in prep.), and are shown for 0.1\Msun\ increments from 0.5 to 1.5 \Msun.  Isochrones are shown for 24 Myr ($\beta$ Pic), 42 Myr (Columba), and the ZAMS.  
Stars grouped by association and age, as indicated. }
\label{hr-diagram}
\end{figure}

With absolute luminosities and effective temperatures, we can infer stellar radii from the Stefan-Boltzmann law.  
We can also use this information to place the stars on an H-R diagram, shown in Fig.~\ref{hr-diagram}.  By comparison with theoretical evolutionary tracks we can estimate the masses of the stars.  

\subsubsection{Evolutionary models}

We used a grid of evolutionary tracks to be published in Amard et al.\ (in prep.). Standard stellar evolution models with masses from 0.5 to 2 M$_\odot$ were computed with the STAREVOL V3.30 stellar evolution code. The adopted metallicity is Z = 0.0134, which corresponds to the solar value when using the \citet{Asplund2009-solar-abun} reference solar abundances. The adopted mixing length parameter $\alpha_{MLT} = 1.702$ is obtained by calibration of a classical (without microscopic diffusion) solar model that reproduces the solar luminosity and radius to $10^{-5}$ precision at 4.57 Gyr. The models include a non-grey atmosphere treatment following \cite{KrishnaSwamy1966}. Mass loss is accounted for starting at the ZAMS following \citet{Reimers1975}. The convective boundaries are fixed by the Schwarzschild criterion, and the local convective velocities are given by the mixing length theory.  This allows us to compute the convective turnover timescale at one pressure scale height above the base of the convective envelope, for each timestep: 
\begin{equation}
\tau_{H_p} = \alpha H_p(r)/V_c(r)
\end{equation}
where $V_c(r)$ is the local convective velocity as given by the mixing length theory formalism at one pressure scale height above the base of the convective envelope, $H_p(r)$ is the local pressure scale height, and $\alpha$ is the mixing length parameter.  This choice of convective turnover timescale is discussed in Appendix \ref{Convective turnover times}.

In order to derive an estimate of the masses and convective turnover timescales (and hence Rossby numbers) for the stars in our sample, we use the maximum likelihood method described in \citet{Valle2014-stellar-params-from-evol-tracks}.  We based our estimates upon \teff\ and Luminosity, and the associated error bars derived from our analysis. 

\subsubsection{Comparison to association isochrones}

From these model evolutionary tracks we computed isochrones for the age of each association.  
Comparing the observed stars positions on the H-R diagram with the model association isochrones, we find that the H-R diagram positions are consistent with the adopted ages for most stars.  This supports the association memberships of those stars.  
However a few stars disagree with their isochrones by more than $2\sigma$, suggesting unrecognized systematic errors, or underestimated uncertainties.  
HII 739 appears to sit well above the association isochrone (at the ZAMS for that \teff), but it is a binary.  HIP 12545 sits somewhat below the association isochrone by slightly more than $2\sigma$, possibly suffering some extinction, but still is well above the ZAMS.  HIP 76768 sits marginally above its isochrone (at the ZAMS), but by less than $2\sigma$.  PELS 031 also sits above its isochrone, by $2\sigma$, however there is no clear evidence that it is a spectroscopic binary.  These discrepancies are not due to metallicity, since the members of one association should have the same metallicity, and there is no observational evidence for significantly non-solar metallicities in our sample.  For the most discrepant cases, HII 739 and HIP 12545, we derive the stellar parameters using \teff\ and the age of their association, rather than luminosity.  For these two stars we also adopt the radii from the evolutionary tracks rather than Stefan-Boltzmann law. 

We use the masses derived from the H-R diagram and the stellar radii to calculate a \lgg\ for the stars.  Comparing this with the spectroscopic \lgg\ we derived earlier shows that the values are consistent. 
For the binary HII 739, this is only true if we use values based on \teff\ and age, rather than luminosity.  
Many of the \lgg\ values from evolutionary radii and masses are formally more precise than the spectroscopic values, however we prefer the spectroscopic values for this study as they have fewer potential sources of systematic uncertainty.

\section{Spectropolarimetric analysis}

\subsection{Longitudinal magnetic field measurements}
\label{longitudinal-magnetic}

Measurements of the longitudinal component of the magnetic field, averaged across the stellar disk, were made from all the individual LSD profiles.  This provides much less information than a full ZDI map, but it depends much less on other stellar parameters (e.g. rotation period, inclination, \vs).  The longitudinal magnetic field was measured using the first order moment method \citep[e.g.][]{Rees1979-magnetic-cog}, by integrating the (continuum normalized) LSD profiles $I/I_{\rm c}$ and $V/I_{\rm c}$ about their center-of-gravity ($v_{\rm 0}$) in velocity ($v$): 
\begin{equation}
B_{\ell}=-2.14\times 10^{11}\ \frac{{\displaystyle \int (v-v_{\rm 0}) V(v)\ dv}}{\displaystyle {\lambda g_{f} c\ \int [1-I(v)]\ dv}}. 
\label{bz-equation}
\end{equation}
Here the longitudinal field $B_{\ell}$ is in Gauss, $c$ is the speed of light, and $\lambda$ (the central wavelength, expressed in nm) and $g_{f}$ (the Land\'e factor) correspond to the normalization values used to compute the LSD profiles (see Sect.~\ref{Least squares deconvolution}).  
The integration range used to evaluate the equation was set to include the complete range of the absorption line in $I$, as well as in $V$. 
The resulting measurements of $B_{\ell}$ are summarized in Table \ref{table-mag-param} and plotted, folded with the stellar rotation periods (see Sect. \ref{rotational-period}), in Figs.~\ref{fig-bz} and \ref{fig-bz2}.

The longitudinal magnetic fields, which vary due to rotational modulation, were used to determine a rotation period for each star.  This was done by generating periodograms, using a modified Lomb-Scargle method, for each star.  The periodograms were generated by fitting sinusoids through the data using a grid of periods, thus producing periodograms in period and $\chi^2$.  The sinusoids were in the form:
\begin{equation}
\sum_{l=0}^{n} a_{l} \sin^{l}(p+\phi),
\end{equation}
where $n$ is the order of the sinusoid used, $a_{l}$ and $\phi$ are free parameters, and $p$ is the period for each point on the grid. This formalism has the advantage of easily accounting for magnetic fields with significant quadrupolar or octupolar components, and reduces to a Lomb-Scargle periodogram when $n=1$.  Searches for a period began with $n=1$, and if no adequate fit to the observations could be obtained, the order was increased to a maximum of $n=3$.

These results usually produced well defined minimum in $\chi^2$, often with harmonics at shorter periods.  However the results for five stars were ambiguous (HIP 12545, PELS 031, HII 296, HII 739, and V447 Lac), with comparable $\chi^2$ minima at multiple periods.  For stars with a well defined minimum, we can use the change in $\chi^2$ from that minimum to provide uncertainties on the period \cite[e.g.][]{numerical-recipes-Fortran}.  The periods found here typically have less precision than the literature photometric period estimates.  This is a consequence of the relatively short time-span over which the observations were collected (typically one to two weeks).

For three stars, our adopted period (see Sect.~\ref{rotational-period} and Appendix \ref{Individual Targets}) produced a particularly poor phasing of the $B_{\ell}$ curves.  For LO Peg, the rotation period is well established both by \citet{Barnes2005-LOPeg-DI} and by our ZDI results.  While there is a lot of apparent scatter in the $B_{\ell}$ curve, the first order fit produces $\chi^2_\nu = 1.1$.  The $B_{\ell}$ curve of HIP 12545 appears, by eye, to indicate a harmonic of the true period.  However examining the phasing of LSD profiles shows a consistent phasing at this period and inconsistent phasings at the possible alternatives, thus this must be the correct period.  
For HII 739, there is very little variation in the longitudinal field curve, with the exception of two observations obtained 10 days earlier than the rest of the data.  There is much clearer variability in the LSD $V$ profiles, and modeling those is what the period was principally based on, however noise is a limiting factor in our analysis of this star.  

We find a wide range of longitudinal magnetic fields.  The strongest star reaches a $B_{\ell}$ of 150 G, most stars at most phases have $B_{\ell}$ of a few tens of gauss, and a few stars have $B_{\ell}$ below 10 G at many phases, as summarized in Table \ref{table-mag-param}.  We consider the maximum observed $B_{\ell}$ as a proxy for the stellar magnetic field strength, to mitigating geometric effects and rotational variability \citep[similar to, e.g.,][]{Marsden2014-Bcool-survey1}.  In the maximum $B_{\ell}$, two stars exceed 100 G, four stars are below 20 G, and the median is $\sim$50 G.  
There is an approximate trend of weakening maximum $B_{\ell}$ with age, the youngest ($\sim$20 Myr) stars have stronger fields than the oldest ($\sim$257 Myr), however there is a very large scatter for the intermediate age stars ($\sim$120 Myr).  There is also a weak trend in rotation rate, with the fastest rotating stars having the strongest fields.  A clearer correlation in decreasing $B_{\ell}$ with Rossby number is found. 
More detailed magnetic results, accounting for magnetic and stellar geometry, are presented in Sect.~\ref{ZDI}.

\begin{figure}
  \centering
  \includegraphics[width=3.0in]{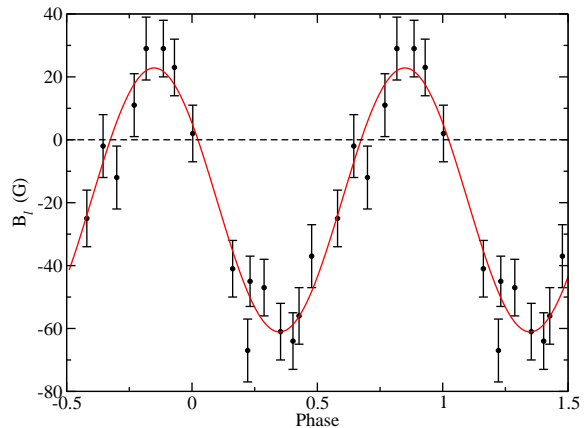}
  \caption{Longitudinal magnetic field measurements for TYC 6349-200-1, phased with the rotation periods derived in Sect.~\ref{rotational-period}.  The solid line is the fit through the observations.  Figures for the full sample can be found in Appendix \ref{Individual Targets}. }
  \label{fig-bz-sample}
\end{figure}

\subsection{Radial velocity}
\label{Radial velocity}

Radial velocities for each observation were measured from the LSD profiles.  These values were measured by fitting a Gaussian line profile to the Stokes $I$ LSD profile, which effectively uses the Gaussian fit to find the centroid of the profile.  The radial velocity variability observed is not a real variation in the velocity of the star (with the partial exception of close binaries), but rather due to distortions in the line profile from spots on the stellar surface.  These distortions are modulated with the rotation period of the star, and thus the apparent radial velocity variation can be used to measure the rotation period of the star.  

\begin{figure}
  \centering
  \includegraphics[width=3.0in]{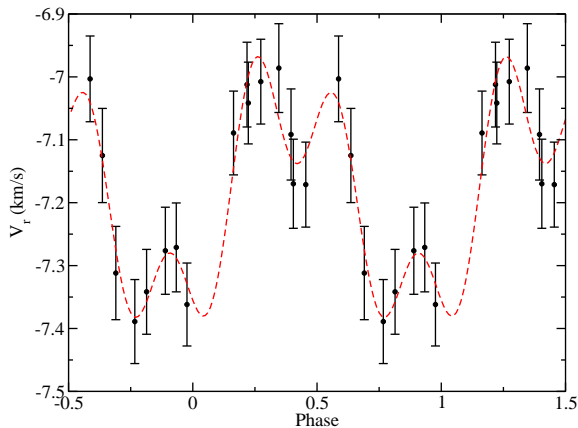}
  \caption{Radial velocity measurements for TYC 6349-200-1, phased with the rotation period derived in Sect.~\ref{rotational-period}.  The dashed line is a second order sine fit through the observations.}
  \label{fig-vr}
\end{figure}

The same modified Lomb-Scargle analysis method that was used to derive a period from the longitudinal field measurements was applied to the radial velocity measurements.  See Fig.~\ref{fig-vr} for an example.  The surface spot distribution is generally more complex than the large-scale magnetic field distribution, and consequently the results of this analysis were more ambiguous than they were for the magnetic field analysis.  Six stars displayed periodograms with multiple ambiguous $\chi^2$ minima (HIP 12545, BD-16351, PELS 031, HII 296, HII 739, and V447 Lac).  However, the periods found from the magnetic analysis were always consistent with one of the stronger minima from radial velocity.  Thus the apparent radial velocity variability supports periods based on the longitudinal magnetic fields, but is not always sufficient to determine a period on its own.  

The observation that radial velocity variability is more ambiguous and less sensitive to rotation periods than magnetic variability has important implications for surveys aimed at characterizing planets around active stars.  This implies that spectropolarimetric observations are much more useful for characterizing the stellar part of the variability than simple spectroscopic observations.

\subsection{Rotation period}
\label{rotational-period}
All of the stars in our sample have literature rotation periods.  However, there is the potential for large systematic errors in these periods.  For example, nearly identical spot distributions on either side of a star can cause a period to be underestimated by a factor of 2.  Indeed a few of the stars in our sample have conflicting literature periods.  Having an accurate rotation period is critical for an accurate ZDI map so we verified and, when possible, re-derived rotation periods for all the stars in the sample.  

Our period search used three methods.  The first technique was based on longitudinal magnetic field measurements, and is described in Sect. \ref{longitudinal-magnetic}.  This method has the advantage of being largely model independent, however the method loses sensitivity for more complex magnetic geometries and for largely toroidal magnetic geometries.  The second method was based on apparent radial velocity variability, as a measure of line profile variability due to spots (c.f.\ Sect.~\ref{Radial velocity}).  Most of the stars are only weakly spotted, and thus in many cases this variability is only weakly detected.  Furthermore the variability in apparent radial velocity is usually more complex than the variability in the longitudinal magnetic field.  Thus the radial velocity variability was only used to confirm the rotation period measurements from the other two methods.  The third method was based on the ZDI analysis, searching for a period that produced the maximum entropy ZDI map \citep[e.g.][]{Petit2008-sunlike-mag-geom}.  
Details of the ZDI procedure are given in Sect. \ref{ZDI}.  Since we use the fitting routine of \citet{Skilling1984-max-entropy-regularisation}, entropy rather than $\chi^2$ is the correct parameter to optimize.  
This rotation period search starts with a grid of rotation periods, and for each period recomputes the phases of the observations, then runs ZDI.  From this we produce a plot of entropy as a function of rotation period, and select the period that maximizes entropy.  This method of period searching is more model dependent than the method based on longitudinal magnetic fields.  However, this method is more sensitive when the magnetic field geometry is complex, since it consistently models all the information available in the Stokes $V$ profiles.  

In most cases, all the period estimates agree with the literature values.  In a few cases, TYC 0486-4943-1, HII 739, and PELS 031, the literature periods were inconsistent with our observations.  The final rotation periods we adopt are in Table \ref{fundimental-param-table}.  Detailed discussions of our analyses and comparisons with literature values for all the stars are given in Appendix \ref{Individual Targets}.

Emission indices were calculated for the Ca {\sc ii} H and K lines, H$\alpha$, and the Ca infrared tripled, summarized in Appendix \ref{Emission indices}.  These quantities usually vary coherently with rotation phase, but do not do so in a simple fashion, and thus were not used for period determination.  

Values for the inclination of the rotation axis with respect to the line of sight were derived using two methods.  When possible, the value was based on \vs\ and the combination of radius and period to determine an equatorial velocity ($v_{\rm eq}$).  However, in a few cases the radius was poorly constrained, either due to an uncertain magnitude in a binary system, or due to an uncertain distance to the star.  In these cases a second method was used, looking for the inclination that maximizes entropy in ZDI.  This was done in the same fashion as the ZDI period search, searching a grid of inclinations and selecting the one with the maximum entropy ZDI solution.  This ZDI based inclination was checked against the inclination from \vs, radius and period, for cases with well defined radii, and a good agreement was consistently found.  The adopted best inclinations are given in Table \ref{fundimental-param-table}, and in cases where the ZDI inclination was used, a discussion of the inclination determination is provided in Appendix \ref{Individual Targets}.

\section{Magnetic mapping}
\label{ZDI}

\begin{figure}
  \centering
  \includegraphics[width=3.0in]{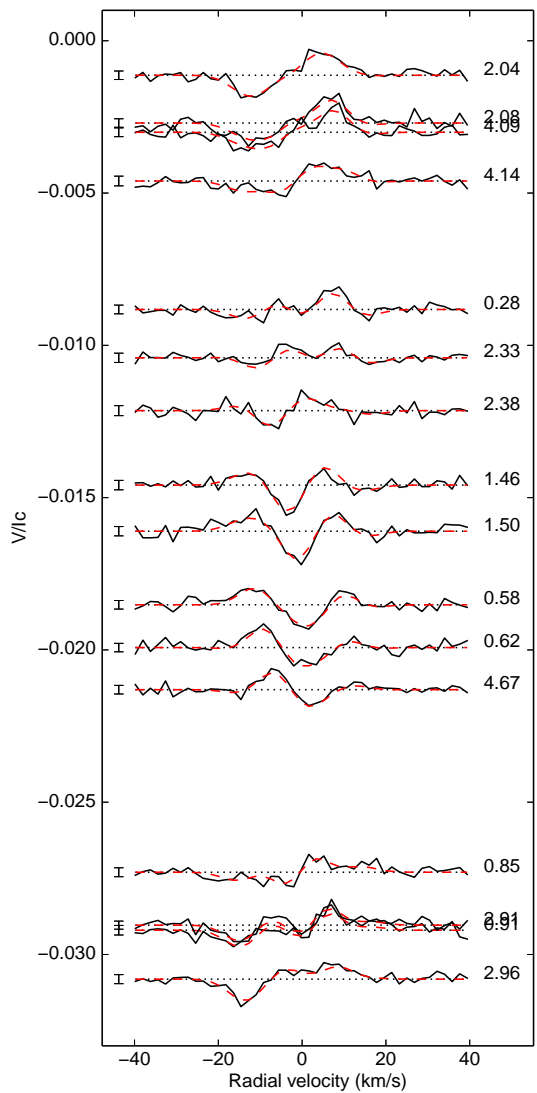}
  \caption{Sample ZDI fit for TYC 6349-200-1.  The solid lines are the observed LSD $V/I_c$ profile and the dashed lines are the fits.  The profiles are shifted vertically according to phase and labeled by rotation cycle.  }
  \label{fig-sample-zdi-fit}
\end{figure}

\subsection{ZDI Model description}

Zeeman Doppler imaging was used to reconstruct surface magnetic field maps for all the stars in this study.  ZDI uses the observed rotationally modulated Stokes $V$ line profiles, and inverts the time-series of observations to derive the magnetic field necessary to generate them.  
We used the ZDI method of \citet{Semel1989-ZDI-intro}, \citet{Donati1997-ZDI-tests} and \citet{Donati2006-tauSco}, which represents the magnetic field as a combination of spherical harmonics, and uses the maximum entropy regularization procedure described by \citet{Skilling1984-max-entropy-regularisation}.  ZDI was performed using the Stokes $V$ LSD profiles, which was necessary to provide sufficiently large S/N. 

ZDI proceeds by iteratively fitting a synthetic line profile to the observations, subject to both $\chi^2$ and the additional constraint provided by regularization, using the spherical harmonic coefficients that describe the magnetic field as the free parameters.  Therefore the model line profile used is of some importance.  
We calculated the local Stokes $V$ line profile, at one point on the stellar surface, using the weak field approximation: 
\begin{equation}
V(\lambda) = - \lambda_{o}^{2} \frac{g_{f} e}{4\pi m_{e} c}  B_{l} dI/d\lambda ,
\end{equation}
where $\lambda_{o}$ is the central wavelength of the line, $g_{f}$ is the mean Land\'e factor, and $B_{l}$ is the line of sight component of the magnetic field.  The local Stokes $I$ profile is approximated as a continuum level minus a Gaussian.  The local line profiles are then weighted by the projected area and brightness of their surface element, calculated using a classical limb darkening law, and Doppler shifted by their rotational velocity.  The local profiles are summed, and finally normalized by the sum of their continuum levels and projected cell areas to produce a final disk integrated line profile.  In this study we have assumed surface brightness variations due to star spots are negligible.  While these brightness variations are detectable in some Stokes $I$ profiles, they are small (1 to 5\% of the line depth), and the impact of this variability would be lost in the noise of the observed Stokes $V$ profiles ($V$ profiles are typically observed with $\sim 5\sigma$ precision per pixel).  

Since we are modeling LSD profiles, the mean Land\'e factor and central wavelength used for the model line were set to the normalization values used for LSD \citep{Kochukhov2010-LSD}.  The width of the Gaussian line profile was set empirically by fitting the line width of the very slow rotator $\epsilon$ Eri \citep{Jeffers2014-epsEri-mag-var}, which has a spectral type typical of the stars in our sample (K2).  LSD was applied to the spectrum of $\epsilon$ Eri with the same normalization and line masks as used for the rest of the stars in this study.  The ZDI model $I$ line profile was then fit to the LSD $I$ line profile of the star, providing a Gaussian line width.  We also checked the line width used against theoretical models. A synthetic line profile was calculated using the {\sc Zeeman} spectrum synthesis program for a star with \teff~$=5000$ K, \lgg~$=4.5$, and 1.2 \kms\ of microturbulence, approximately average for the stars in this study, but no rotation.  The atomic data for this model line were taken to be the average of the atomic data used for the lines in the LSD line mask.  Line broadening included the quadratic Stark, radiative, and van der Walls effects, as well as thermal Doppler broadening.  
This detailed synthetic line profile was then fit with the Gaussian ZDI line model, to find a theoretical best width for the ZDI line profile.  
Good agreement was found, with the theoretical width and the empirical width from $\epsilon$ Eri differing by less than 10\%, 
thus a full width at half maximum of 7.8 \kms\ ($1\sigma$ width of 3.2 \kms) 
was adopted for the ZDI model line.  Line depths were set individually for each star in the study, by fitting the central depth of the ZDI Stokes $I$ line to the central depth of the average LSD $I$ line profile. 

For the stellar model we used in ZDI, \vs\ was taken from the spectroscopic analysis in Sect.~\ref{spectrum-fitting}.  A linear limb darkening law was used with a limb darkening coefficient of 0.75, typical of a K star at our model line wavelength \citep{Gray2005-Photospheres}.  The ZDI maps are largely insensitive to the exact value of the limb darkening parameter \citep[e.g.][]{Petit2008-sunlike-mag-geom}.  The inclination of the stellar rotation axis to the line of sight was determined from stellar radius and \vs\ from Sect.~\ref{fund-param} and the rotation period used for the star was derived in Sect.~\ref{fund-param}.  
Differential rotation was assumed to be negligible, however this will be investigated further in the next paper in this series.  The exception to this is LO Peg, where a reliable literature differential rotation measurement exists.  

The model star was calculated using 2000 surface elements of approximately equal area, and the spherical harmonic expansion was carried out to the 15th order in $l$.  For our spectral resolution and local line with, and a typical \vs\ of 10 \kms, \cite{Morin2010-Mdwarfs} suggest that only the first $\sim$8 harmonics should carry any useful information (at a \vs\ of 15 \kms\ that becomes the first 10 harmonics).  This matches our results, as in all cases the magnetic energy in the coefficients drops rapidly by fifth order, often sooner, with the coefficients being driven to zero by the maximum entropy regularization.  

\subsection{ZDI Results}

A sample ZDI fit is presented in Fig.\ \ref{fig-sample-zdi-fit}.  The resulting magnetic maps are presented in Figs.\ \ref{fig-zdi-maps} and \ref{fig-zdi-maps2}.  Several parameters describing the magnetic strength and geometry are given in Table \ref{table-mag-param}.  The mean magnetic field ($\langle B \rangle$) is the global average strength of the (unsigned) large-scale magnetic field over the surface of the star (i.e. the magnitude of the magnetic vector averaged over the surface of the star).  The field is broken into poloidal and toroidal components \citep[as in][]{Donati2006-tauSco}, into axisymmetric ($m=0$ spherical harmonics) and non-axisymmetric components, and the fraction of the magnetic energy (proportional to $B^2$) in different components is given.  Note that in some cases, the values are considered as fractions of the total magnetic energy, and in some cases they are fractions of one component, such as the fraction of poloidal energy in the dipolar mode.

We find a wide range of magnetic strengths and geometries.  Mean magnetic field strengths vary from 14 to 140 G, with some dependence on age and rotation rate.  The magnetic field geometries vary from largely poloidal to largely toroidal.  The majority (12/15) of the stars have the majority of their magnetic energy in poloidal modes, however there are significant toroidal components found in many stars.  There is a large range of observed axisymmetry, and the majority of the stars have the majority of their energy in non-axisymmetric components.  There is also a wide range of complexity (dominant $l$ order) to the fields.  However, none of the observed stars are entirely dipolar or entirely axisymmetric.  This diversity of magnetic properties is qualitatively typical of stars with radiative cores and convective envelopes \citep[e.g.][]{Donati2009-ARAA-magnetic-fields}.

\begin{table*}
\centering
\caption{Derived magnetic properties for the stars in our sample.  The maximum disk integrated longitudinal magnetic field is in column 2, and the amplitude of variability in the longitudinal field is in column 3.  The surface averaged large-scale magnetic field strength from the ZDI map is in column 4, and the maximum field value from the ZDI map is in column 5.  The remaining columns present the percent of the magnetic energy in different components of the field.  }
\begin{sideways}
\begin{tabular}{lcccccccccccccc}
\hline\hline
Star & $B_{l, {\rm max}}$ & $B_{l, {\rm range}}$ & $\langle B \rangle$ & $B_{\rm peak}$ & pol. & tor.  & dip.  & quad.  & oct. & axisym. & axisym.  & axisym. & axisym. \\
     & (G) & (G) & ZDI (G) & ZDI (G) & (\%tot) & (\%tot) & (\%pol) & (\%pol) & (\%pol) & (\%tot) & (\%pol) & (\%tor) & (\%dip) \\ 
\hline
TYC 6349-0200-1 &  65 &  95 &   59.8 &  184.6 &  77.9 &  22.1 &  68.2 &   8.5 &   6.3 &  30.2 &  22.6 &  57.1 &  28.9 \\
HIP 12545       &  68 &  40 &  115.7 &  418.4 &  57.0 &  43.0 &  57.0 &  16.4 &  11.6 &  59.6 &  39.6 &  86.0 &  64.6 \\
TYC 6878-0195-1 &  58 &  92 &   55.3 &  198.2 &  69.0 &  31.0 &  71.5 &  10.7 &   6.3 &  36.0 &  17.8 &  76.6 &  21.1 \\
BD-16351        &  59 & 116 &   49.0 &  209.3 &  61.8 &  38.2 &  55.2 &  24.9 &   8.9 &  41.1 &   8.9 &  93.1 &  10.5 \\
LO Peg          & 150 & 150 &  139.6 &  793.4 &  63.2 &  36.8 &  35.4 &   9.4 &   7.3 &  42.5 &  38.9 &  48.6 &  80.4 \\
PW And          & 125 & 175 &  125.8 &  503.6 &  75.7 &  24.3 &  60.4 &  11.8 &   4.4 &  25.6 &  17.0 &  52.4 &  24.8 \\
HIP 76768       &  80 &  66 &  112.8 &  400.7 &  37.4 &  62.6 &  78.5 &   5.3 &   6.7 &  83.7 &  69.3 &  92.3 &  84.1 \\
TYC 0486-4943-1 &  28 &  48 &   25.0 &   71.5 &  75.7 &  24.3 &  39.3 &  27.9 &  20.2 &  24.8 &  11.5 &  66.4 &   6.5 \\
TYC 5164-567-1  &  43 &  46 &   63.9 &  145.5 &  89.4 &  10.6 &  74.2 &  10.4 &   8.5 &  67.3 &  70.6 &  40.0 &  88.8 \\
HII 739         &  15 &  28 &   15.4 &   48.3 &  71.1 &  28.9 &  26.8 &  16.5 &  23.7 &  26.7 &  19.1 &  45.2 &  51.1 \\
PELS 031        &  26 &  37 &   44.1 &  124.2 &  31.2 &  68.8 &  21.5 &  15.9 &  17.9 &  75.6 &  31.6 &  95.6 &  58.2 \\
HII 296         &  50 &  50 &   80.4 &  273.6 &  89.5 &  10.5 &  62.7 &  14.7 &  10.8 &  42.1 &  42.8 &  35.9 &  60.1 \\
V447 Lac        &  13 &  14 &   39.0 &   98.8 &  15.0 &  85.0 &  29.9 &  35.2 &  21.6 &  92.3 &  58.0 &  98.4 &  75.3 \\
DX Leo          &  18 &  36 &   29.1 &   83.0 &  83.1 &  17.0 &  71.2 &  16.1 &   4.6 &   8.0 &   1.9 &  37.8 &   2.0 \\
V439 And        &  13 &  11 &   13.9 &   40.2 &  59.4 &  40.6 &  72.3 &  16.2 &   5.2 &  80.0 &  73.0 &  90.4 &  83.3 \\
\hline\hline
\end{tabular} 
\end{sideways}
\label{table-mag-param} 
\end{table*}

\section{Discussion}

This discussion focuses on the large-scale magnetic properties of our sample, and trends in those properties with the physical parameters of age, rotation period, mass, and Rossby number.  We then compare our results to those for younger T Tauri stars and older field stars, in order to provide a synthetic description of the magnetic evolution of solar-type stars from the early pre-main sequence to the end of the main sequence.  

\subsection{Magnetic trends in young stars}
\label{Magnetic trends in young stars}

\begin{figure*}
  \centering
  \includegraphics[width=3.3in]{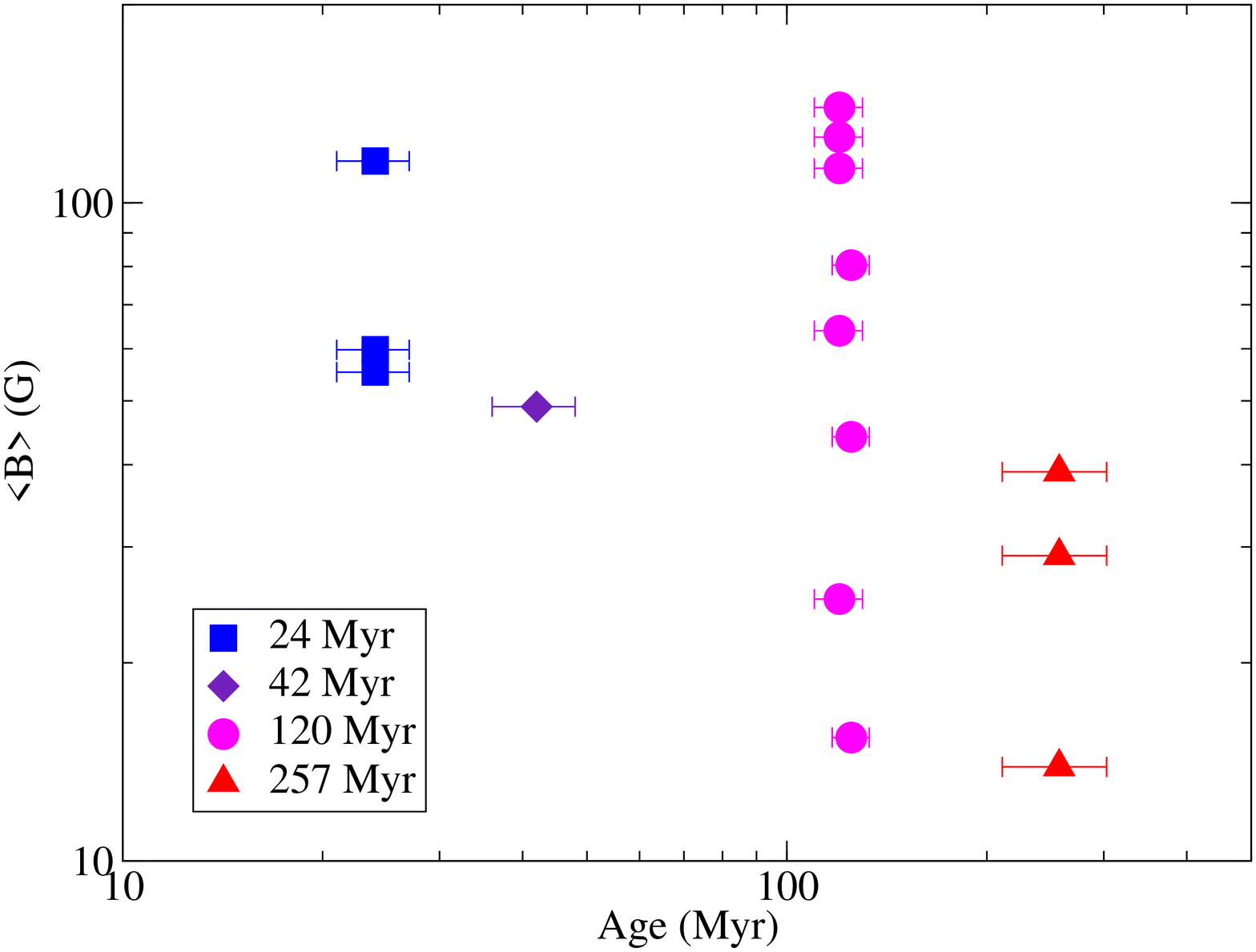}
  \includegraphics[width=3.3in]{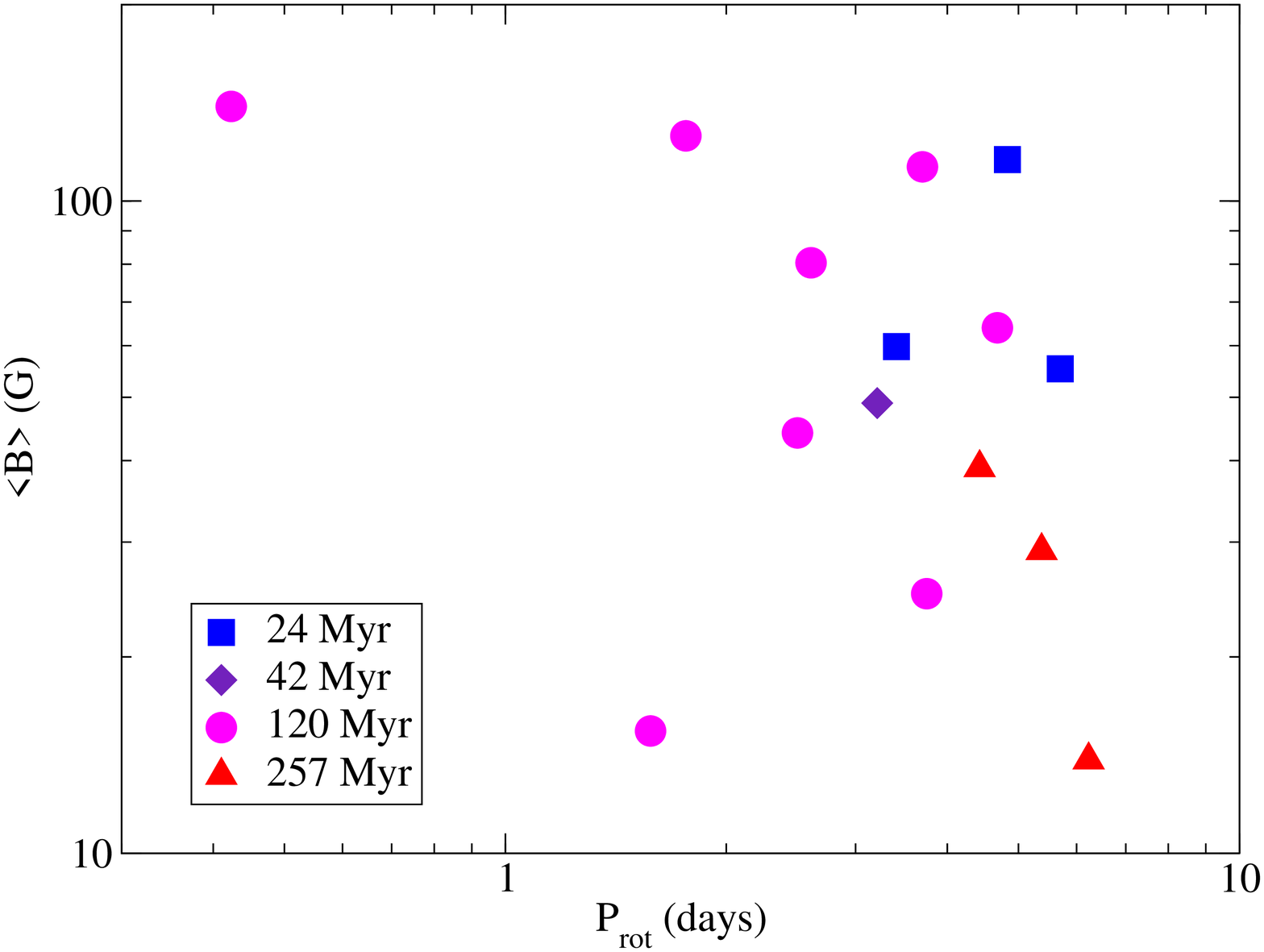}
  \includegraphics[width=3.3in]{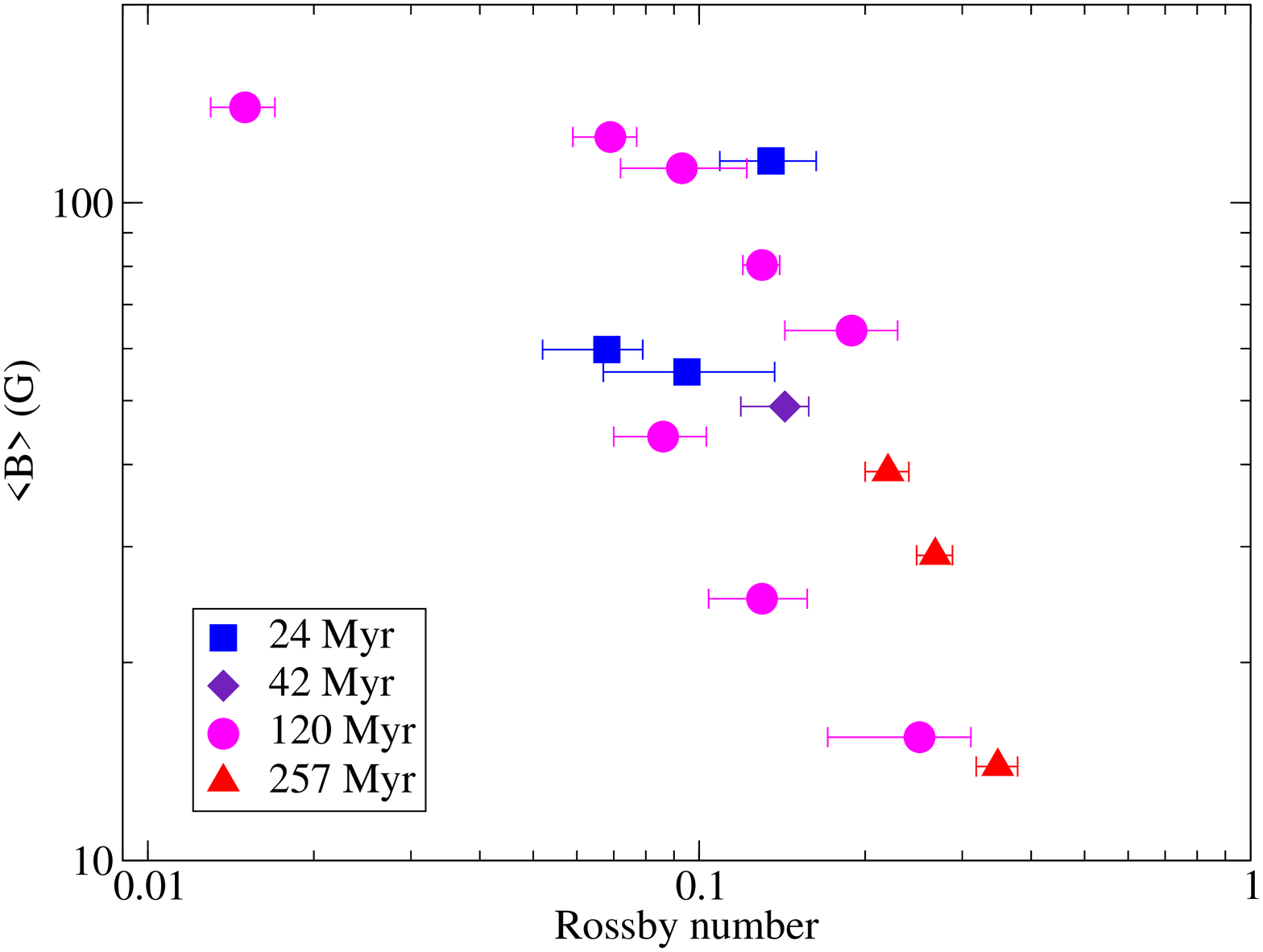}
  \caption{Trends in global mean large-scale magnetic field strength with age, rotation period, and Rossby number for the stars in our sample.  Different symbols correspond to different age bins. }
  \label{fig-B-trends-HMS}
\end{figure*}

\subsubsection{Trends in magnetic strength}

Several trends are apparent from our sample, that are illustrated in Figure~\ref{fig-B-trends-HMS}. 
We find a global decrease in the mean large-scale magnetic field strength with age from 20 to 250~Myr, even though a large scatter is seen at intermediate ages at around 120~Myr.  120 Myr is also the age with maximum scatter in rotation period.  No clear trend between mean magnetic field strength and rotation period is seen in our limited sample.  Similarly, we do not find any clear trend in magnetic field strength with just convective turnover time.  
However, conventional ($\alpha$-$\Omega$) dynamo generation is thought to be due to the combination of rotation and convection, which can be parameterized by the Rossby number of the star (the ratio of the rotation period to the convective turnover time: $R_{o} = P_{\rm rot} / \tau_{\rm conv}$).  
We do find a significant trend in decreasing magnetic field strength with Rossby number, which appears to take the form of a power law.  Fitting a power law (and excluding LO Peg, the fastest rotator) we find $\langle B \rangle \propto R_{o}^{-1.0 \pm 0.1}$.  This is based on a $\chi^2$ fit, and accounts for uncertainty in $R_{o}$ but not the systematic uncertainty in $\langle B \rangle$, which is largely driven by long term variability, thus the uncertainty on the power law may be underestimated.  LO Peg, the outlier, has by far the lowest Rossby number in our sample, but has a comparable field strength to the other strongly magnetic stars in our sample.  This suggests that we might be seeing evidence for the saturation of the global magnetic field strength and hence of the stellar dynamo.  Saturation of Ca {\sc ii} H and K emission \citep{Noyes1984-CaHK-Rossby}, and X-ray flux \citep{Pizzolato2003-Xray-saturation-rossby}, is well established and typically happens around a Rossby number of 0.1\footnote{The exact value of the Rossby number is model dependent, since it depends on the prescription for convective turnover time. Thus it can vary by a factor of a few (see Appendix \ref{Convective turnover times}).}.  Zeeman broadening measurements \citep{Saar1996-maybe-saturation-zeeman-broad,Saar2001-maybe-saturation-zeeman-broad, Reiners2009-ZeemanBroad-saturation-Mdwarfs} have found some evidence for the saturation of the small scale magnetic field, again around a Rossby number of 0.1, particularly for M-dwarfs.  
Evidence for saturation of the global magnetic field is good for fully convective M-dwarfs \citep{Donati2008-mag-Mdwarfs, Morin2008-Mdwarf-topo, Vidotto2014-magnetism-age-rot}. 
However, direct evidence for the saturation of the global magnetic field in stars with radiative cores remains tentative. 
\citet{Vidotto2014-magnetism-age-rot} studied a set of published ZDI results, which includes the data reported here, and found evidence for the global dynamo saturating near the same Rossby number as the X-ray flux.  LO Peg adds a significant extra data point to support this trend.  
It is interesting that the X-ray flux, which is only indirectly related to the magnetic field, the small-scale magnetic field, and the large-scale magnetic field all show qualitatively similar behavior with Rossby number.

\begin{figure*}
  \centering
  \includegraphics[width=3.3in]{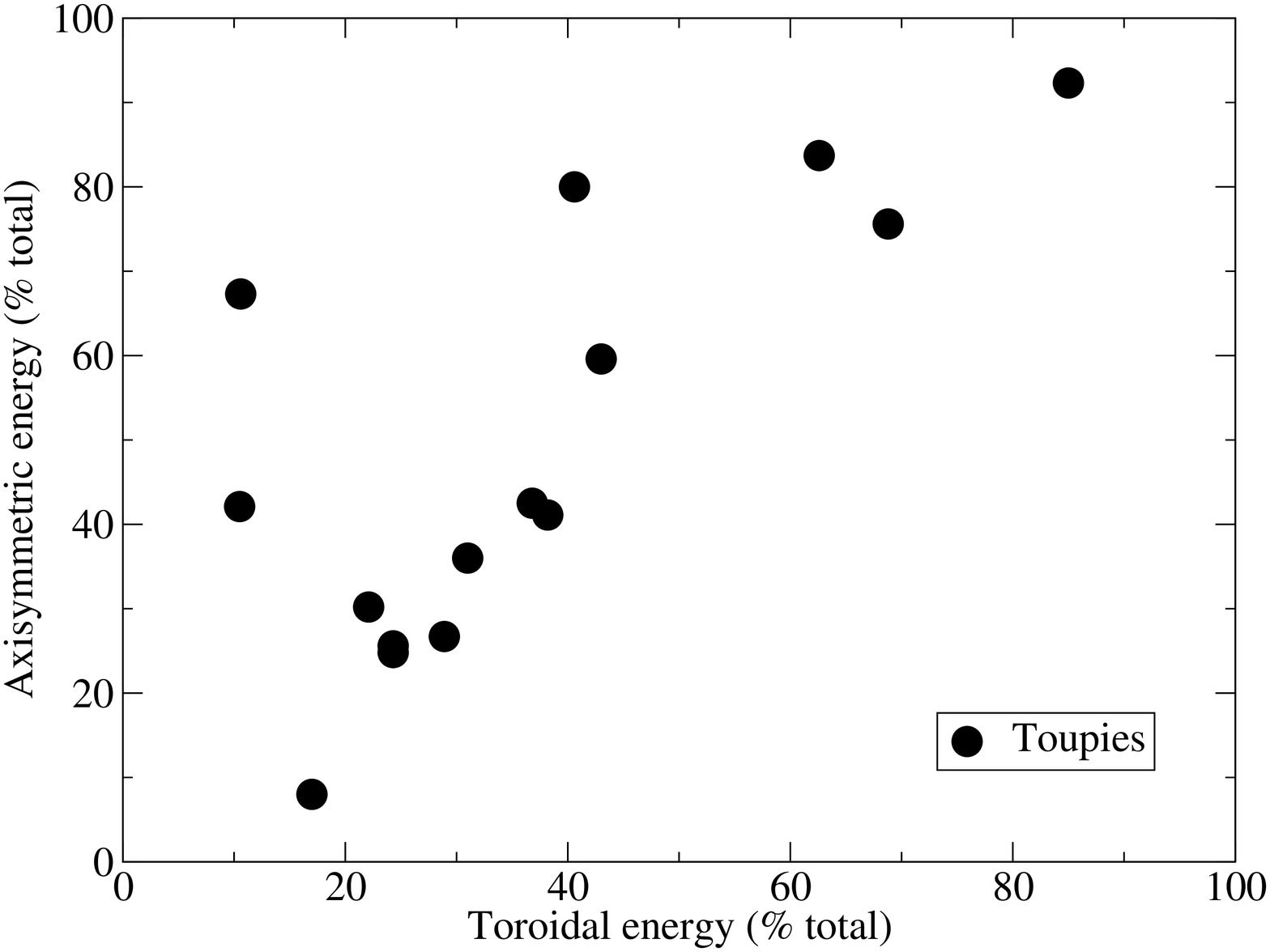}
  \includegraphics[width=3.3in]{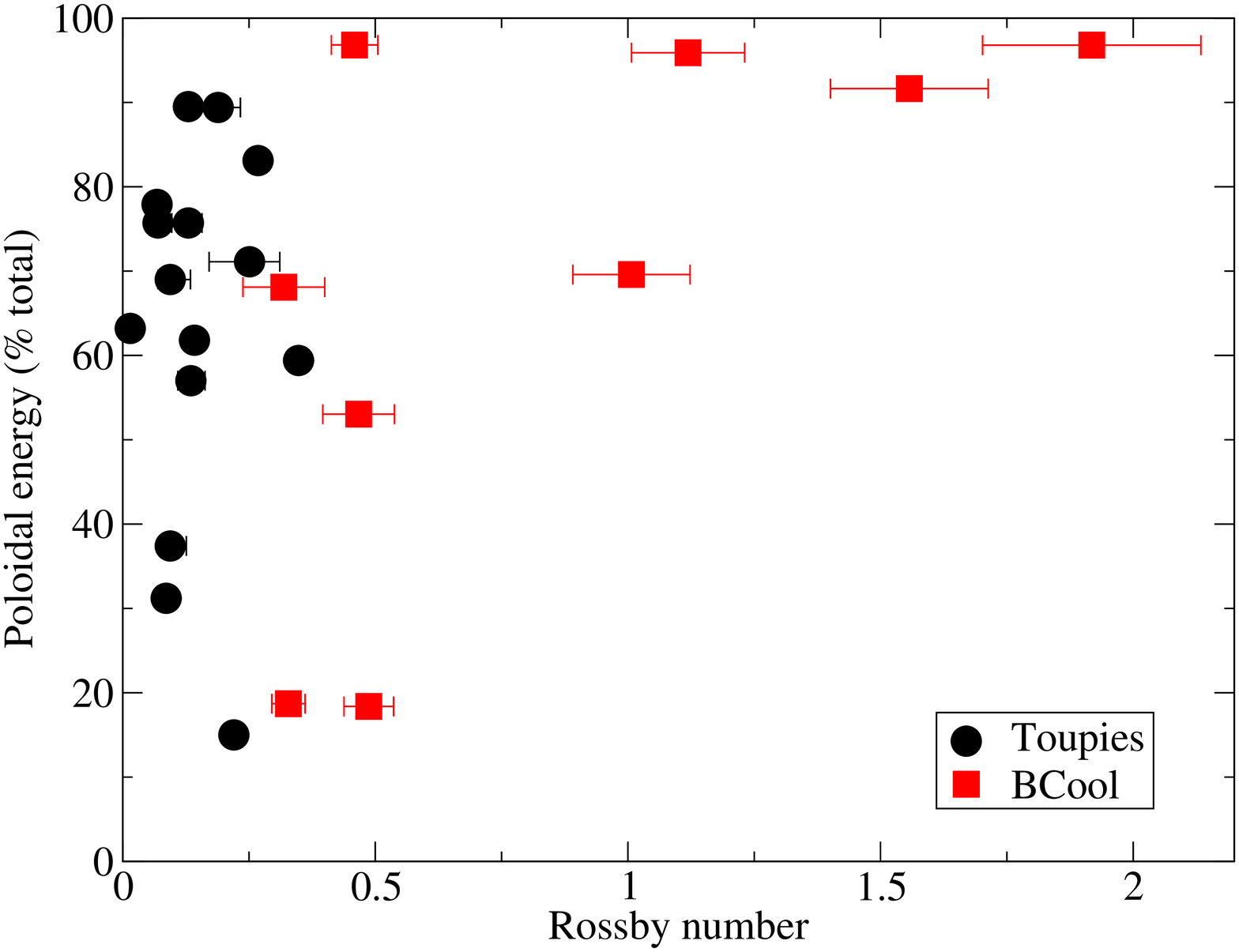}
  \includegraphics[width=3.3in]{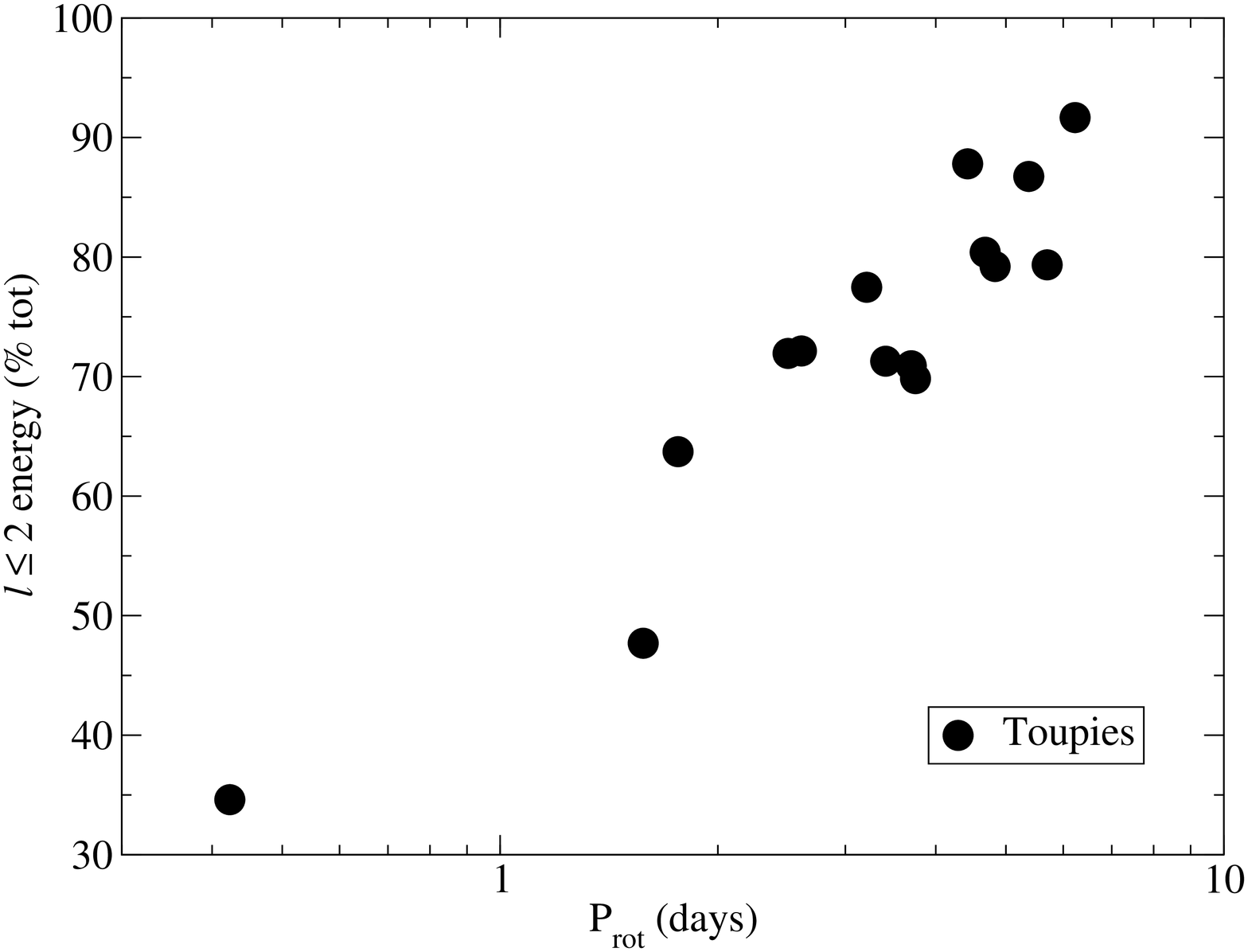}
  \caption{Trends in magnetic geometry.  Left panel: the fraction of magnetic energy in axisymmetric modes as a function of the fraction of magnetic energy in toroidal modes (our data only).  Right panel: the fraction of magnetic energy in poloidal modes as a function of Rossby number  (our data and BCool data).  A shift towards dominantly poloidal fields, from mixed toroidal-poloidal fields, is found for stars with long rotation periods.  
Bottom panel: fraction of magnetic energy in spherical harmonics of order $l \leq 2$, as a function of rotation period (our data only).  
}
  \label{fig-tor-trends}
\end{figure*}

\subsubsection{Trends in magnetic geometry}

Turning to magnetic topology, we find that stars with largely toroidal magnetic fields in our sample have largely axisymmetric geometries, as shown in Fig.~\ref{fig-tor-trends}.  More specifically the toroidal components of the magnetic field are largely axisymmetric, and become more axisymmetric as they become more dominant.  However, the axisymmetry of the poloidal component is independent of how toroidal the field is, thus the trend towards axisymmetry is driven by the toroidal component.  This is in line with similar results reported for a larger sample of late-type dwarfs by \citet{See2015-toroidal-axisymm}, which also includes the results presented here.
We also examined the fraction of magnetic energy contained in the poloidal component as a function of rotation period, shown in Fig.~\ref{fig-tor-trends}. No clear trend is seen in our sample, which encompasses a limited range of rotation periods from about 2 to 6 days.  Indeed, within this period range, the fraction of magnetic energy contained in the poloidal component ranges from 15 to 90\%.  In order to enlarge the period range, we combine our sample with the slowly rotating stars reported by \citet{Petit2008-sunlike-mag-geom}, and with additional slow rotators from the BCool sample (\citealt{Petit201-Bcool-ZDI-in-prep}, see Sect.~\ref{Comparison with older field stars}). \citet{Petit2008-sunlike-mag-geom} reported that slowly rotating stars have dominantly poloidal fields and rapidly rotating stars have dominantly toroidal fields.  The stars of our HMS sample all lie in the fast portion of their parameter space and exhibit a large range of poloidal to toroidal ratios. If we consider the additional data from the BCool project, we do indeed find that very slow rotators are dominated by poloidal fields, but faster rotators have a wide range of mixed geometries.  The transition from mixed topologies to dominantly poloidal fields seems to occur at roughly $P_{\rm rot} \approx 10-15$ days, in agreement with \citet{Petit2008-sunlike-mag-geom}.  In Rossby number, this transition occurs roughly between 0.5 and 1.0 (cf. Fig.~\ref{fig-tor-trends}). 

We examined the complexity of the reconstructed large-scale magnetic field, by considering the magnetic energy in all spherical harmonic modes with $l \leq 2$.  This includes dipolar and quadrupolar modes, and their corresponding toroidal modes.  We find a trend towards decreasing complexity with increasing rotation period, illustrated in Fig \ref{fig-tor-trends}, and a similar trend with increasing Rossby number.  Thus it may be that faster rotators, with stronger dynamos, have more complex magnetic fields.  This is in contrast to the fully convective T Tauri stars that often have simple magnetic field geometries.  However, the spatial resolution of ZDI is a function of the \vs\ of a star, thus there is a potential systematic effect that could impact on this result.  The correlation we find appears to be stronger with rotation period than \vs, and all our stars should have maps with a resolution higher than an $l$ order of 2, thus this trend appears to be real.  Never the less, we caution the reader that this result is tentative.  A more detailed investigation, with an evaluation of potential systematics, is needed and planned for a forthcoming paper. 

\subsubsection{Implications for models}
 
\citet{Barnes2003-rotation-C-I-sequence} identified the C and I sequences of, respectively, fast and slow rotating young main sequence stars.
In an attempt to reproduce these sequences \citet{Brown2014-metastable-dynamo-model-rotation} proposed the `Metastable Dynamo Model' of rotational evolution.  
In this model, stars are initially very fast rotators and are weakly coupled to their winds.  Eventually, stars {\it randomly} switch to being strongly coupled to their wind, and then quickly spin down.  The difference between the weakly and strongly coupled modes is ascribed to different magnetic topologies, hypothetically corresponding to different dynamo modes. 
However, within our sample we find no strong differences in magnetic geometry between very fast rotators ($ P_{\rm rot} < 2$ days) and moderate rotators ($ P_{\rm rot} > 2$ days), which would essentially correspond to the transition between the C and I sequences of \citet{Barnes2003-rotation-C-I-sequence}.  There is a significant transition to dominantly poloidal fields at large Rossby numbers or equivalently large rotation periods (Fig.~\ref{fig-tor-trends}), but that occurs at rotation periods of $\sim$10-15 days (c.f. Table \ref{fundimental-param-table}), which is much beyond the line between the C and I sequences.  We do find a general trend towards magnetic geometries with more energy in higher spherical harmonics for shorter rotation periods (Fig.~\ref{fig-tor-trends}).  These more complex fields could be of interest for producing fast rotators that are more weakly coupled to their wind for their global magnetic field strength.  But this trend appears to be continuous over a wide range of rotation, so it is not clear how it would produce a bimodal distribution, and this result is very tentative, as discussed above.  
Our current results thus do not provide clear evidence to support the Metastable Dynamo Model.  Additional planned observations of both faster and slower rotators, with ages extending up to 600 Myr, may provide additional constraints.

\subsection{Comparison with older field stars}
\label{Comparison with older field stars}

\begin{figure*}
  \centering
  \includegraphics[width=3.3in]{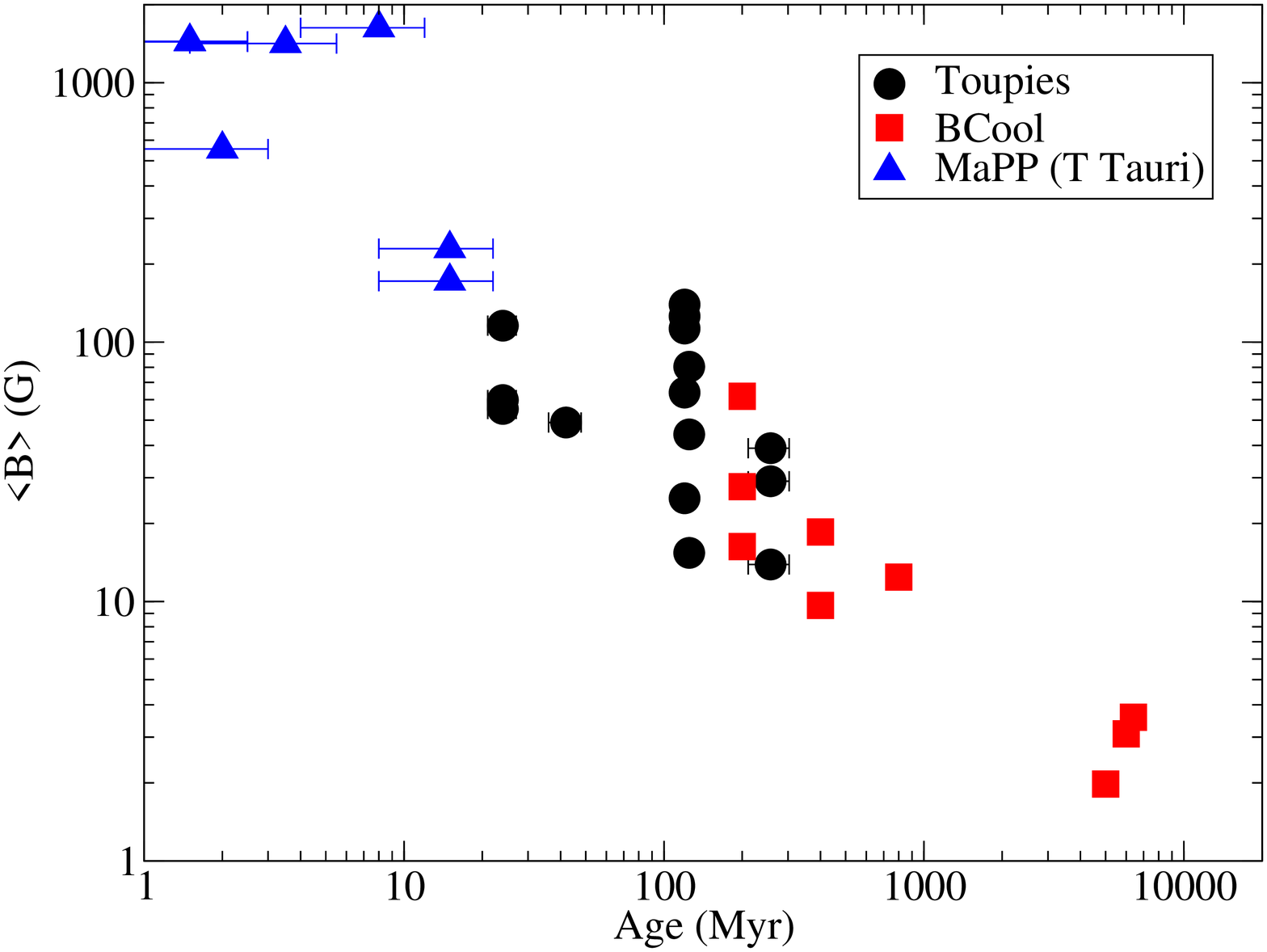}
  \includegraphics[width=3.3in]{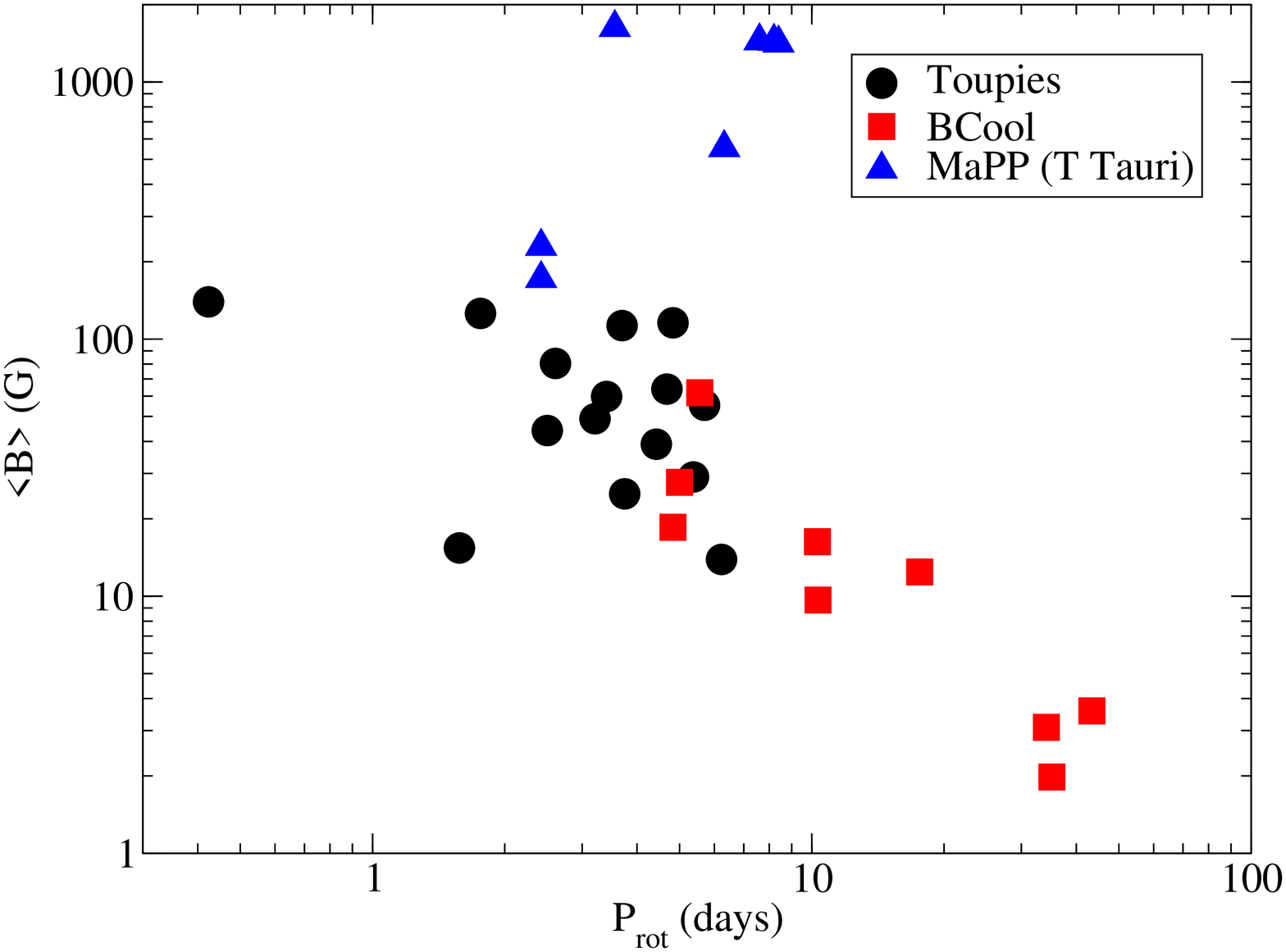}
  \includegraphics[width=3.3in]{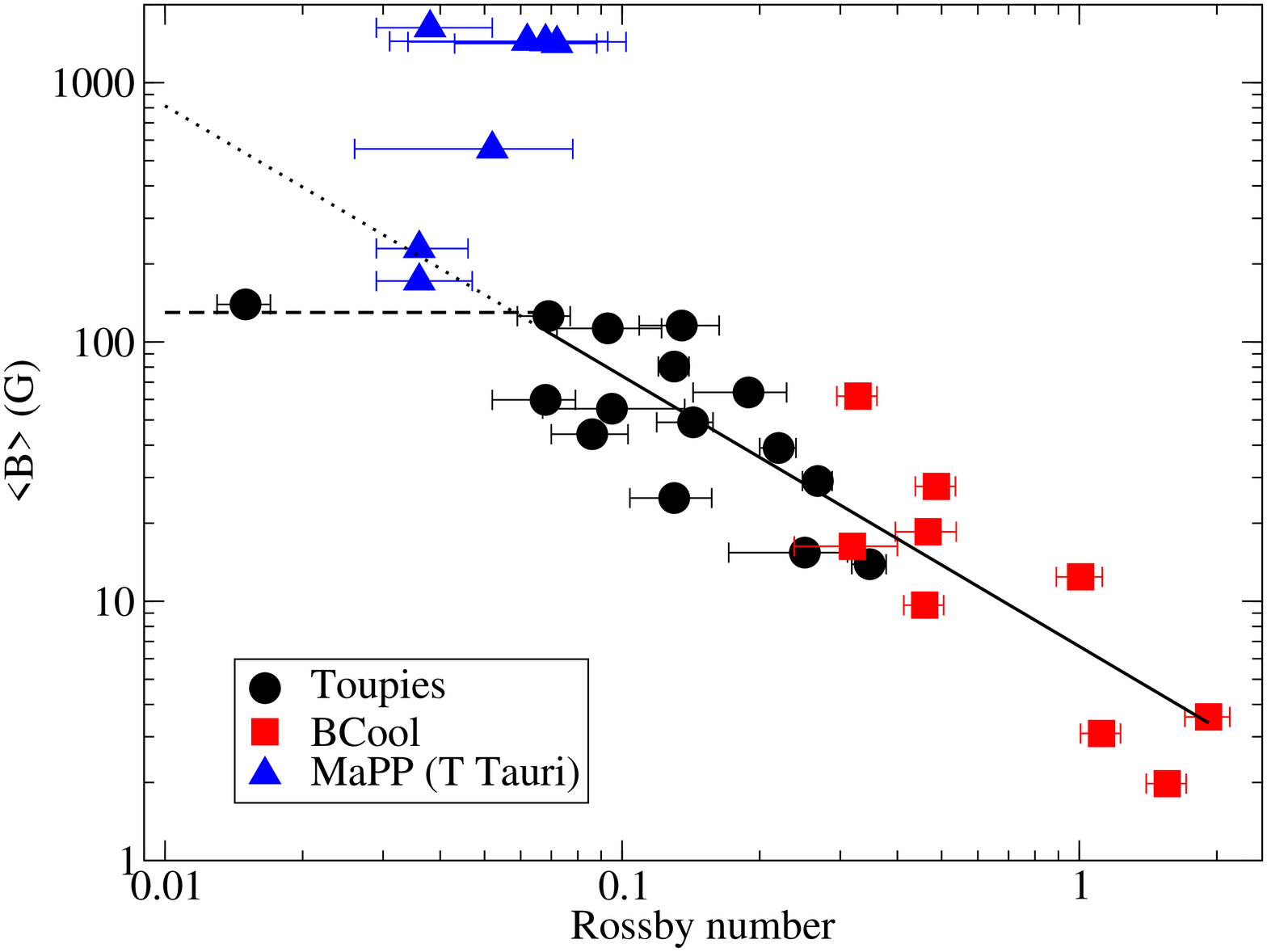}
  \caption{Trend in mean large-scale magnetic field strength with age, rotation period, convective turnover time, and Rossby number.  Black circles are from our sample (the Toupies observational project), while red squares are additional field stars we consider from the BCool project.  Blue triangles are T Tauri stars from the MaPP project.  In the bottom panel, the solid line is the best fit power law, the dotted line is an extrapolation of this fit, and the dashed line is a hypothetical saturation value for the large-scale magnetic field.  The T Tauri stars fully convective, and are exceptions to the trends in rotation period and Rossby number.  }
  \label{fig-B-trends}
\end{figure*}

One of the limitations of our current sample is that it is still a modest size, and is focused mostly on young quickly rotating stars.  Stronger conclusions can be drawn by adding older more slowly rotating stars from previous studies.  We selected stars from the BCool sample \citep[see][for the first major paper in the series]{Marsden2014-Bcool-survey1}, which are mostly field main sequence stars with ages of a few Gyr.  This is an excellent comparison sample, as the observations were obtained with the same instruments and observing strategy as our observations, and the same analysis techniques were used to derive magnetic maps for the stars.  We specifically focus on stars in a similar mass range as our sample: 0.7 to 0.9 $M_{\odot}$, using the objects 
HD 22049 ($\epsilon$ Eri; \citealt{Jeffers2014-epsEri-mag-var}), 
HD 131156A ($\xi$ Boo A; \citealt{Petit2005-xiBoo-ZDI}; \citealt{Morgenthaler2012-xiBoo-mag-var}), 
HD 131156B ($\xi$ Boo B; \citealt{Petit201-Bcool-ZDI-in-prep}),  
HD 201091 (61 Cyg A; Boro Saikia et al.\ in prep.; \citealt{Petit201-Bcool-ZDI-in-prep}), 
HD 101501, HD 10476, HD 3651, HD 39587, and HD 72905 \citep{Petit201-Bcool-ZDI-in-prep}.  
Ages for the stars are take from \citet{Mamajek2008-ages-chromospheric} based on chromospheric activity.  Since these stars are on the main sequence, and typically the younger half of the MS, H-R diagram ages are highly uncertain.  The exception to this is HD 201091, where we use the more precise age of \citet{Kervella2008-61Cyg-params}, based on an interferometric radius.  

With these added stars, we see a much clearer trend in large-scale magnetic field strength with rotation period, as illustrated in Figure~\ref{fig-B-trends}, while the trends in magnetic field strength with age and Rossby number are further supported.  There is still no clear correlation between magnetic field strength and convective turnover time, largely because the BCool stars have the same range of turnover times as our sample.  The added stars have a much stronger correlation between rotation period and age, since by these older ages the rotation rates of the stars have largely converged to a single sequence.  The correlation of magnetic field strength with age and with rotation rate agree with the results from \citet{Vidotto2014-magnetism-age-rot}, however they considered a much larger mass range (from $\sim$0.2 $M_\odot$ to $\sim$1.3 $M_\odot$), and some of our early results were included in that paper.

\begin{figure*}
  \centering
  \includegraphics[width=6.0in]{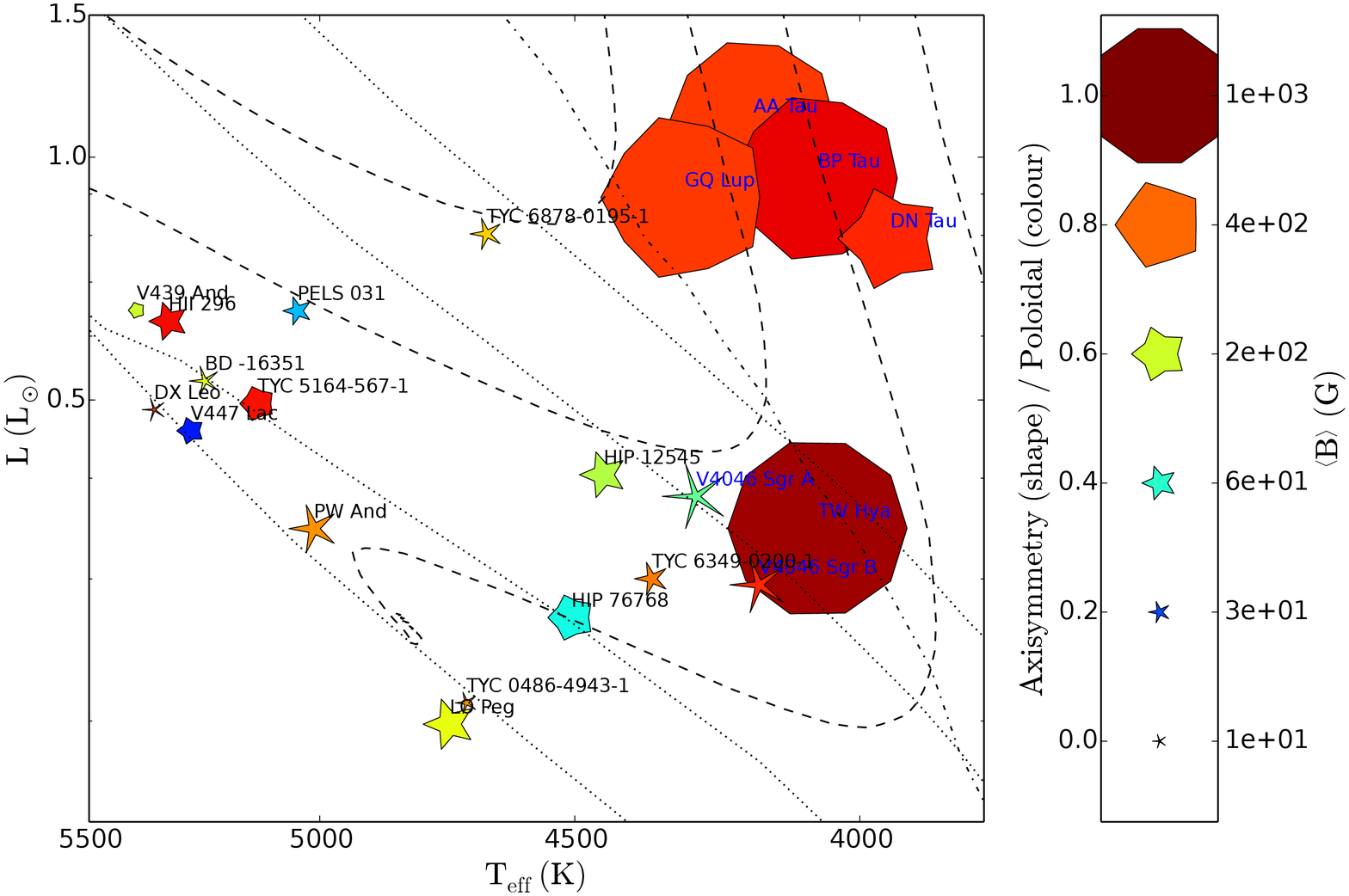}
  \caption{Magnetic parameters plotted for different physical parameters.  Stars labeled in blue are from the MaPP project, stars labeled in red are from the BCool project, and stars labeled in black are from our study.  Symbol size indicatesmean magnetic strength, symbol colour indicates how poloidal the magnetic field is (red is more poloidal and blue is more toroidal), and symbol shape indicates how axisymmetric the poloidal component of the magnetic field is (more circular is more axisymmetric).  
Dashed lines are evolutionary tracks for 1.2, 1.0, 0.8 and 0.6 $M_\odot$, the dotted lines are isochrones for 10, 20, 50 and 100 Myr (the ZAMS), and the dash-dotted line indicates where a significant convective core has formed ($> 50$\% mass), from Amard et al.\ (in prep.).  }
  \label{fig-zdi-confusog1}
\end{figure*}

\begin{figure*}
  \centering
  \includegraphics[width=6.0in]{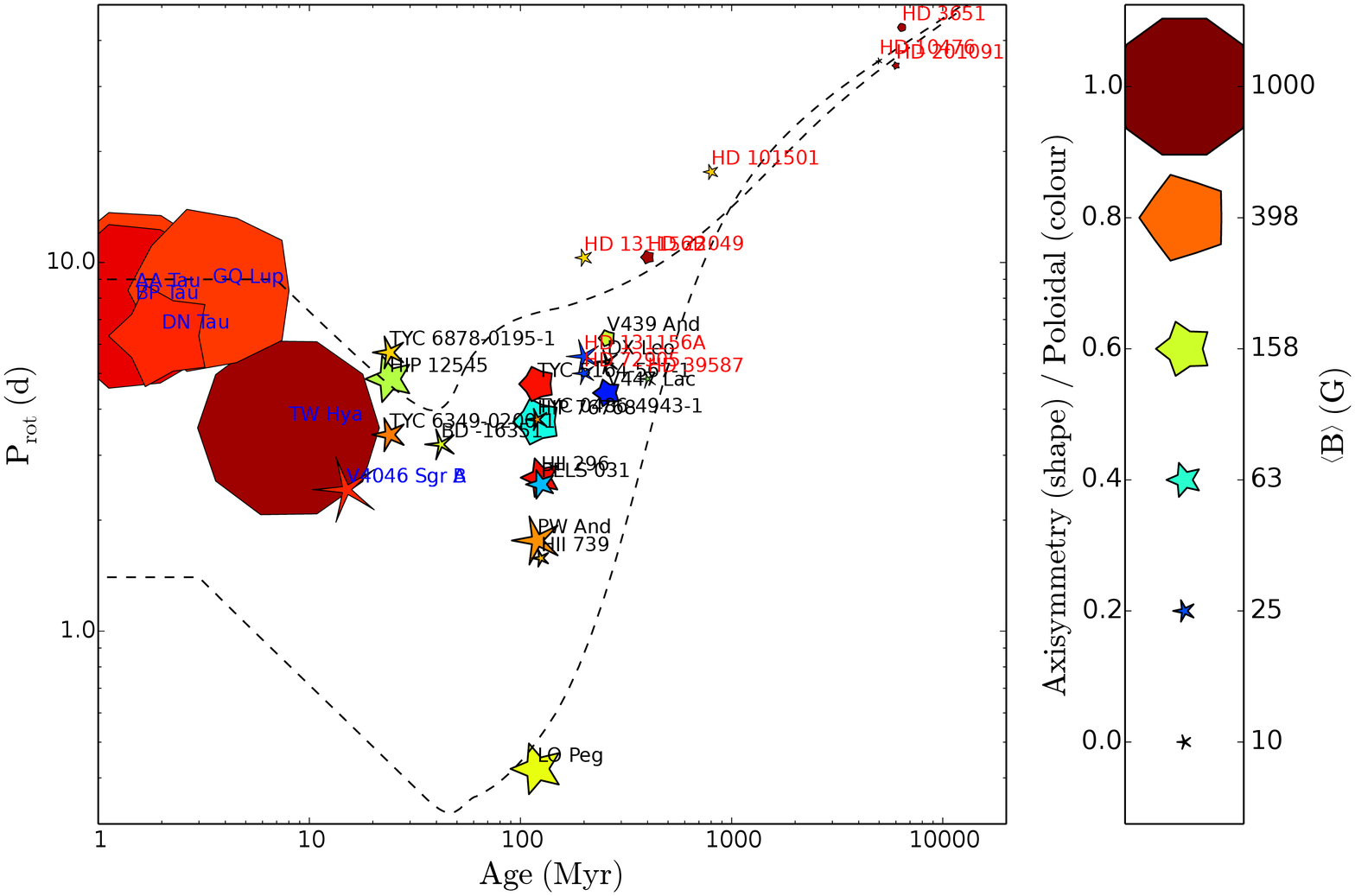}
  \includegraphics[width=6.0in]{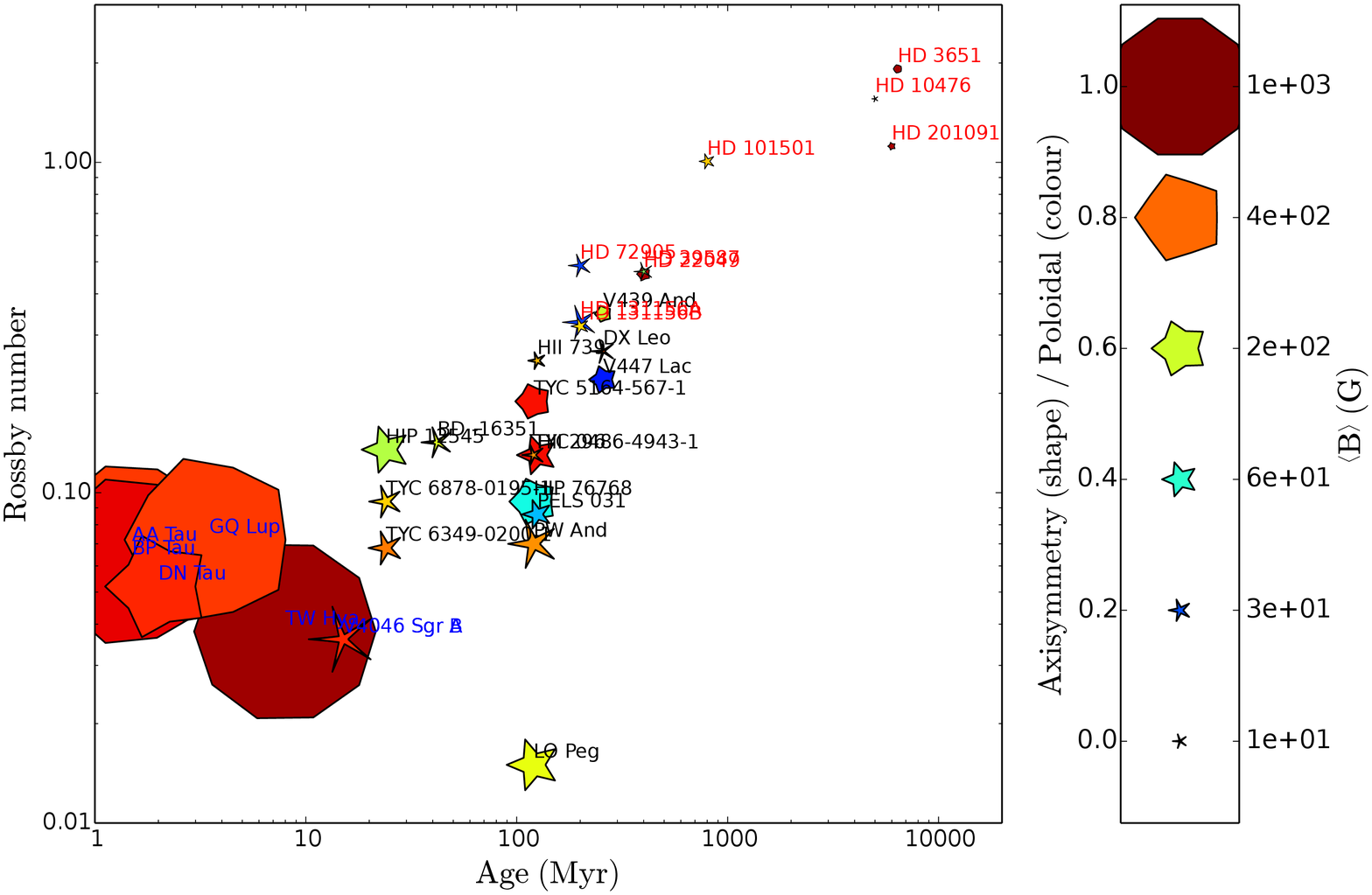}
  \caption{Magnetic parameters plotted for different physical parameters.  Stars labeled in blue are from the MaPP project, stars labeled in red are from the BCool project, and stars labeled in black are from our study.  Symbol size indicates mean magnetic strength, symbol colour indicates how poloidal the magnetic field is (red is more poloidal and blue is more toroidal), and symbol shape indicates how axisymmetric the poloidal component of the magnetic field is (more circular is more axisymmetric).  
In the upper panel, dashed lines are rotational evolutionary tracks for fast and slow rotators at 0.8 \Msun\ from \citet{Gallet2015-Bouvier-ang-mom-evol2}. }
  \label{fig-zdi-confusog23}
\end{figure*}

\subsection{Comparison with T Tauri stars}
\label{Comparison with T Tauri stars}

Younger T Tauri stars (TTS), with an age of a few Myr, also make an interesting comparison to our sample.  We consider seven stars from the `Magnetic Protostars and Planets' (MaPP) project: TW Hya \citep{Donati2011-TWHya-ZDI}, AA Tau \citep{Donati2010MNRAS-AATau-ZDI}, BP Tau \citep{Donati2008-BPTau-ZDI}, GQ Lup \citep{Donati2012-GQLup-ZDI}, DN Tau \citep{Donati2013-DNTau-ZDI}, V4046 Sgr A \citep{Donati2011-V4046Sgr-ZDI}, and V4046 Sgr B \citep{Donati2011-V4046Sgr-ZDI}.  These stars are earlier on the pre-main sequence than our youngest stars, and are expected to still be accreting.  These stars are chosen to be in a similar mass range as our objects (0.7 to 1.0 $M_\odot$).   

From Fig.~\ref{fig-B-trends} it is clear that TTS largely do not follow the trends seen for ZAMS and MS stars.  Their mean large-scale magnetic field strength is about 10 times stronger than that of ZAMS stars, which qualitatively extends the trend of decreasing magnetic field strength with age, from the early PMS through the ZAMS and down to the MS.  However, the rotation periods of TTS are similar to those of ZAMS stars.  Hence, the difference in magnetic strength cannot result from rotational properties.  Being mostly convective, TTS have much longer convective turnover timescales than older stars. Thus, at a given rotation period, TTS will have smaller Rossby numbers than ZAMS stars.  This is shown in Fig.~\ref{fig-B-trends}, where TTS have both lower Rossby numbers and stronger magnetic fields than ZAMS and MS dwarfs. Yet, the location of TTS in this plot is well above the trend seen for older stars.  TTS seem to have excessively strong large-scale magnetic fields for their Rossby number compared to the extrapolation of the magnetic field vs.\ Rossby number relationship seen for ZAMS and MS dwarfs.  It is possible, although as yet unobserved, that the energy in very small scale magnetic structures is the same between TTS and ZAMS stars, but the large-scale component of the magnetic field clearly undergoes a dramatic transition.  Hence, another parameter must come into play when comparing the magnetic properties of PMS stars to those of ZAMS and MS stars.  Most notably, the large difference in internal structure between young TTS and main sequence stars of the same mass must play a role.  This appears to be analogous to the transition of magnetic properties between M-dwarfs and more massive stars \citep{Morin2008-Mdwarf-topo,Morin2010-Mdwarfs, Donati2008-mag-Mdwarfs}.

\subsection{A synthetic view of magnetic field evolution in young stars} 
\label{A synthetic view of magnetic field evolution }

The trends seen in the magnetic properties of stars in the mass range 0.7-1.2~M$_\odot$, as they evolve from the T Tauri phase through the ZAMS and onto the MS, are summarized in Figs.~\ref{fig-zdi-confusog1} and \ref{fig-zdi-confusog23}. 

Fig.~\ref{fig-zdi-confusog1} shows the pre-main sequence and ZAMS stars in the H-R diagram, and clearly illustrates the difference between the magnetic properties of the T Tauri stars and the older pre-main sequence stars in our sample.  The T Tauri star large-scale magnetic fields are much stronger, and consistently dominated by a poloidal field aligned with the rotation axis (with the marginal exception of V4046 Sgr A and B).  The T Tauri magnetic fields are also generally simpler, mostly heavily dominated by a dipolar component.  
This is in strong contrast to the older pre-main sequence stars and main sequence stars in our sample. 
This large difference in magnetic properties may be a consequence of the large differences in internal structure between the early and late pre-main sequence.  The very young T Tauri stars are almost entirely convective, while our later pre-main sequence stars have large radiative cores.  This implies a significantly different form of dynamo is acting in the T Tauri stars.   
\citet{Donati2011-V4046Sgr-ZDI} proposed the development of a radiative core to explain the difference between the younger T Tauri stars and the slightly older V4046 Sgr A and B.  This is consistent with our observation that these two stars have magnetic properties closer to our sample than the rest of the T Tauri stars.  
\citet{Gregory2012-TTauri-B-structure} consider this hypothesis in detail, using a sample of T Tauri stars, and compare T Tauri stars to largely convective M-dwarfs.  Another important difference between the older PMS stars in our sample and the classical T Tauri stars is that the latter are still accreting significant amounts of material from their disk, while the former are well beyond the main accretion phase.  Whether the accretion process, and particularly the magnetic star-disk interaction in TTS, impacts on their surface magnetic properties is still to be investigated \citep[e.g.][]{Donati2015-V819Tau-WTTs-mag}.  

The upper panel of Fig.~\ref{fig-zdi-confusog23} shows the various samples in a rotational evolution scheme, where rotation period is plotted as a function of age. At similar periods, the clear differences between TTS and ZAMS stars, discussed above, still remain. Rotational convergence is seen to occur at ages larger than about 1 Gyr, and the decrease of rotation rate for older main sequence stars is clearly associated with weaker large-scale magnetic fields.  However, the geometry of the magnetic field does not seem to evolve significantly between the ZAMS and the older MS.  This supports the idea that it is primarily dictated by the star's internal structure, at least for rotation periods up to 10-15 days (see Sect.~\ref{Magnetic trends in young stars} above). 

The lower panel of Fig.~\ref{fig-zdi-confusog23} show the samples in a Rossby number vs.\ age plot, which illustrates the evolution of the magnetic dynamo with time. As discussed above (see Sect.~\ref{Comparison with T Tauri stars})
unlike the rest of our sample, the T Tauri stars do not follow the same trend of large-scale magnetic field strength with Rossby number.  
They have much stronger magnetic fields for their nominal Rossby number than the rest of the stars.  Indeed these magnetic fields are much stronger than the apparent saturation value from LO Peg.  This further supports the hypothesis that the magnetic dynamo operating in these stars is significantly different than in older stars with radiative cores.  

There is evidence of 2 groups of stars around the ZAMS with distinct magnetic properties. One group with ages ranging from 20 to 120 Myr seems to have systematically smaller Rossby numbers and higher large-scale magnetic field strengths than the second group of post-ZAMS stars at about 250 Myr.   Hence, in a relatively short time frame around the ZAMS, the magnetic properties of young stars appear to evolve significantly (this evolution is likely continuous but we do not yet have the observations to confirm this).  This is most likely driven by the rapid rotational evolution they experience over a few 100 Myr at the start of the main sequence evolution \citep[cf.][]{Gallet2013-Bouvier-ang-mom-evol, Gallet2015-Bouvier-ang-mom-evol2}.  Indeed, from Fig.~\ref{fig-zdi-confusog23} it appears that the magnetic properties of stars evolve much less from 250 Myr to about 2 Gyr than they do in the first 250 Myr.

The main goal of this study was to investigate how the magnetic properties of solar-type stars evolve with time, especially during the dramatic change their rotation rates experience around the ZAMS. Indeed, recent models predict that, as solar-type stars land on the ZAMS, their radiative core should spin much faster than their outer convective envelope \citep[e.g.][]{Irwin2007-NGC2516-rot-evol-need-2zone, Spada2011-rotational-evol-2zone}. Furthermore, the velocity gradient at the tachocline depends on the lifetime of the accretion disk during the pre-main sequence, and slow ZAMS rotators are predicted to have {\it more} radial differential rotation than fast rotators \citep[e.g.][]{Bouvier2008-rad-diff-rot-Li-mixing}. This could conceivably impact the internal dynamo process, and could therefore be reflected at the stellar surface through a variety of magnetic properties. At this point in our study, given our limited sample size and relatively narrow range of rotation periods investigated so far, we cannot fully assess whether the richness of magnetic properties seen in our ZAMS sample reveals such a trend. We find no clear differences in large-scale magnetic field strength for a single Rossby number at different ages, as illustrated in Fig.~\ref{fig-zdi-confusog23}.  
Thus, we do not have clear evidence for an evolution in magnetic properties with age distinct from variations with Rossby number. The strong exception to this being the T Tauri stars, which likely reflect an evolution in internal structure, as discussed above.  
The rotational evolution models of \citet{Gallet2013-Bouvier-ang-mom-evol} and \citet{Gallet2015-Bouvier-ang-mom-evol2} predict that there is enhanced radial differential rotation for stars just reaching the ZAMS, near 100 Myr for stars around 0.8 to 1.0 $M_{\odot}$.  
We do not see clear evidence for this enhanced differential rotation in the surface magnetic properties of our stars. However, there is evidence for a significant scatter in the magnetic properties of ZAMS solar-type stars at a given Rossby number, most notably regarding the field geometry (cf. Fig.~\ref{fig-zdi-confusog23}). This scatter  remains to be accounted for and may be a signature of different processes occurring within the star at a given mass, age, and surface rotation rate around the ZAMS.  Surface, latitudinal, differential rotation is likely detectable in several of these stars \citep[e.g.][]{Petit2002-diff-rot-DI}, and this will be investigated in a forthcoming paper.  Intrinsic variability in large-scale magnetic fields, on timescales of a year or greater, introduces further uncertainty to this study.  This can be dealt with in a statistical fashion, by observing several stars to characterize the range of intrinsic variability, however that requires an expanded sample.

\section{Conclusions}

We have measured the large-scale magnetic field strengths and topologies for 15 solar-type stars close the zero-age main sequence, with ages ranging from 20 to 250 Myr. The stars have a range of complex magnetic geometries, with global average strengths from 14 G to 140 G. This fills the gap between younger T Tauri stars and older main sequence stars whose magnetic properties have been derived elsewhere. These new results  thus provide us with a continuous picture of the evolution of magnetic field in solar-type stars from the pre-main sequence, through the ZAMS, and down onto the MS. We find that the evolution of magnetic properties at young ages, from the PMS to the ZAMS, is primarily driven by structural changes in the stellar interior, as a radiative core develops in initially fully convective stars.  This is analogous to differences observed between K and M-dwarfs.  Once on the ZAMS, however, the subsequent evolution of magnetic properties is largely driven by the stars rotational evolution. Indeed, we find a tight relationship between the magnitude of the large-scale mean magnetic field and Rossby number in our ZAMS sample, which extends to the more mature MS sample as well.  While the combination of structural changes during the PMS and rotational evolution up to and past the ZAMS accounts for the global evolution of the magnetic properties of solar-type stars, a significant scatter is nevertheless observed at each age for a given mass and rotation period. Whether this residual scatter calls for an additional parameter impacting the magnetic properties of young stars, such as internal differential rotation and its relationship with PMS disk lifetimes, is difficult to assess from our limited sample. We will investigate this issue further in a forthcoming paper by enlarging our stellar sample to a wider range of rotation periods and ages up to 600 Myr.

\section*{Acknowledgments} 
We thank Florian Gallet for helpful discussions, as well as providing rotational evolution tracks.  
This study was supported by the grant ANR 2011 Blanc SIMI5-6 020 01 ``Toupies: Towards understanding the spin evolution of stars'' ({http://ipag.osug.fr/Anr\_Toupies/}). 
S.V.J. acknowledges research funding by the Deutsche Forschungsgemeinschaft (DFG) under grant SFB 963/1, project A16.
A.A.V. acknowledges support by the Swiss National Science Foundation through an Ambizione Fellowship.

\bibliographystyle{mnras}
\bibliography{massivebib.bib}{}

\appendix
\section{Individual Targets}
\label{Individual Targets}

\begin{figure*}
  \centering
  \includegraphics[width=6.8in]{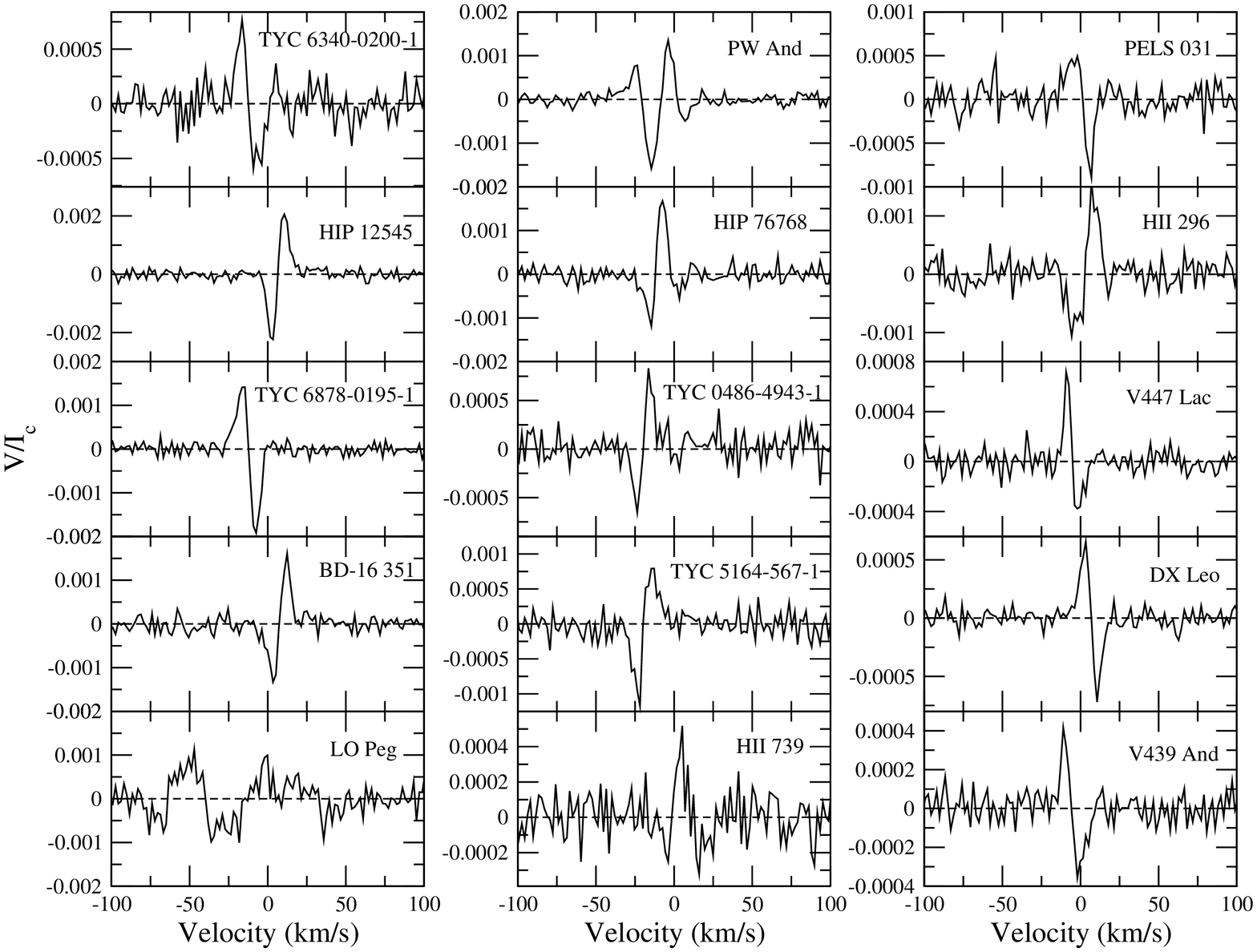}
  \caption{Sample LSD $V$ profiles for the stars in this study.  }
  \label{fig-lsd-grid}
\end{figure*}

\begin{figure*}
  \centering
  \includegraphics[width=2.8in]{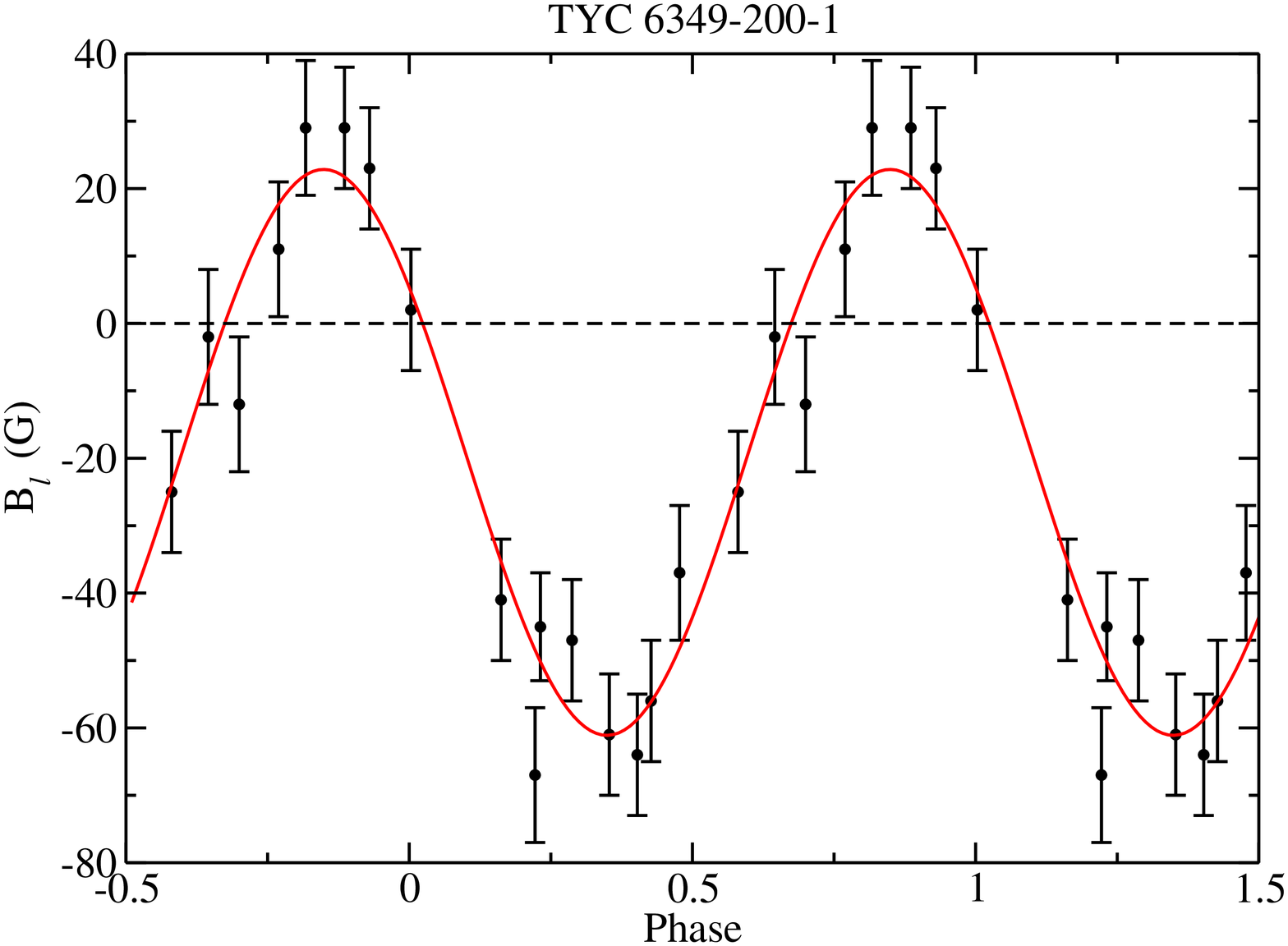}
  \includegraphics[width=2.8in]{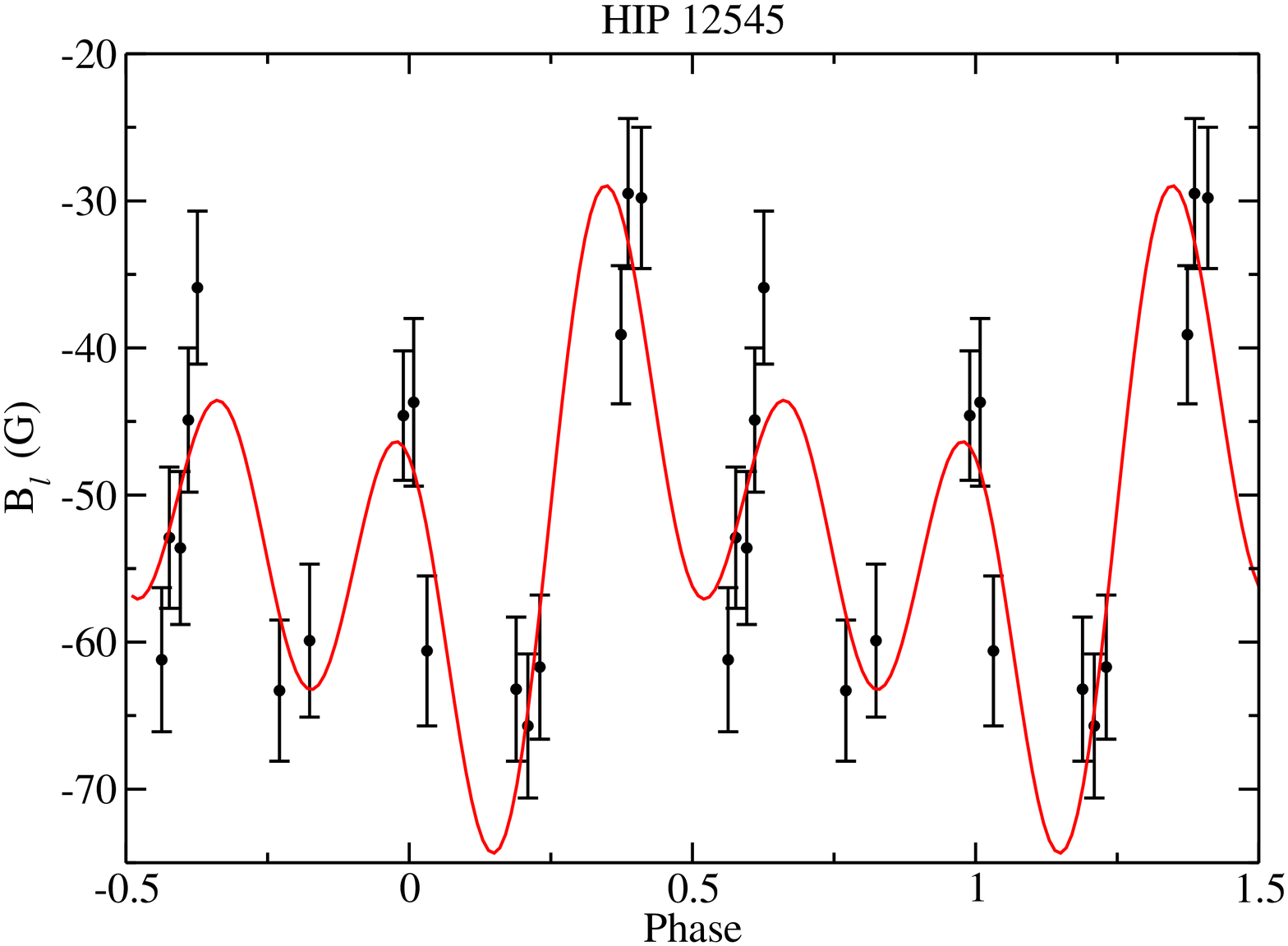}
  \includegraphics[width=2.8in]{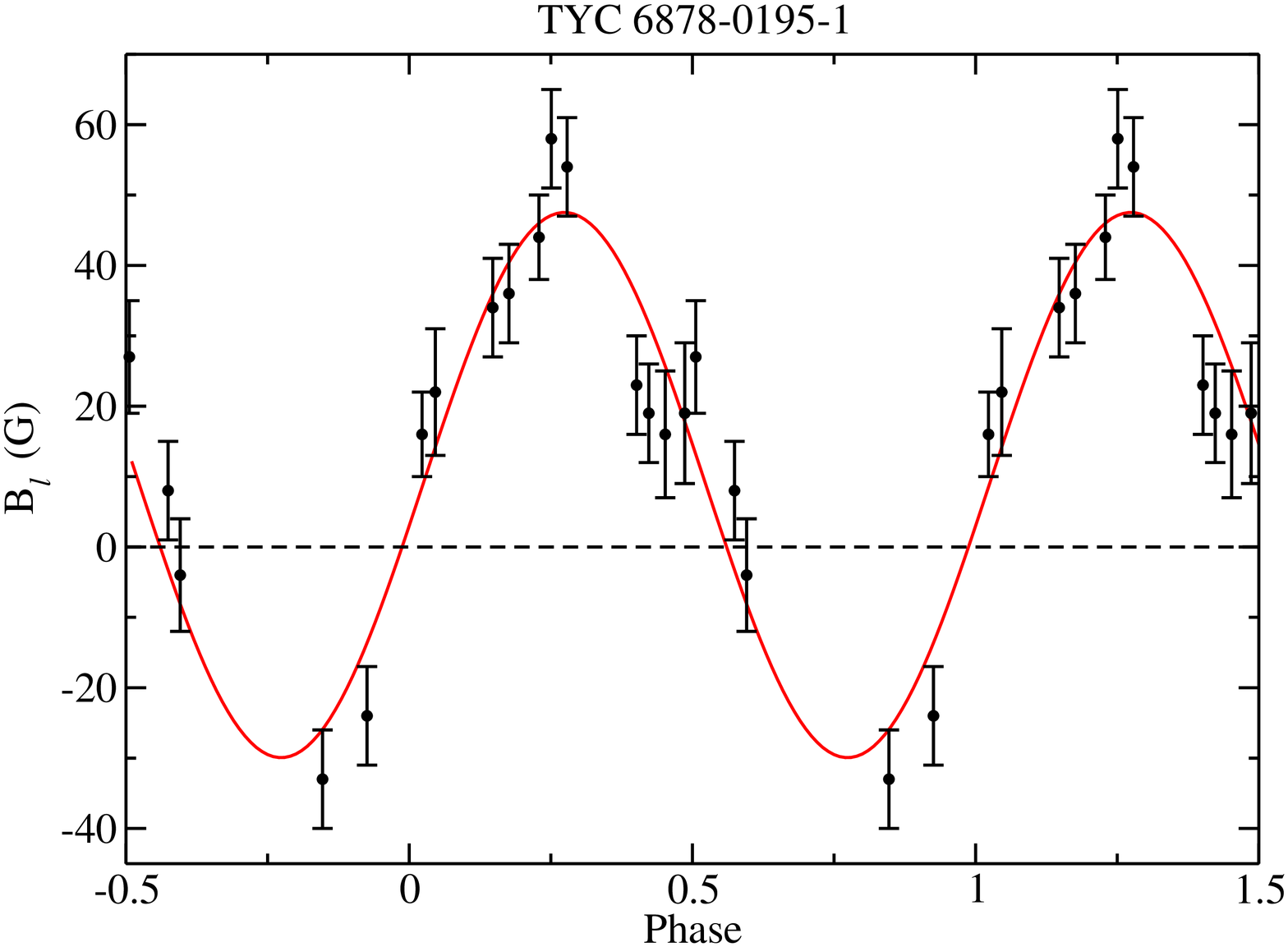}
  \includegraphics[width=2.8in]{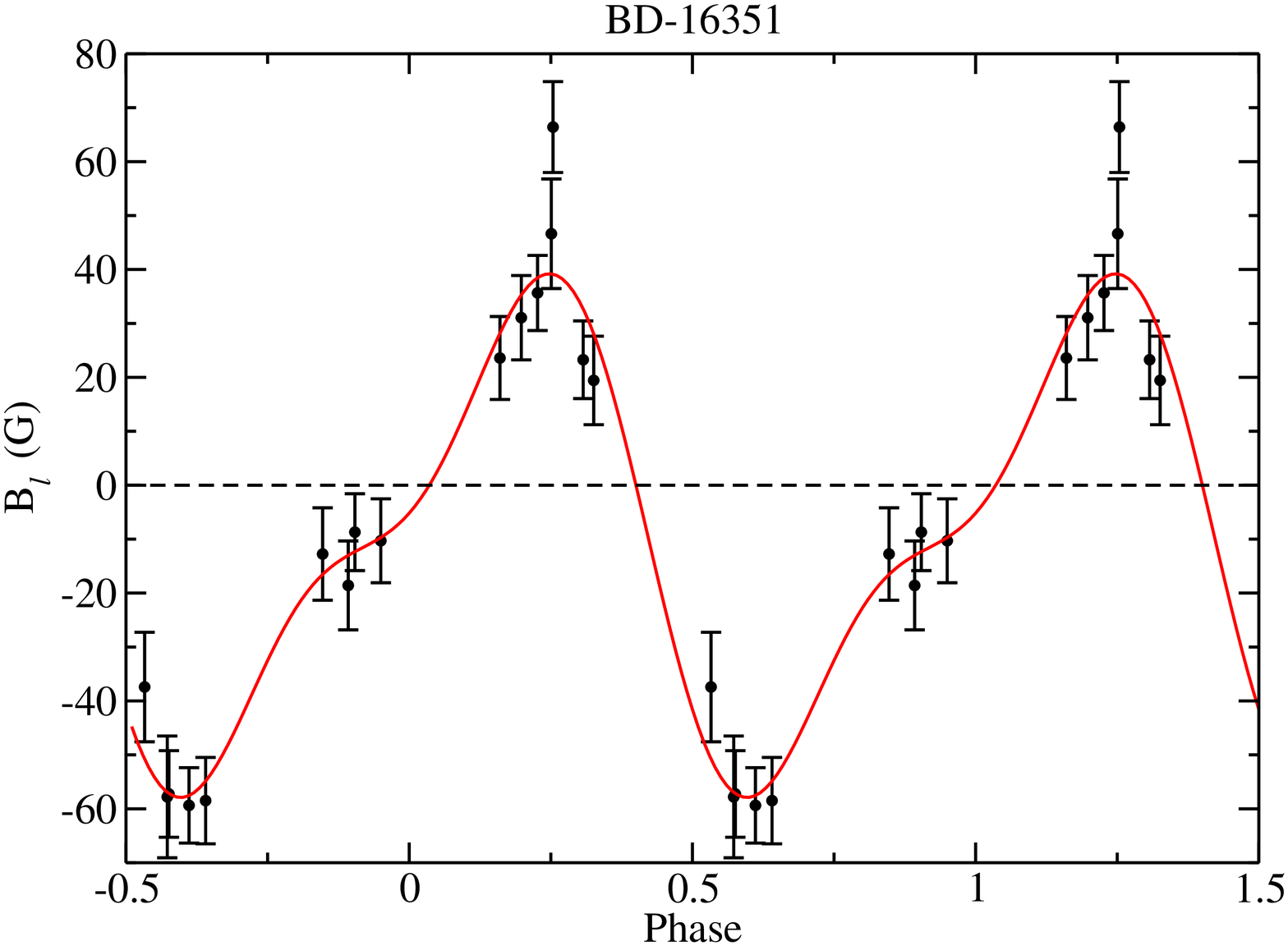}
  \includegraphics[width=2.8in]{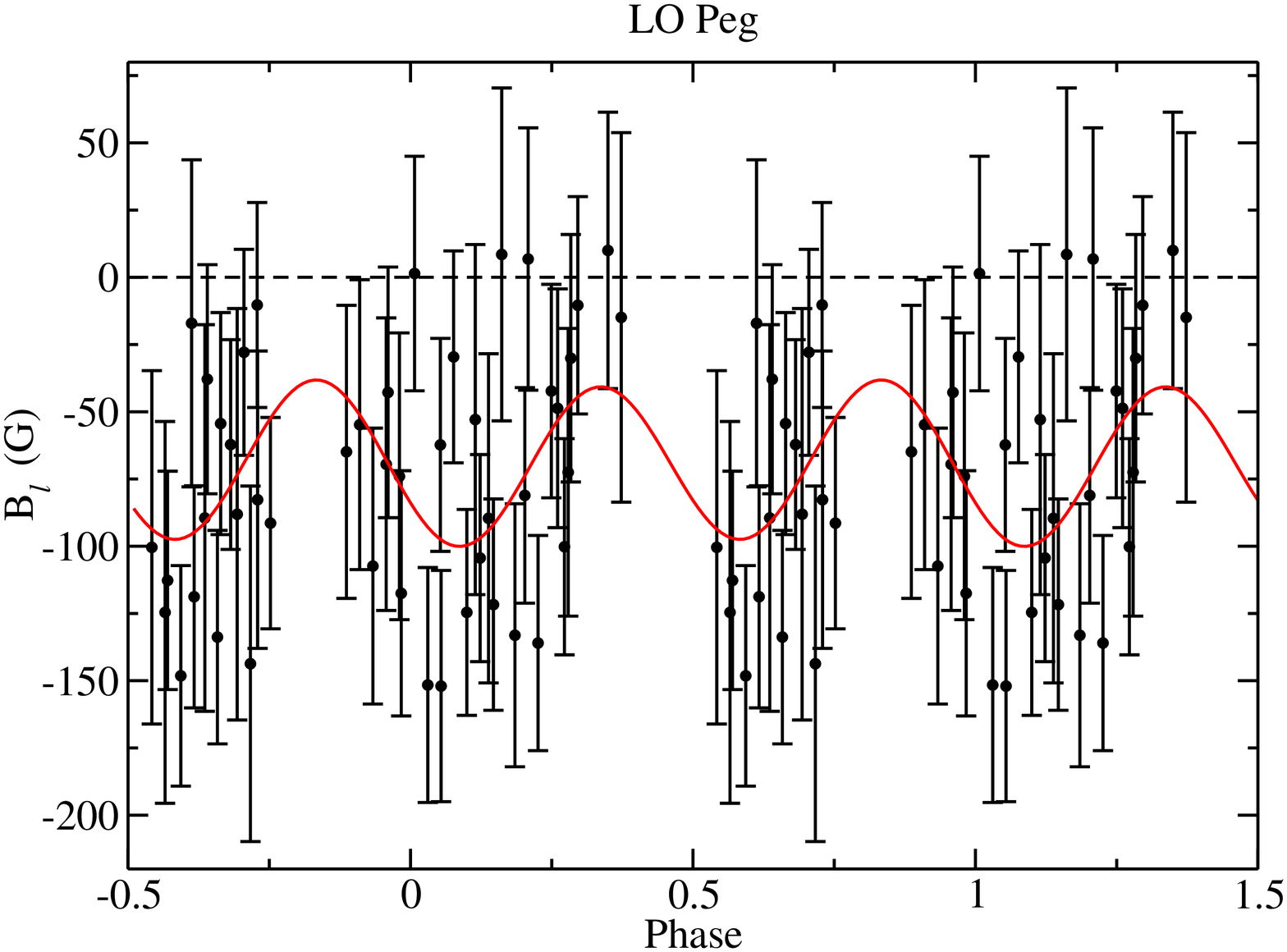}
  \includegraphics[width=2.8in]{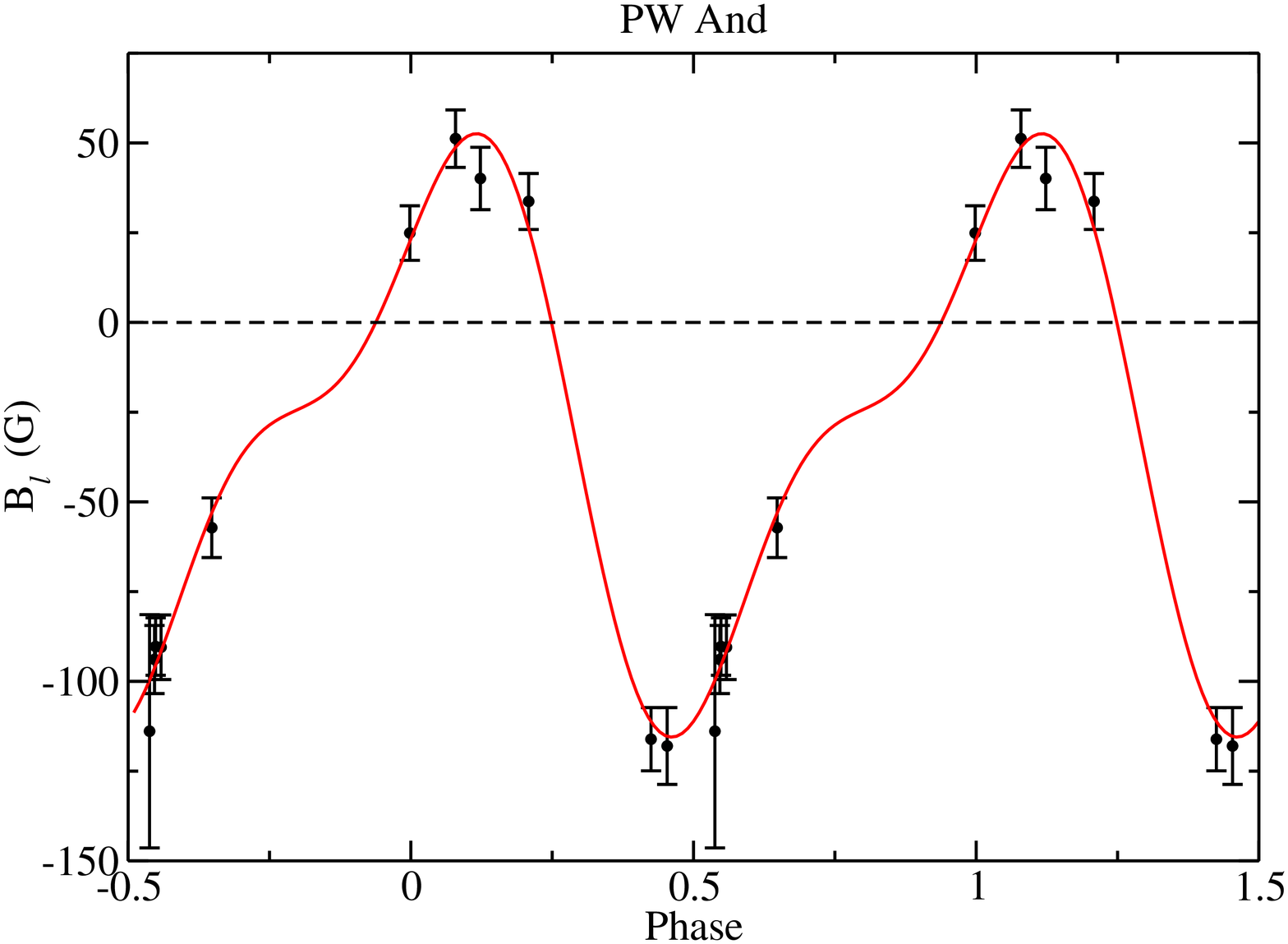}
  \includegraphics[width=2.8in]{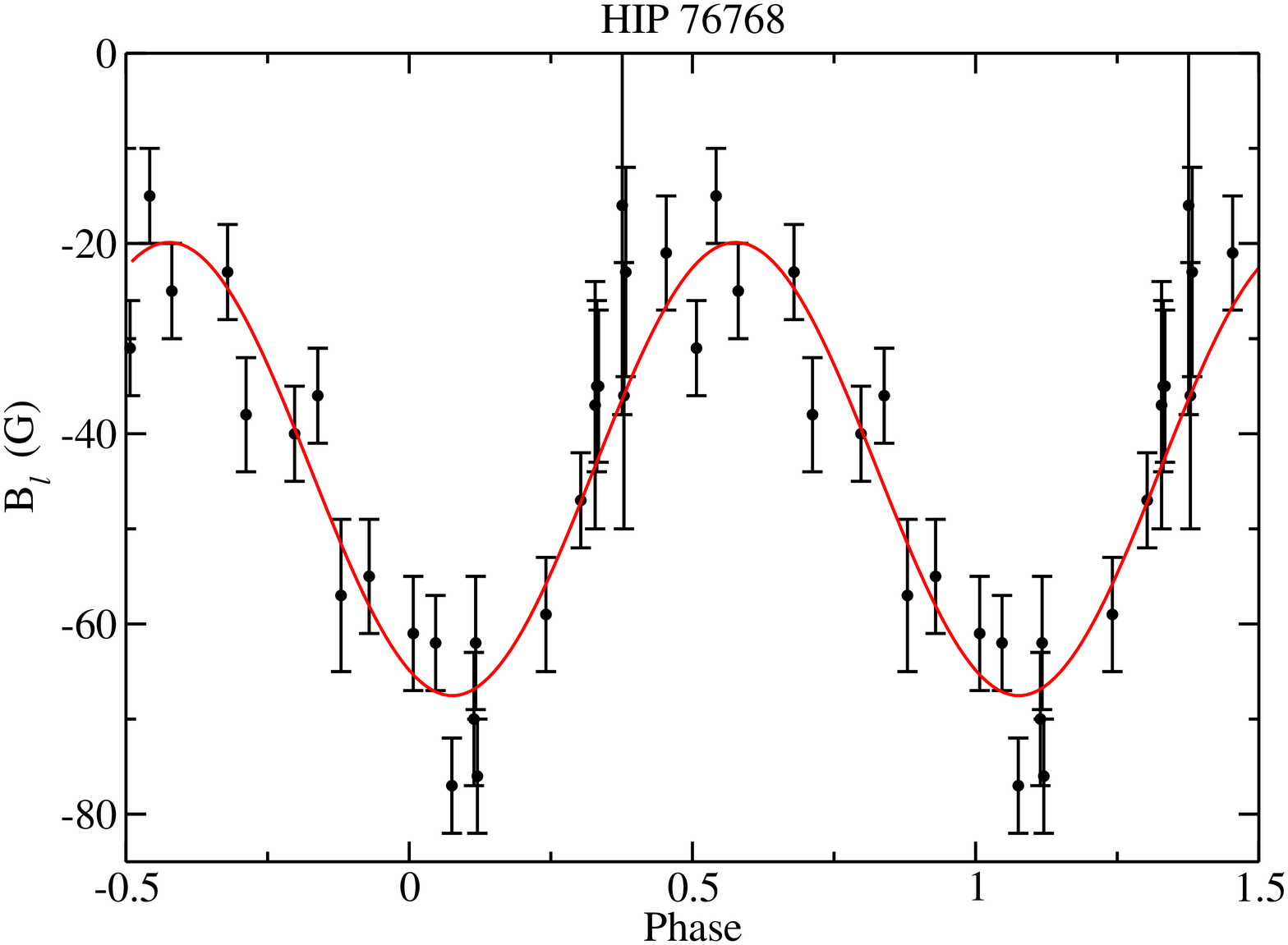}
  \includegraphics[width=2.8in]{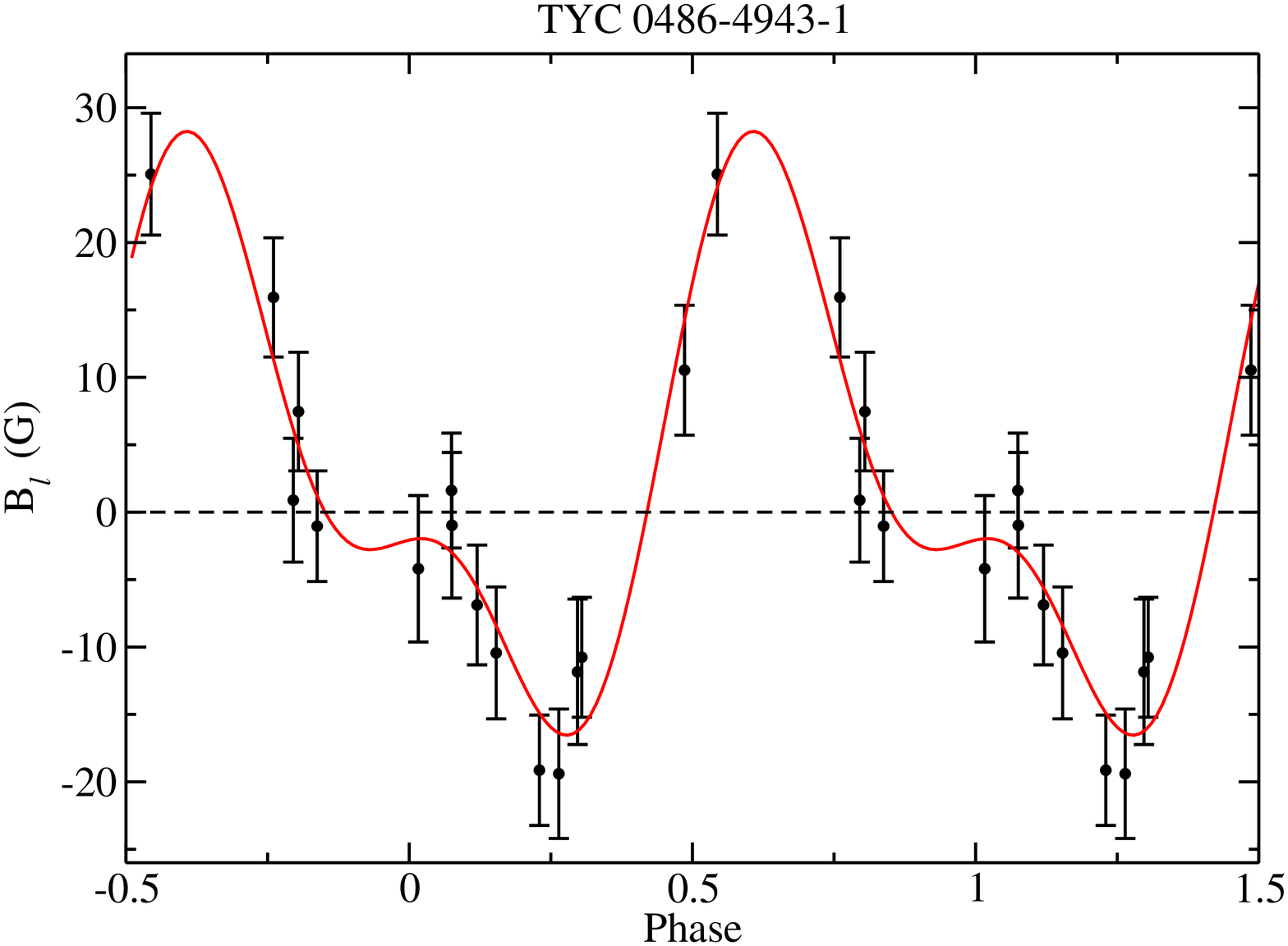}
  \caption{Longitudinal magnetic fields measured for stars in our sample, phased with the rotation periods derived in Sect.~\ref{rotational-period}.  The solid line is the fit through the observations.   }
  \label{fig-bz}
\end{figure*}

\begin{figure*}
  \centering
  \includegraphics[width=2.8in]{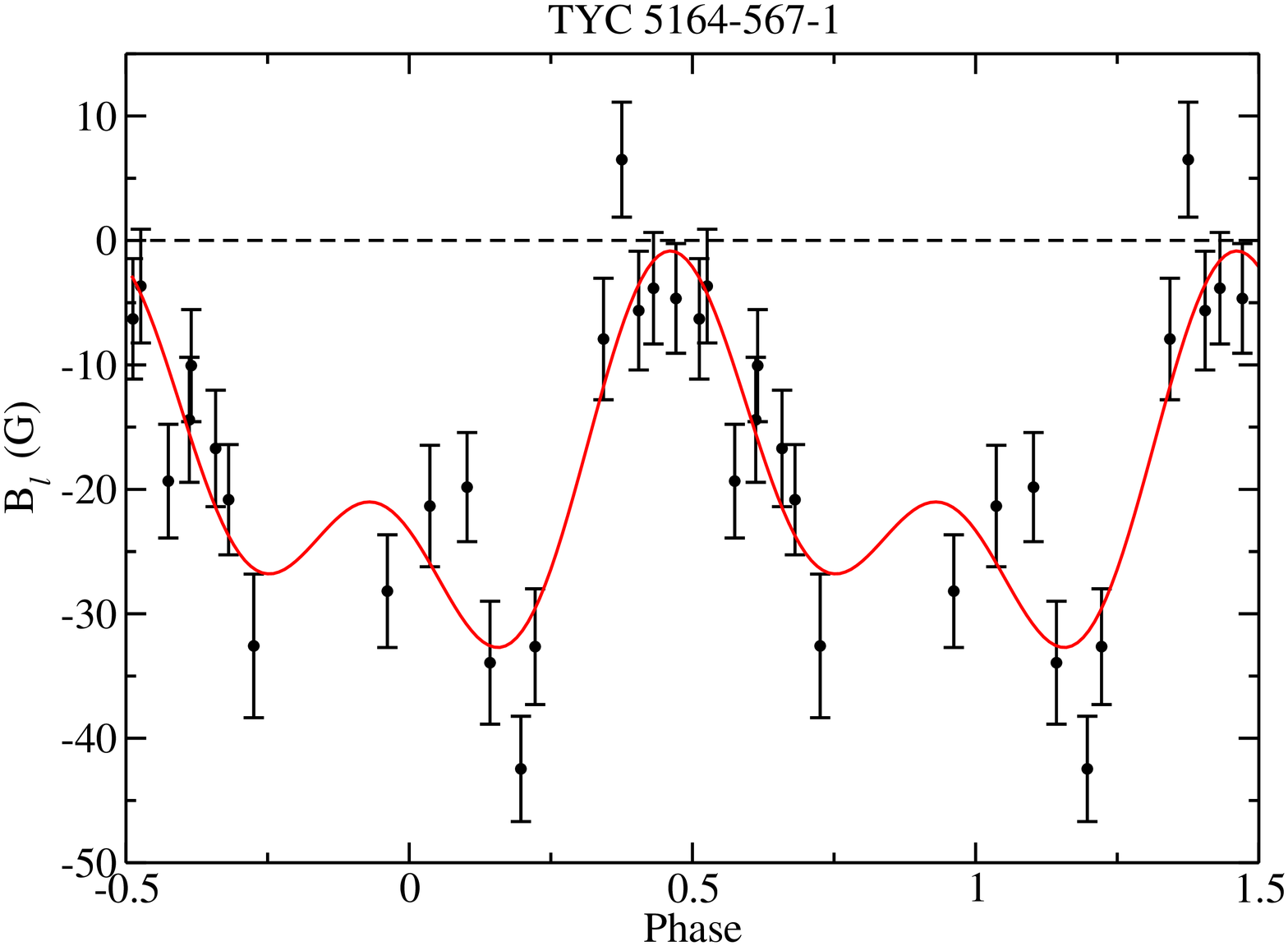}
  \includegraphics[width=2.8in]{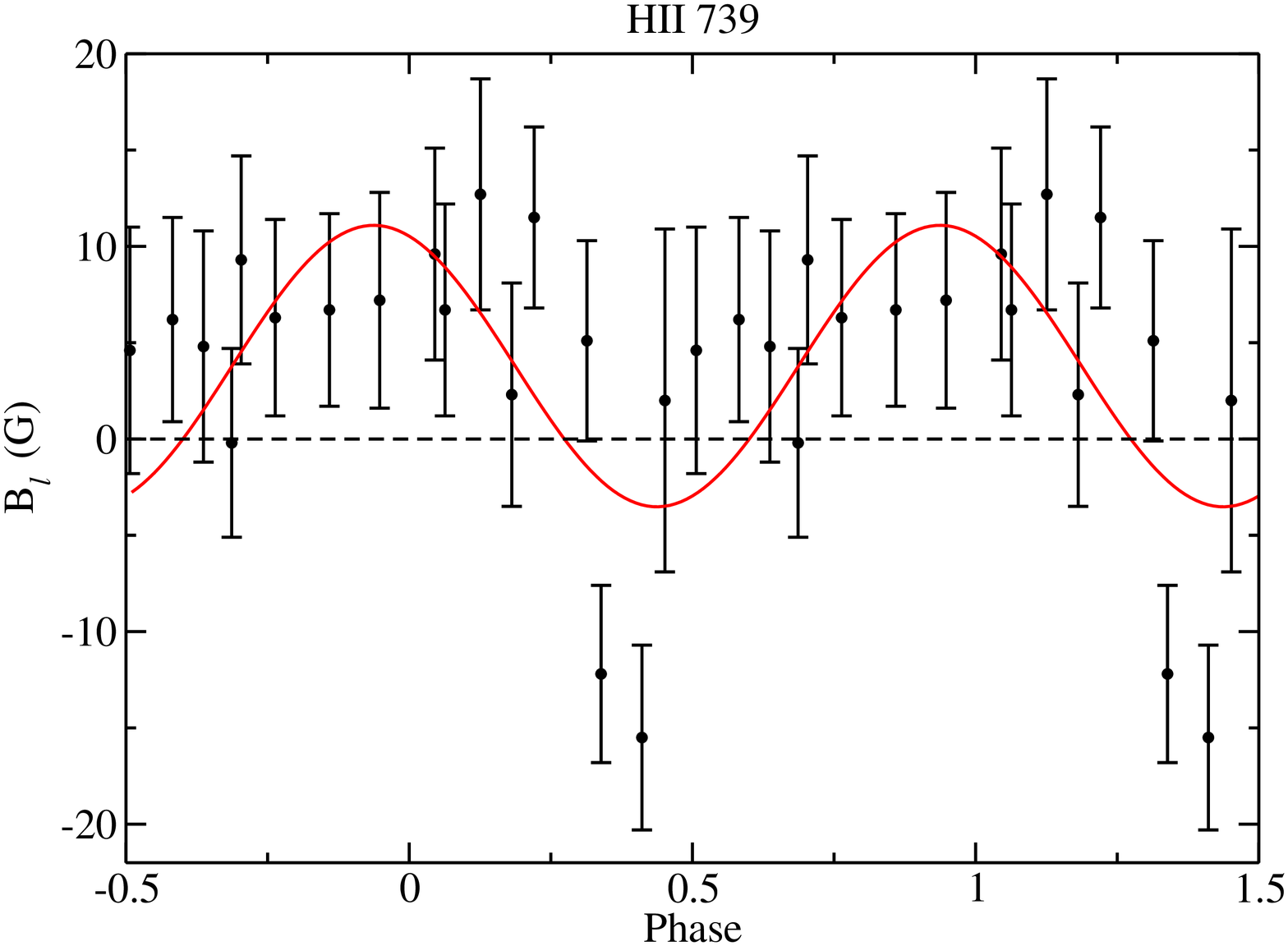}
  \includegraphics[width=2.8in]{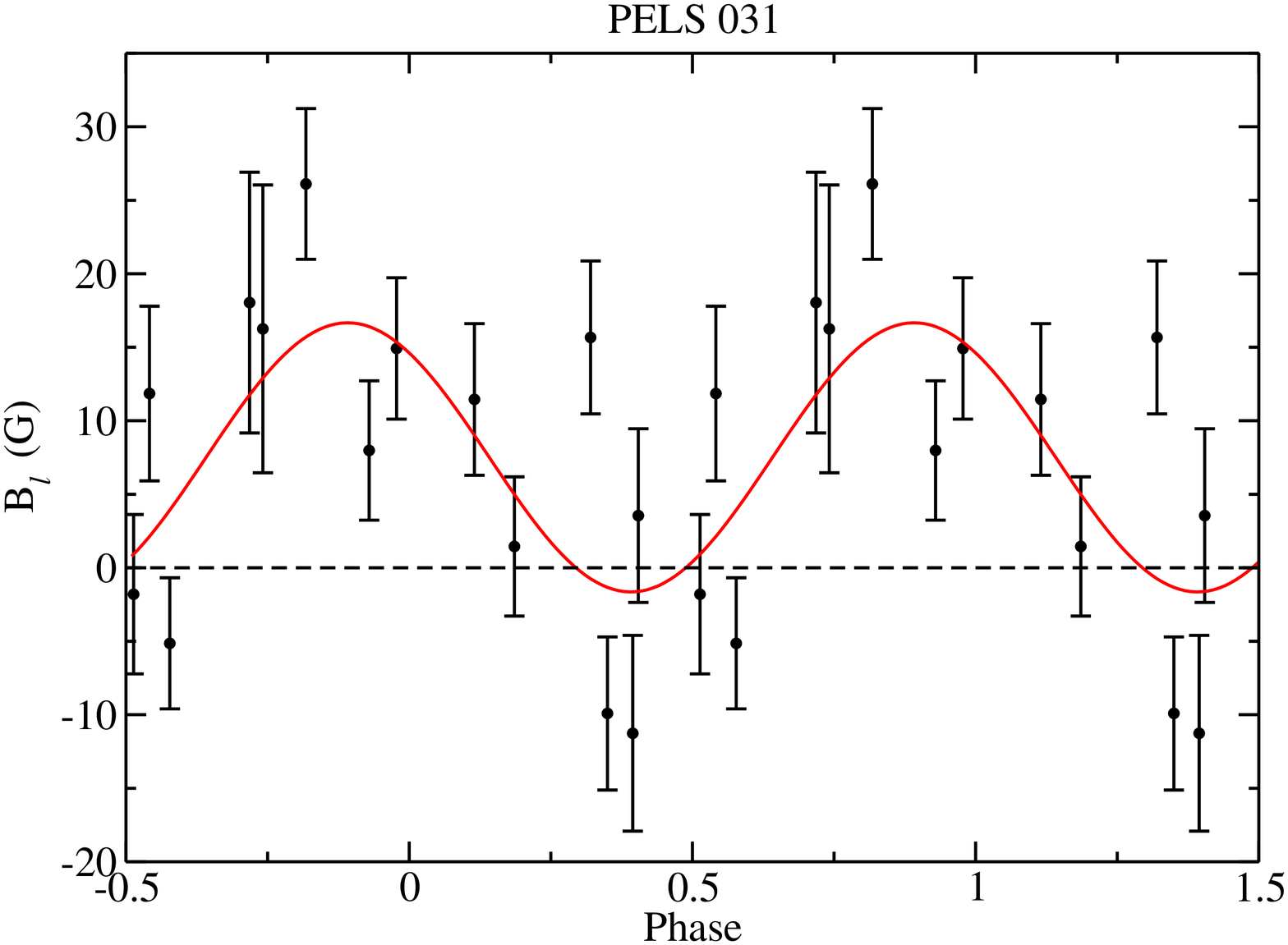}
  \includegraphics[width=2.8in]{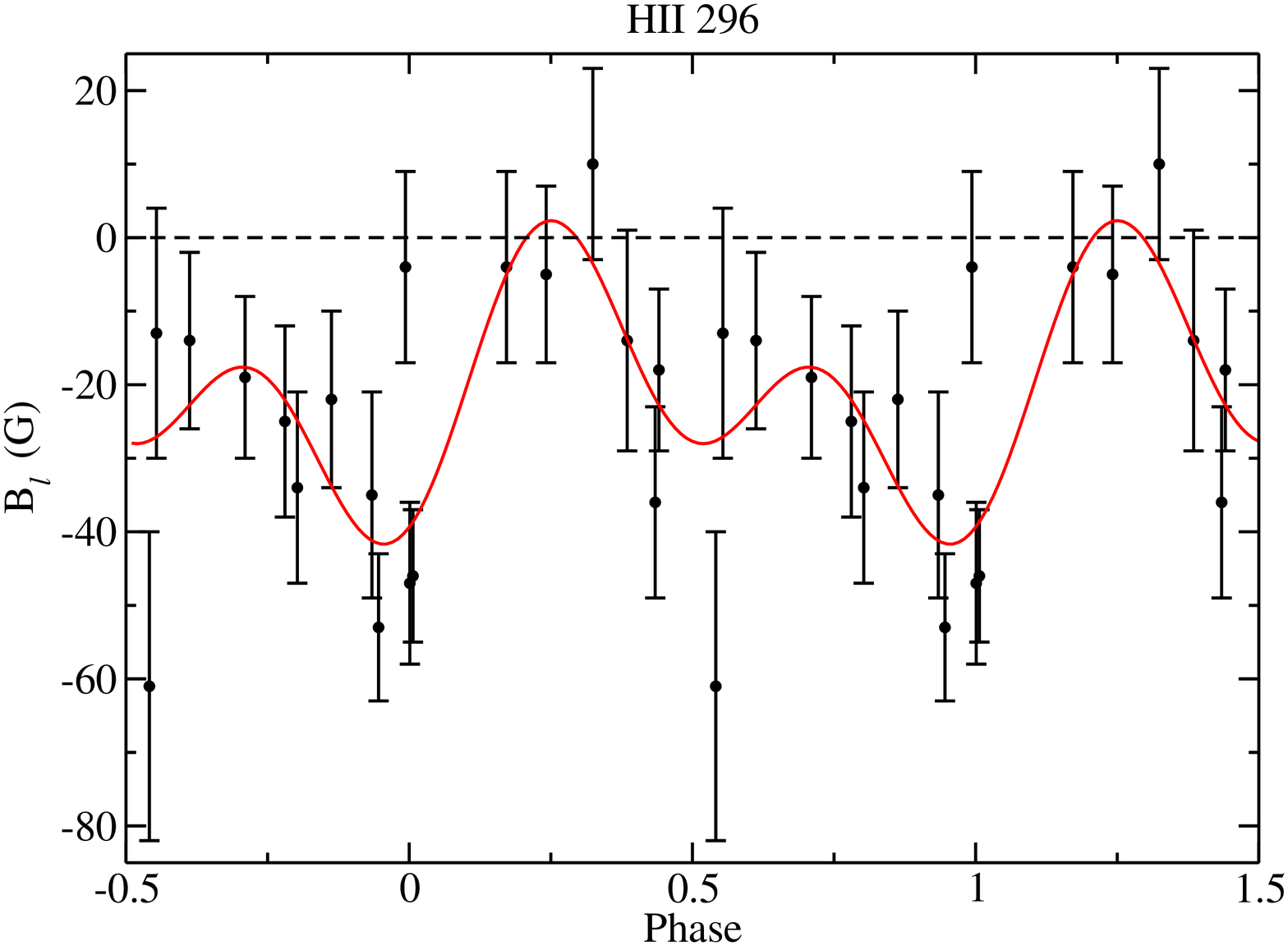}
  \includegraphics[width=2.8in]{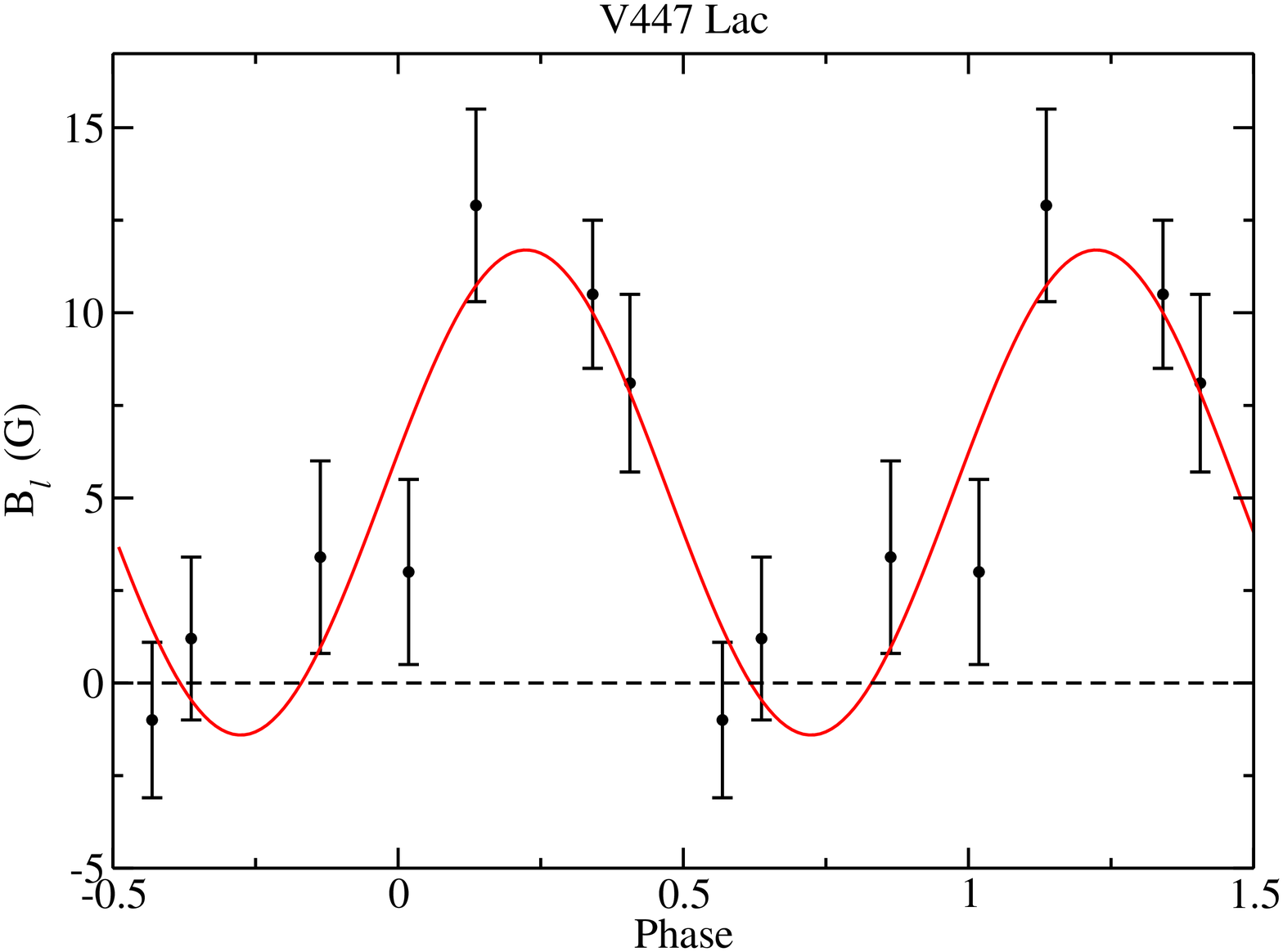}
  \includegraphics[width=2.8in]{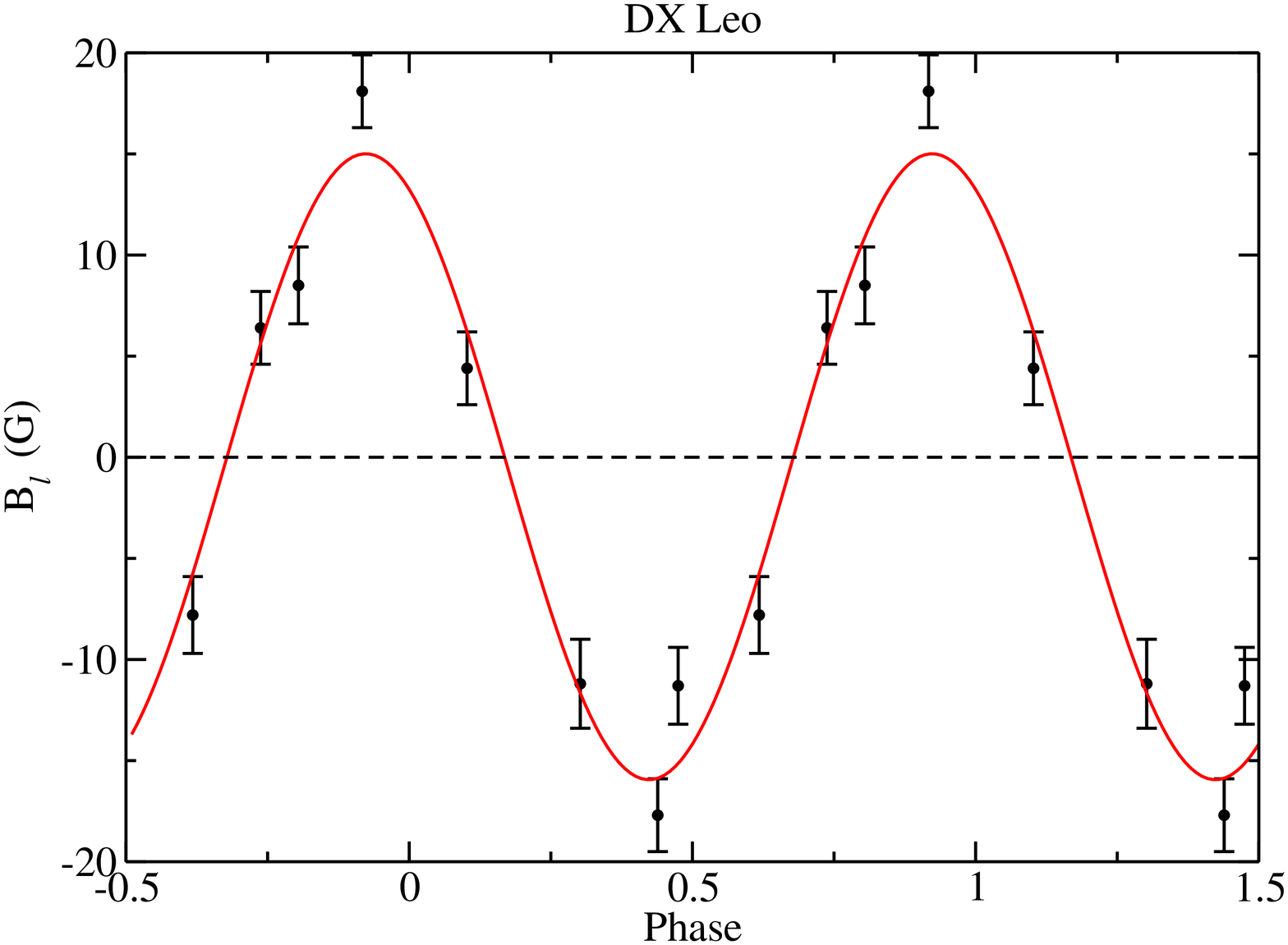}
  \includegraphics[width=2.8in]{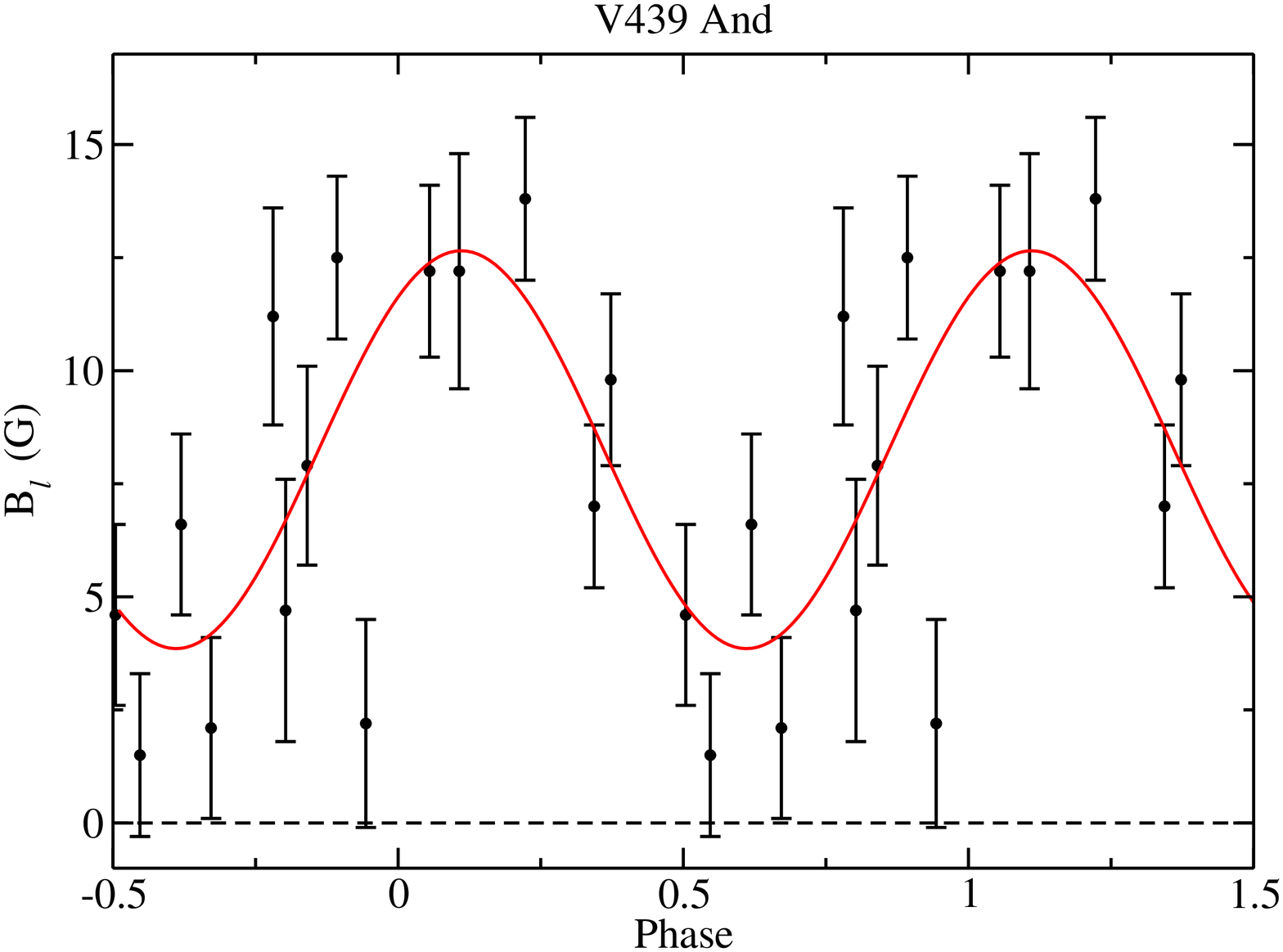}
  \caption{Longitudinal magnetic fields measured for stars in our sample phased with rotation period, as in Fig.~\ref{fig-bz}.  }
  \label{fig-bz2}
\end{figure*}

\begin{figure*}
  \centering
  \includegraphics[width=2.0in]{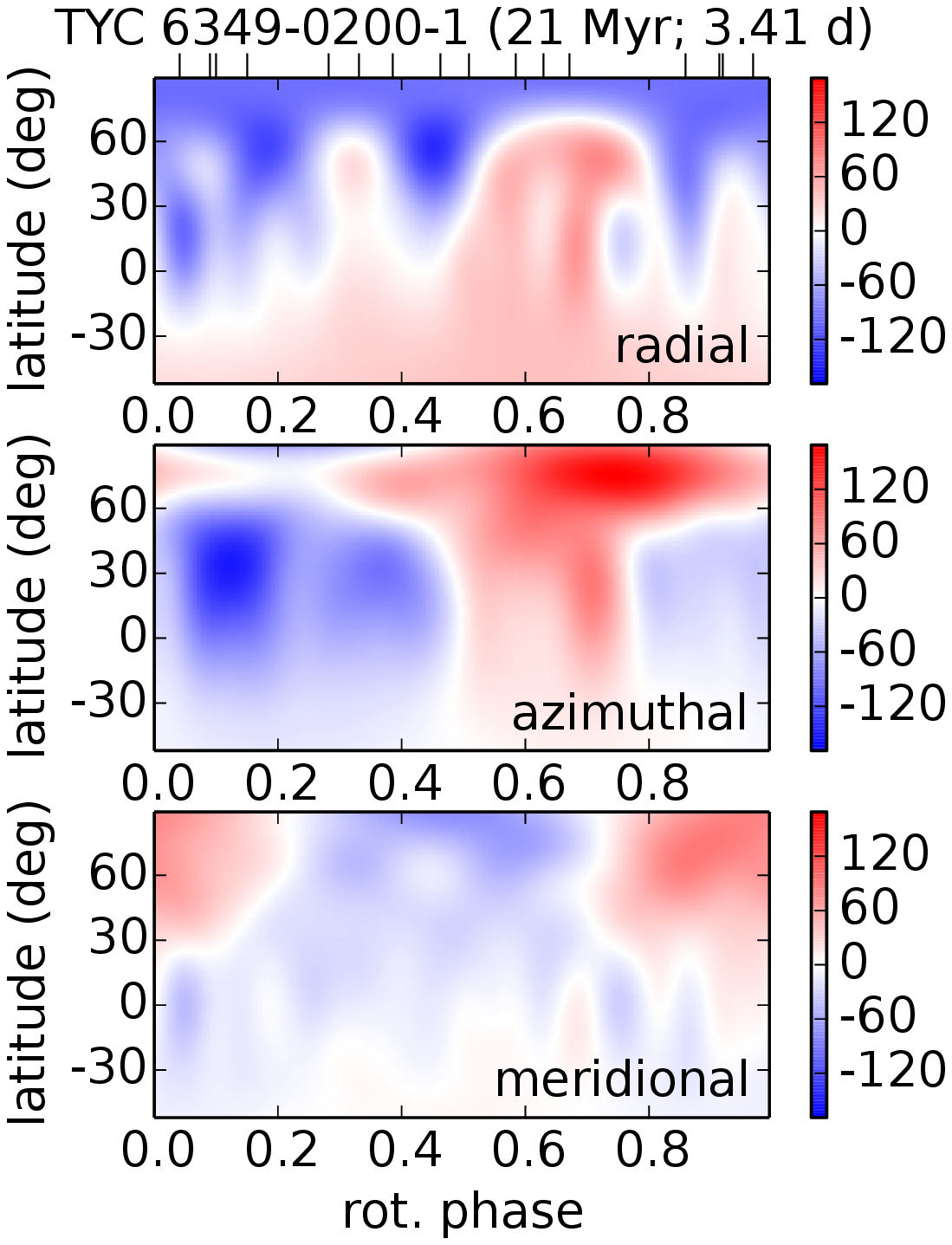} 
  \includegraphics[width=2.0in]{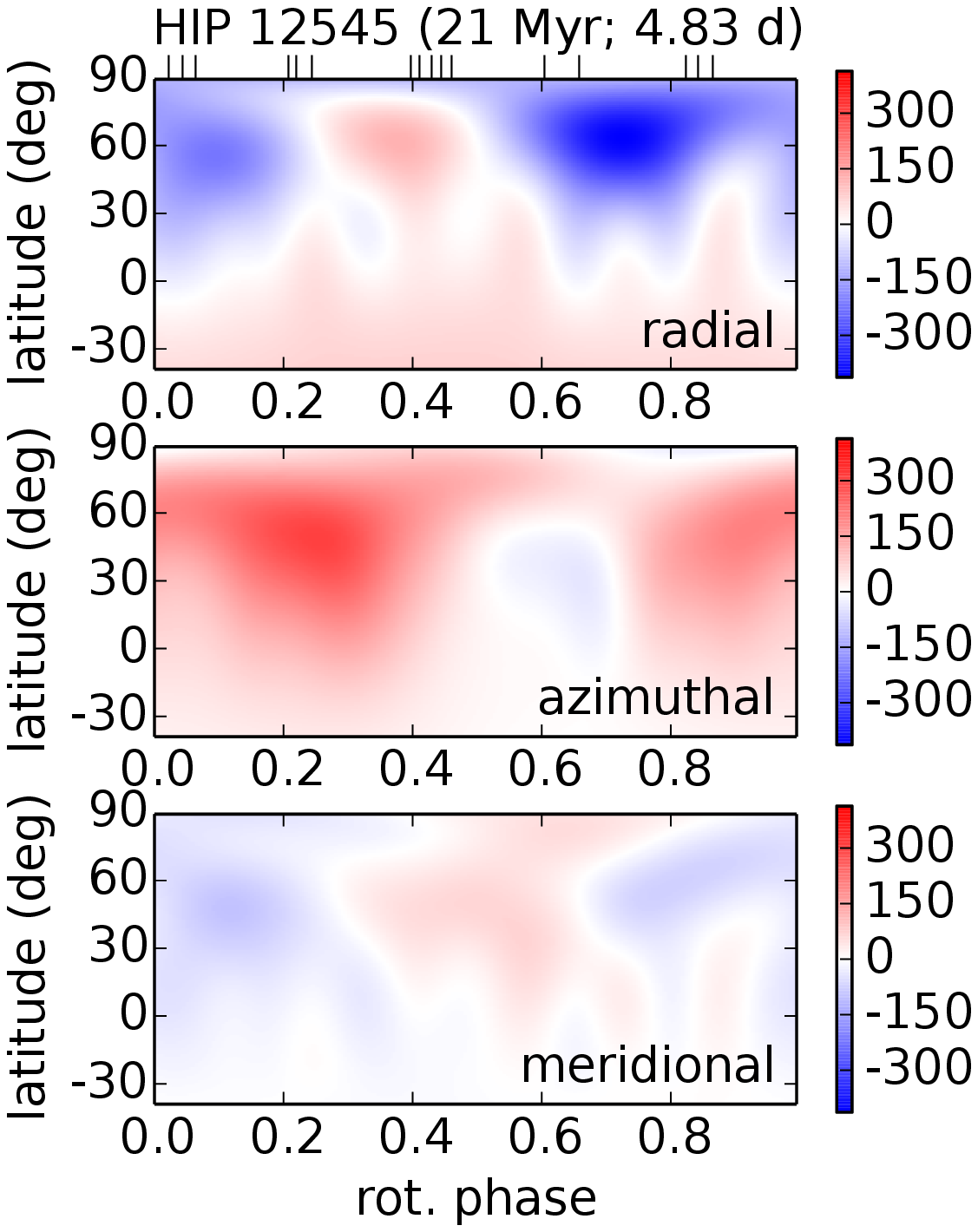} 
  \includegraphics[width=2.0in]{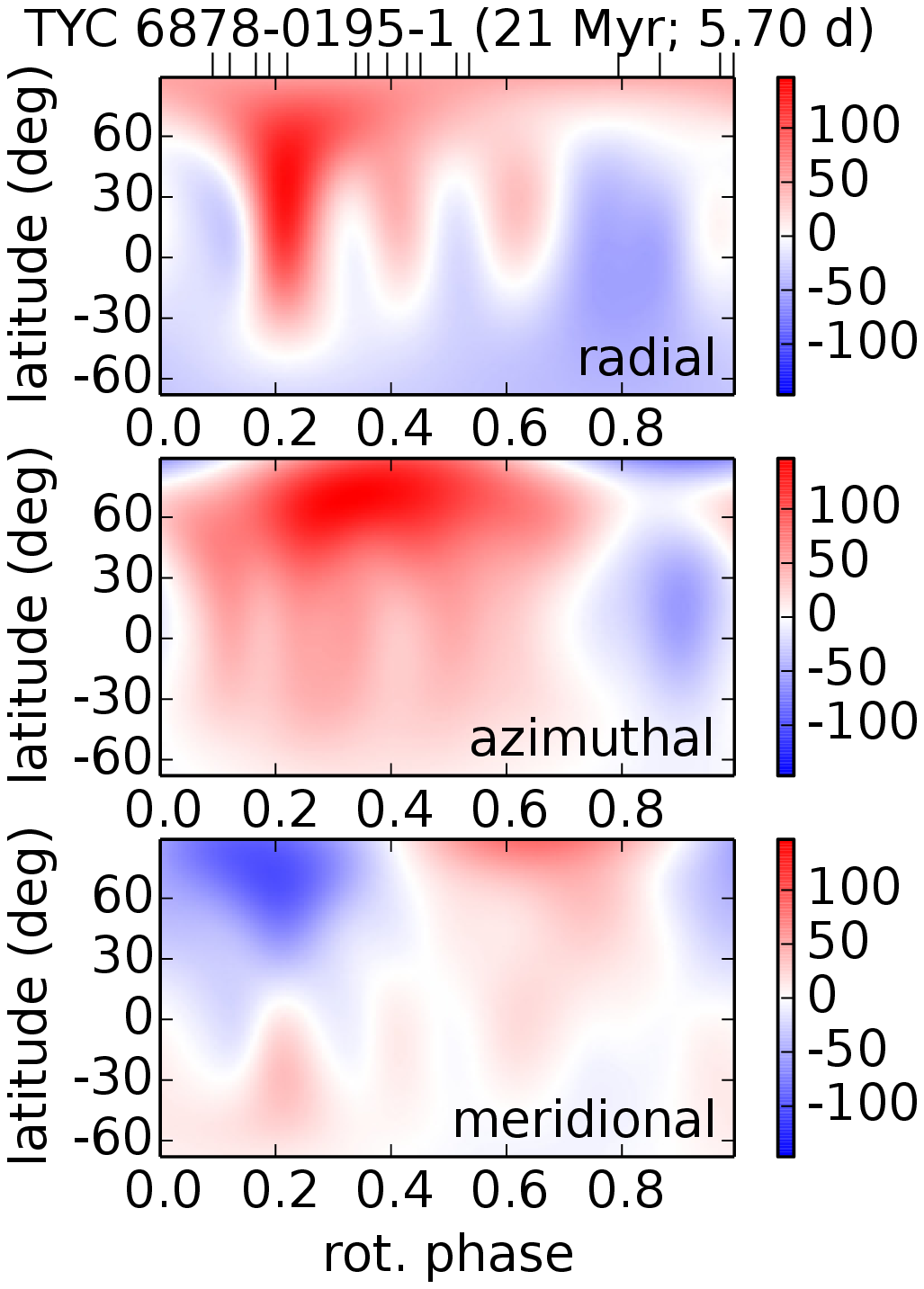}
  \includegraphics[width=2.0in]{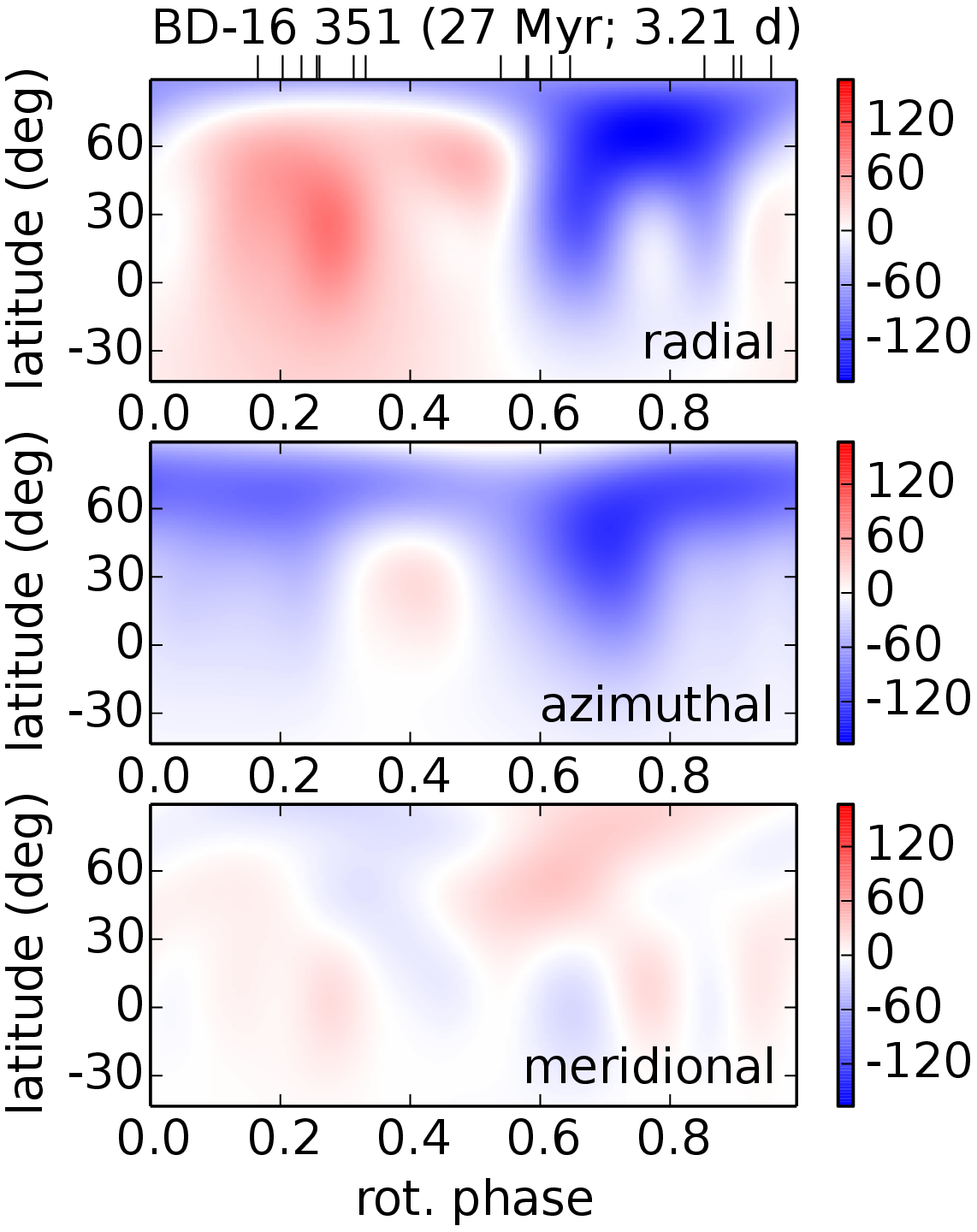} 
  \includegraphics[width=2.0in]{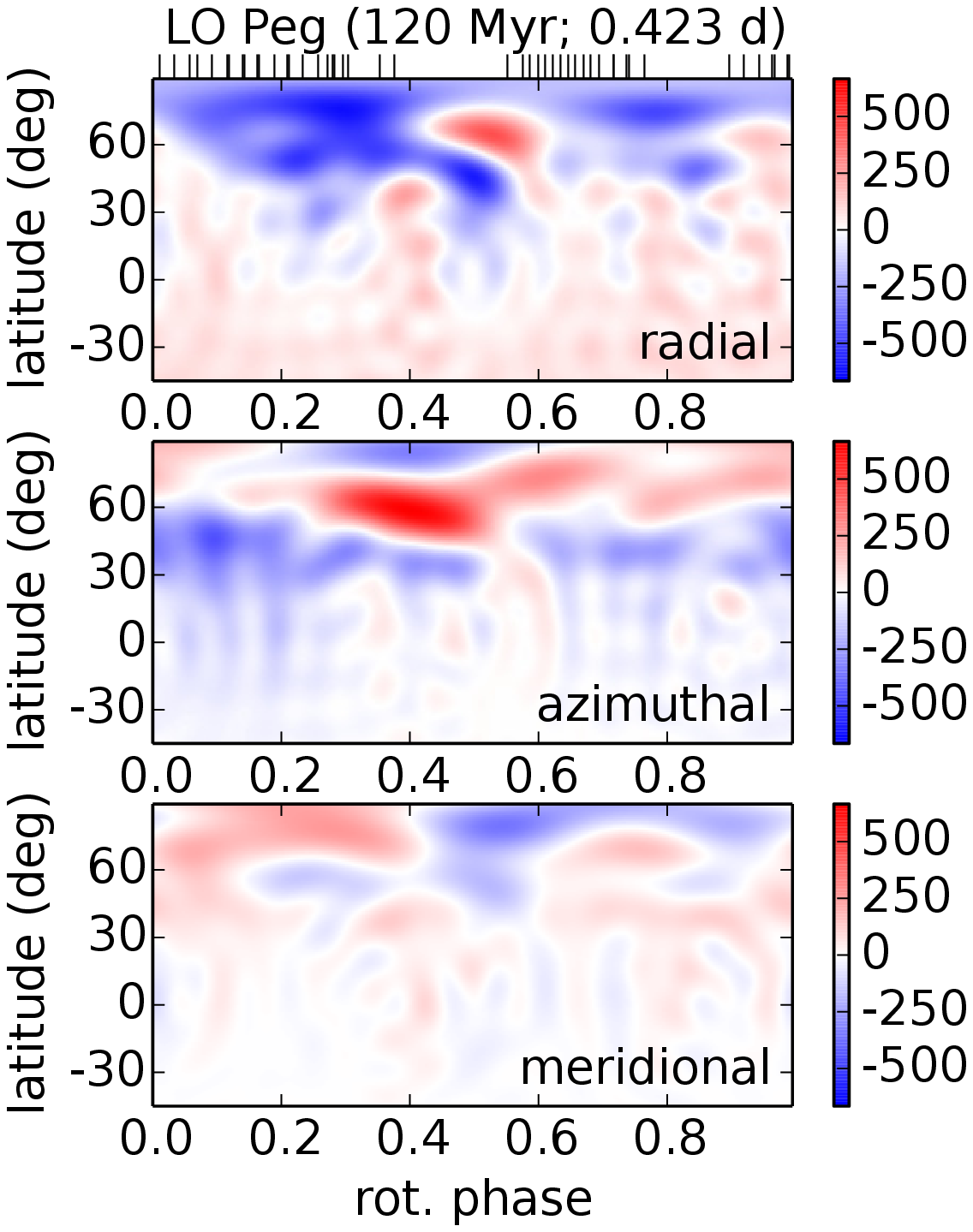} 
  \includegraphics[width=2.0in]{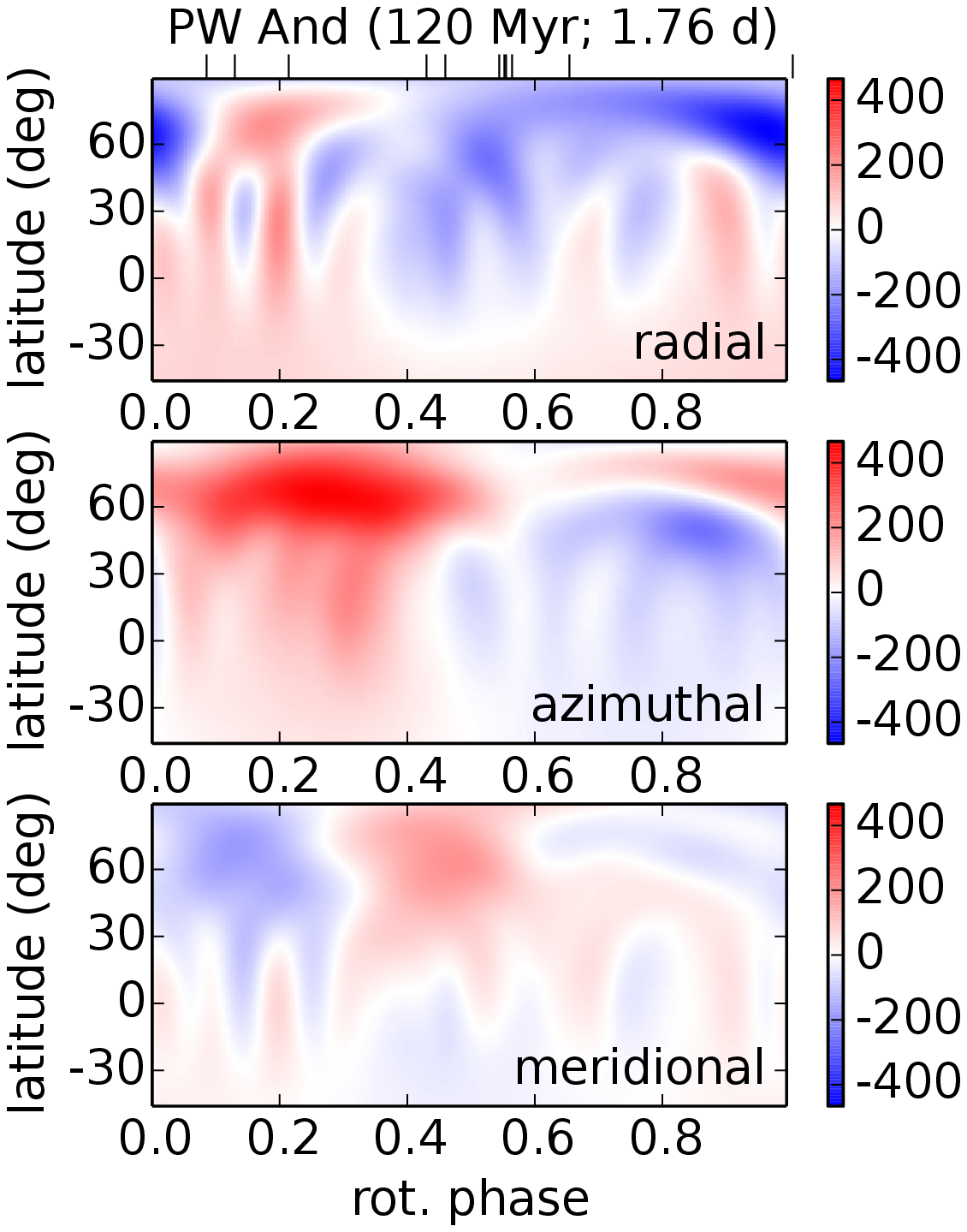} 
  \includegraphics[width=2.0in]{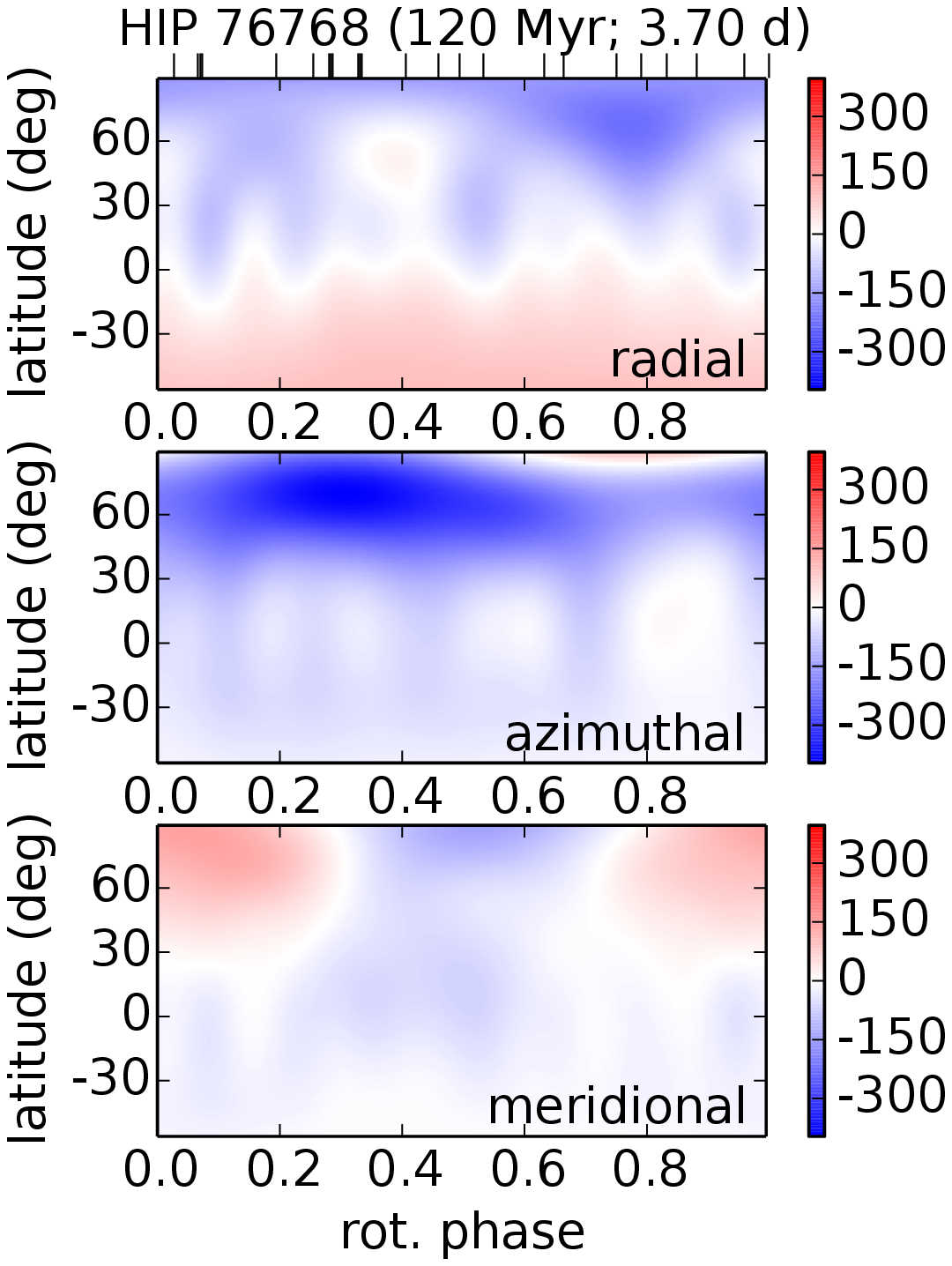} 
  \includegraphics[width=2.0in]{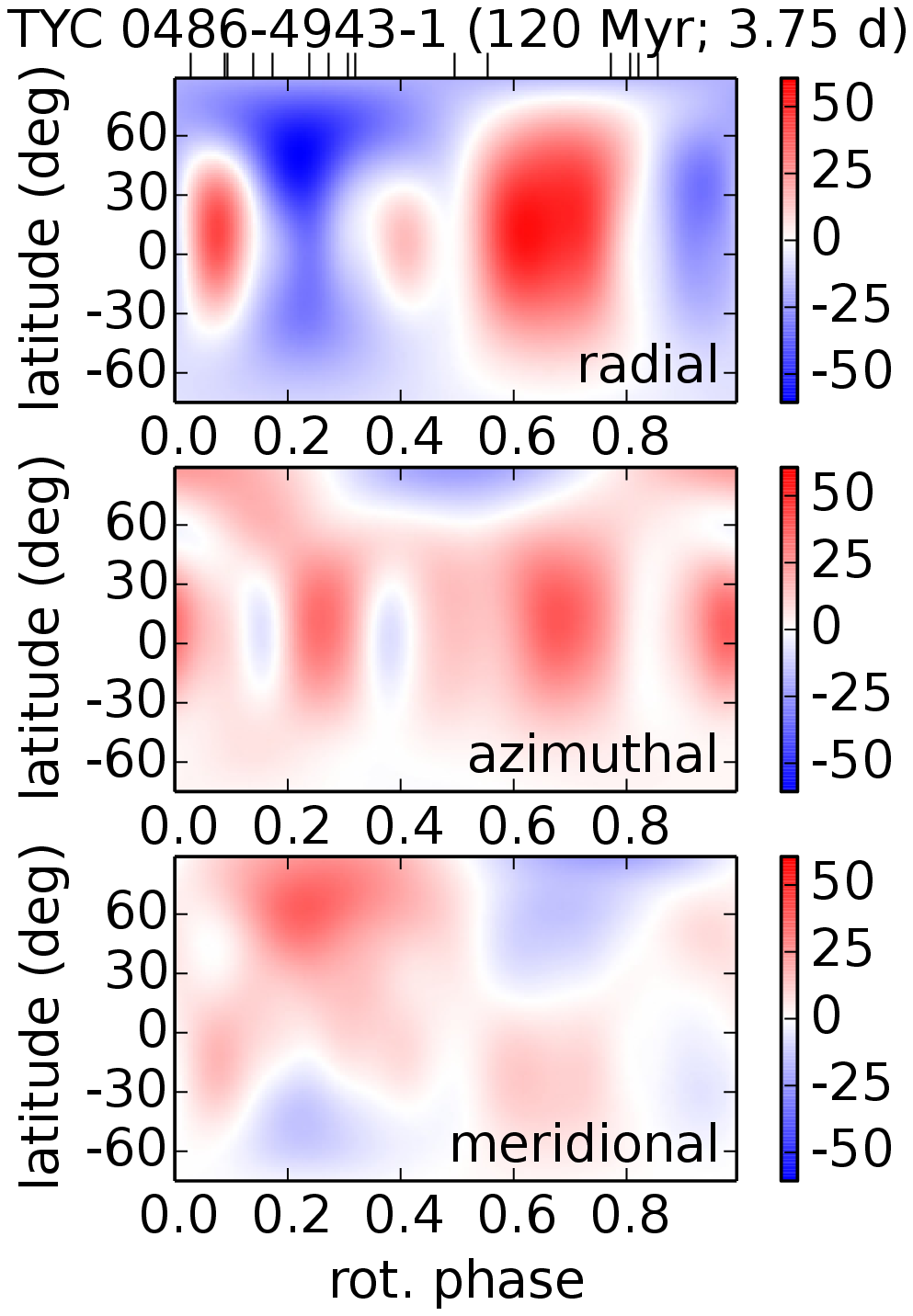} 
  \includegraphics[width=2.0in]{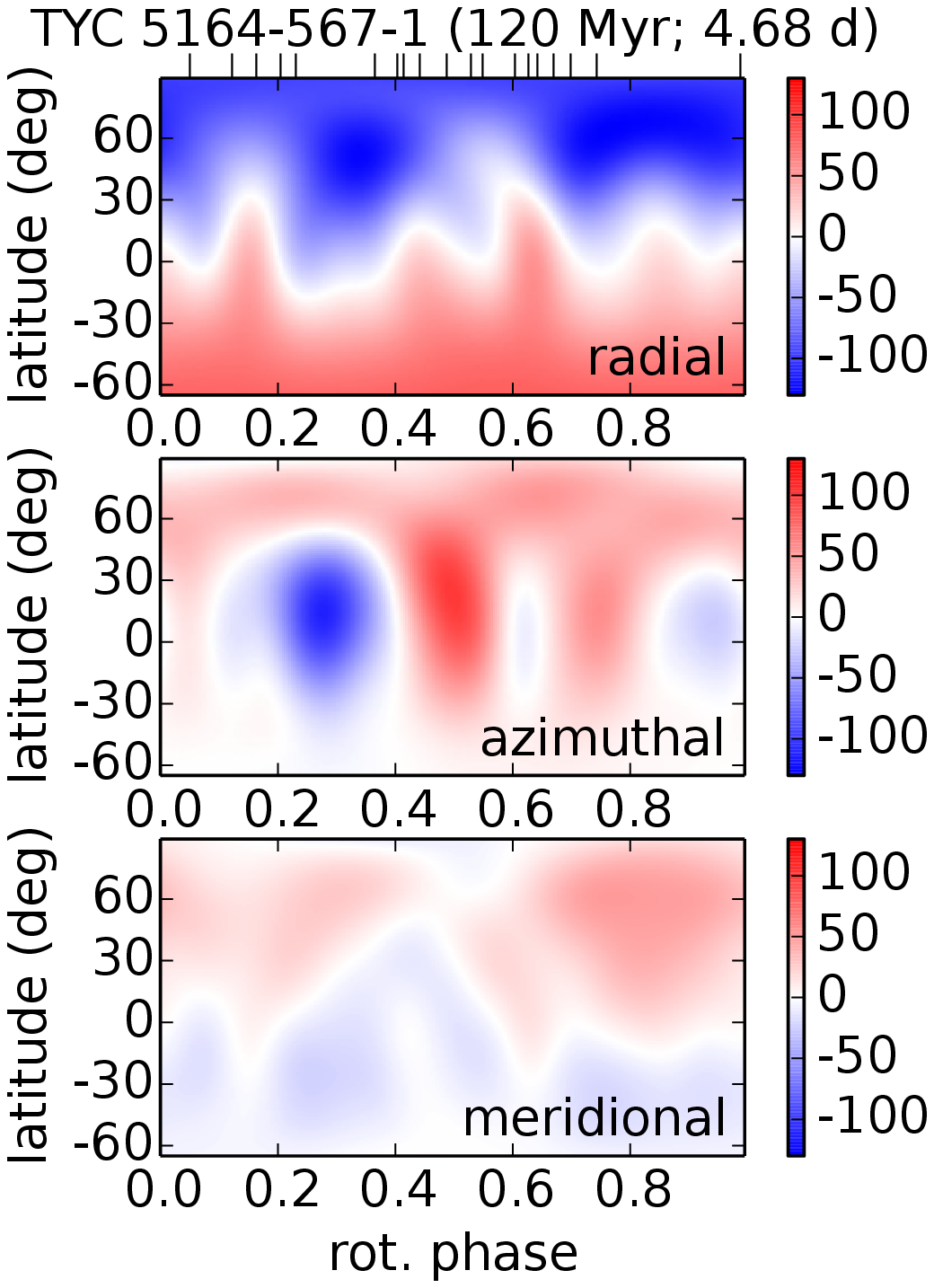} 
  \caption{Maps of the derived magnetic fields for the stars in this study.  
  Plotted are the radial (top), azimuthal (middle), and meridional (bottom) components of the magnetic fields.  Sub-figures are labeled by the name of the star, followed by its age and rotation period.  Tick marks at the top of the figure indicate phases at which observations were obtained.  }
  \label{fig-zdi-maps}
\end{figure*}

\begin{figure*}
  \centering
  \includegraphics[width=2.0in]{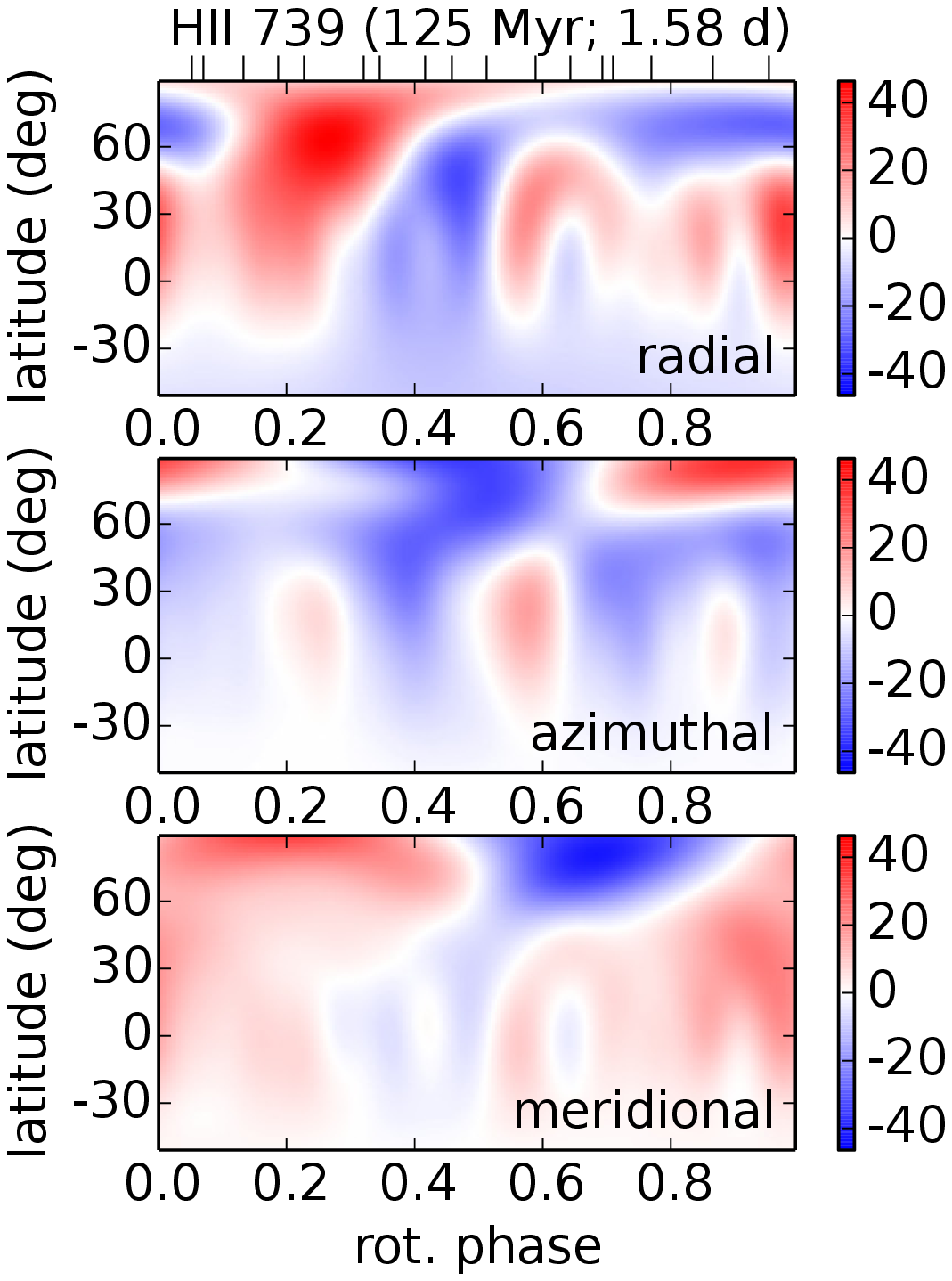} 
  \includegraphics[width=2.0in]{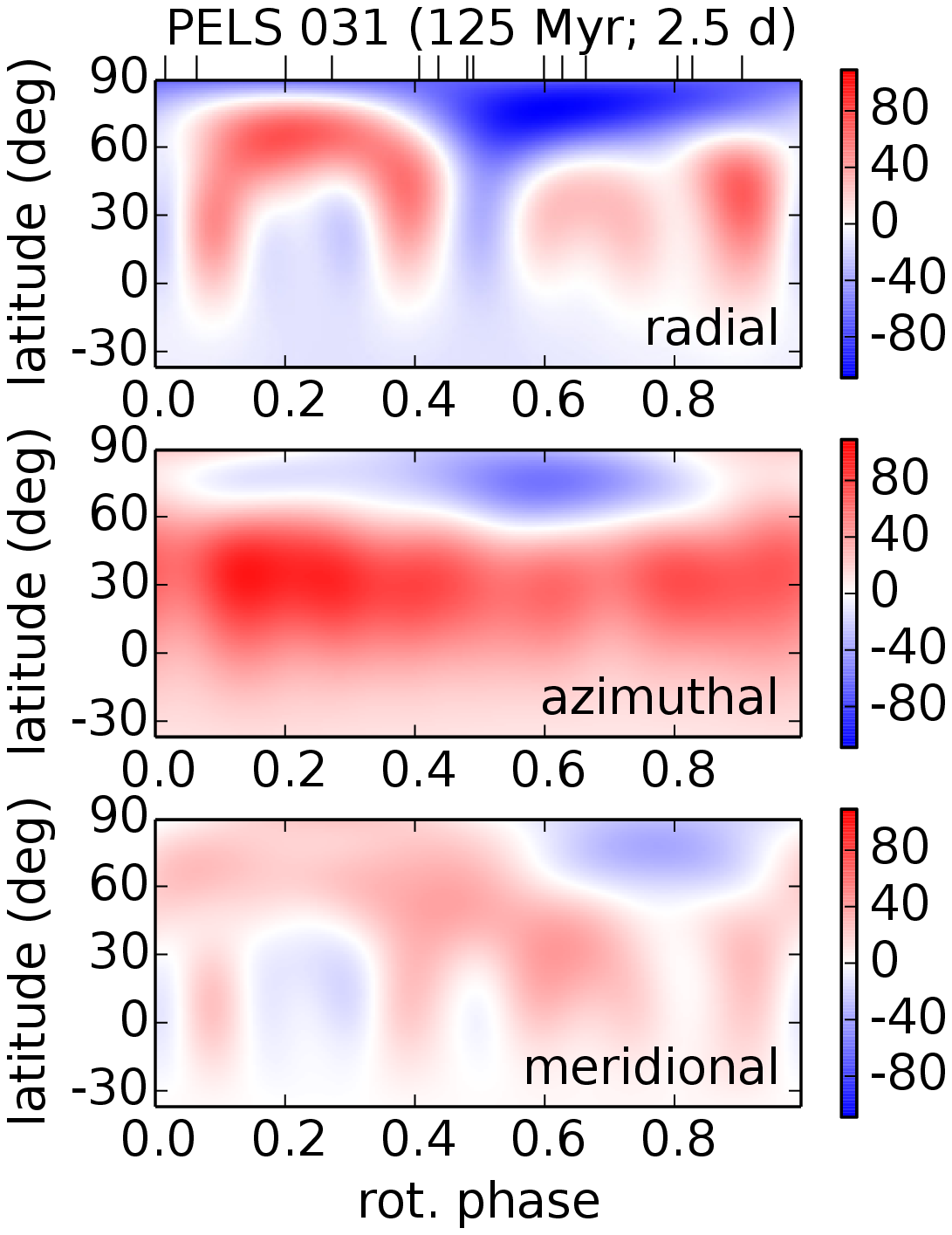} 
  \includegraphics[width=2.0in]{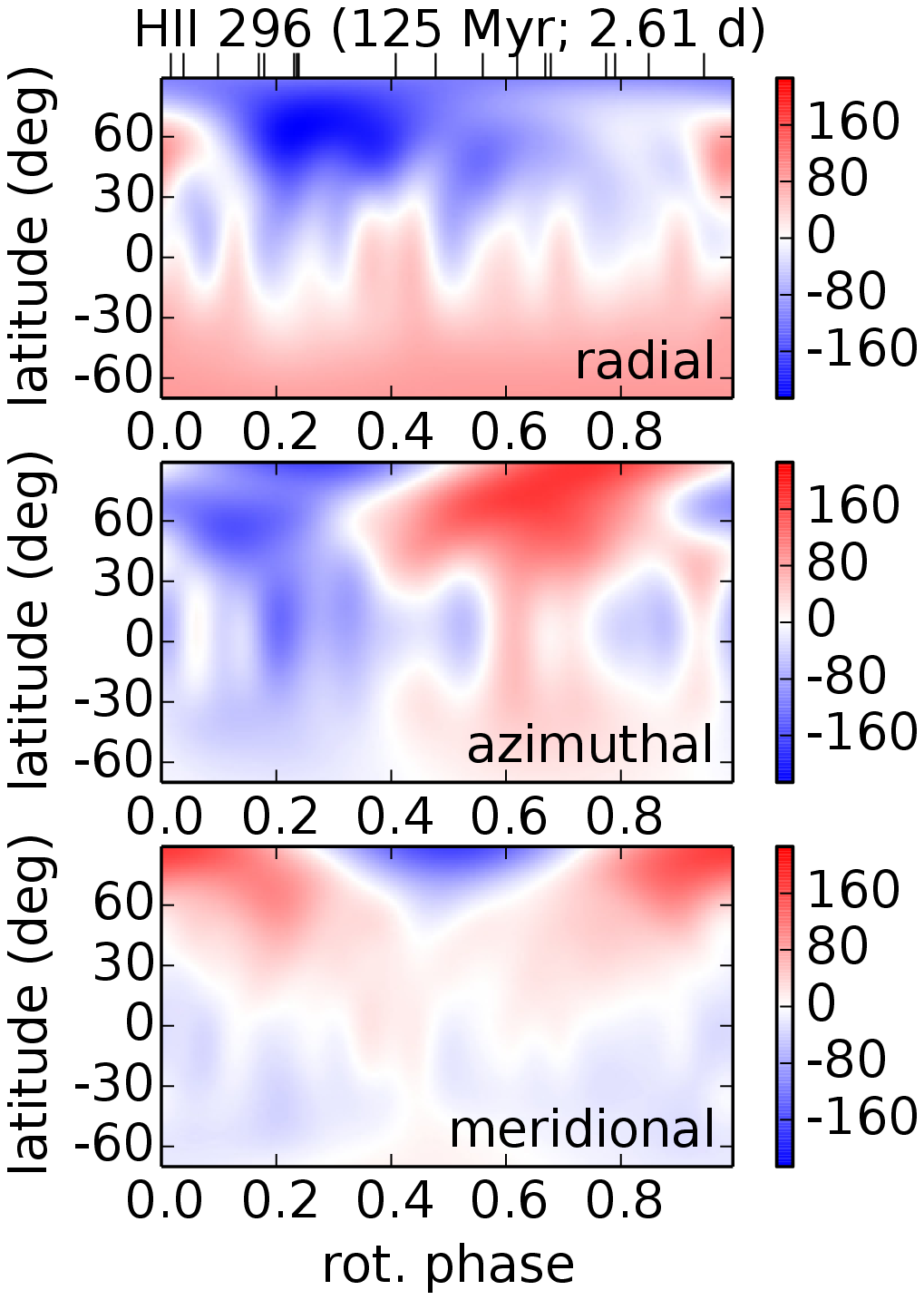} 
  \includegraphics[width=2.0in]{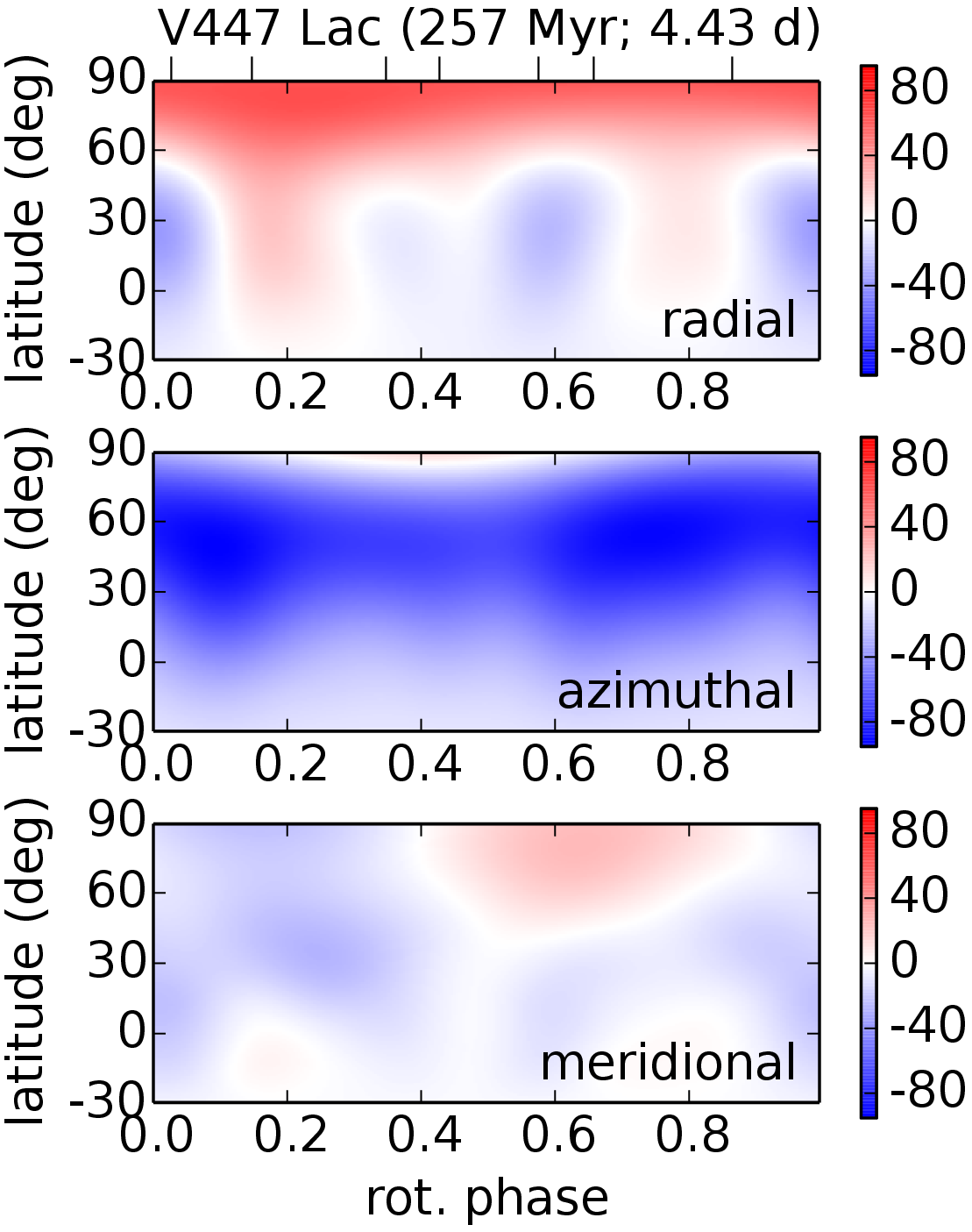} 
  \includegraphics[width=2.0in]{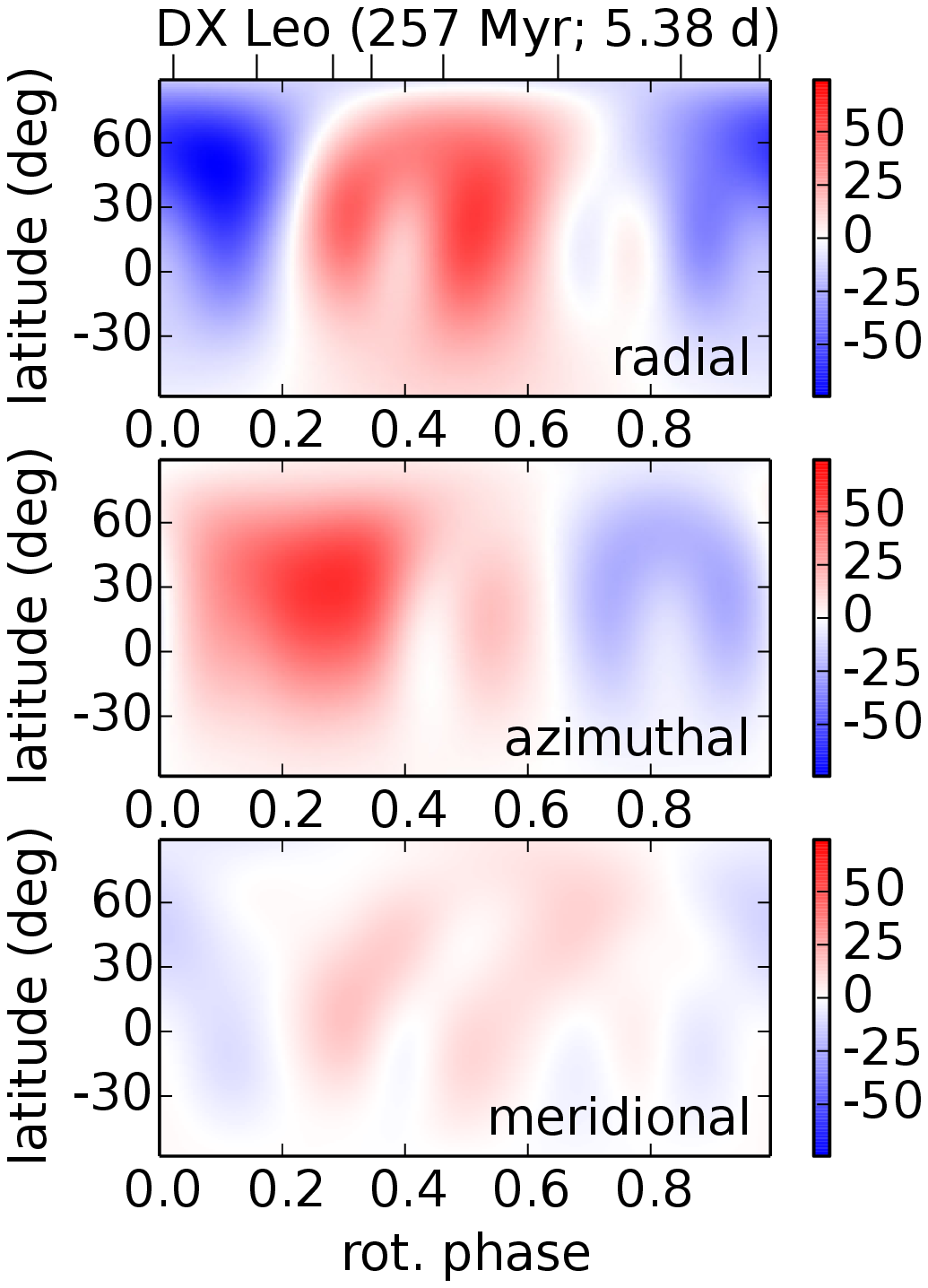} 
  \includegraphics[width=2.0in]{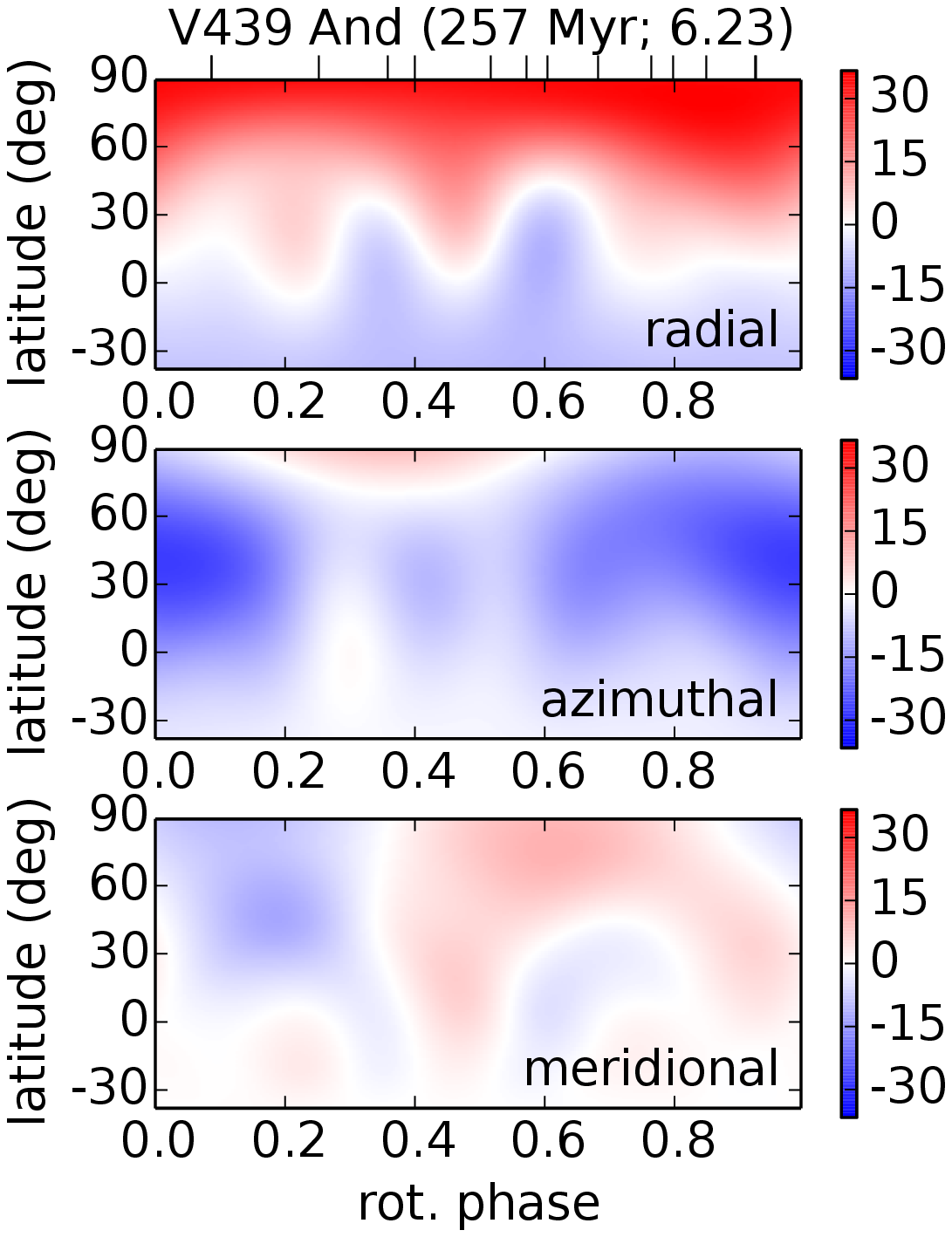} 
  \caption{Maps of the derived magnetic fields for the stars in this study, as in Fig.~\ref{fig-zdi-maps}.   }
  \label{fig-zdi-maps2}
\end{figure*}

\subsection{TYC 6349-200-1}
TYC 6349-200-1 (AZ Cap, HD 358623, BD-17 6128) is a member of the $\beta$ Pic association \citep{Zuckerman2004-young-nearby-assoc-rev, Torres2006-search-youngAssocaitions, Torres2008-youngNearbyAssoc}.  \citet{Messina2010-RACE-OC-periods} find a photometric rotation period of P = $3.41 \pm 0.05$ days.  \citet{Kiraga2012-photo-periods} find a photometric rotation period of P = 3.403 days, confirming this value.  We find periods consistent with these values from our search with longitudinal magnetic fields, and our search with ZDI.  Therefore we adopt a value of P = $3.41 \pm 0.05$ days.

This period is marginally inconsistent with our estimated radius ($0.96 \pm 0.07$ $R_\odot$) and \vs\ ($15.8 \pm 0.5$ \kms), with the radius being too small for the period and \vs\ by $\sim 1.5\sigma$.  This leads to a poorly constrained inclination.  \citet{Torres2006-search-youngAssocaitions} note a visual secondary at 2.2'' separation, but it is 2 mag fainter and thus has a minimal impact on the derived luminosity.  The presence of a second star would only serve to reduce the radius estimate leading to a worse discrepancy.  There is some evidence for extinction towards the star, with $E(B-V) = 0.16$.  Therefore, we determined an inclination by searching for the value that gives us a ZDI map with the maximum entropy.  This is the same procedure as that used to determine periods from ZDI.  We find a maximum entropy inclination of $52 \pm 20^\circ$.  This inclination would imply a radius of $1.35 \pm 0.45$ $R_\odot$.  We adopt the maximum entropy inclination, but the photometric radius ($0.96 \pm 0.07$ $R_\odot$) since it is formally more precise and obtained in a fashion consistent with the rest of our sample.

\subsection{HIP 12545}
HIP 12545 (BD+05 378, TYC 53-30-1) is a member of the $\beta$ Pic association \citep{Zuckerman2004-young-nearby-assoc-rev, Torres2006-search-youngAssocaitions, Torres2008-youngNearbyAssoc}.
A photometric rotation period for the star was determined by \citet{Messina2010-RACE-OC-periods}, finding P = $4.83 \pm 0.01$ days (note: there is a misprint in one of their Table 4, which has the wrong value for this period).  \citet{Kiraga2012-photo-periods} found a period of 4.831 days, which confirms this value.  

Our attempts to measure a rotation period from longitudinal magnetic field measurements produced ambiguous results, due to the rather complicated variability of the longitudinal field.  This can be seen in the $V$ LSD profiles, which become complicated at some phases, but never reverse sign.  The period measurement for radial velocity produces three ambiguous values, however one is consistent with the literature period.  Our measurement of the rotation period through ZDI produced a best period of P = 4.83 days.  Therefore we can confirm the literature rotation period.  

However, this period seems to be inconsistent with our derived \vs\ and radius (based on luminosity and \teff).  These values imply a period P $< 3.75$ days, which is inconsistent with the literature values and our best ZDI period.  The \vs\ should be very accurate, as the observed line profiles are well fit by the synthetic spectrum with rotational broadening.  With a Hipparcos parallax the distance to the star should also be accurate.  However, using this distance and the 2MASS photometry to derive a luminosity, the star falls well below the association isochrone on the H-R diagram.  This suggests the luminosity, and the derived radius, are underestimated.  The star is still young ($\sim$24 Myr), and there is some evidence for extinction towards the star, with an $E(B-V)$ of 0.17 (based on the intrinsic colours of \citealt{Pecaut2013-PMS-BC-withJ}). This implies $\sim$0.5 magnitudes of extinction (assuming $A_V = 3.1 E(B-V)$).  Therefore we adopt a mass, radius, and convective turnover time by assuming the star lies on the association isochrone.  

We derive an inclination for HIP 12545 by searching for the value that produces a ZDI map with maximum entropy.  This produces a value of $i = 39 \pm 20$ degrees.  From this inclination, our \vs, and rotation period, we can infer a radius and then with our \teff\ a luminosity.  Using this luminosity to place the star on the H-R diagram, we find a value consistent with the association isochrone, which supports our inclination.  This also argues against the photometric luminosity, which is inconsistent with the association isochrone as well as the \vs\ and period.

\subsection{TYC 6878-195-1}
TYC 6878-195-1 (CD-26 13904) is a member of the $\beta$ Pic association \citep{Torres2006-search-youngAssocaitions, Torres2008-youngNearbyAssoc}.  \citet{Messina2010-RACE-OC-periods} find a photometric rotation period of $5.65 \pm 0.05$ days.  \citet{Messina2011-RACE-OC-2} updated this period to $5.70 \pm 0.06$ days.  Our longitudinal magnetic field period is in good agreement with these values, as is our ZDI period.  Therefore we adopt P = $5.70 \pm 0.06$ days.

TYC 6878-195-1 has a companion with a separation of 1.1'' and a magnitude difference of 3.5 in $V$ \citep{Torres2006-search-youngAssocaitions}.  This faint secondary is not clearly visible in our spectrum of TYC 6878-195-1.  However, in our LSD profile of the star there is a constant asymmetry to the $I$ line profile in all rotational phases.  This likely is a weak contribution from the secondary, with a radial velocity close to that of the primary.  TYC 6878-195-1 falls above the association isochrone on the H-R diagram, but by less than $1\sigma$, so there is no clear evidence for the system being a photometric binary.   Since the secondary is only marginally visible in Stokes $I$, we assume that it is negligible in Stokes $V$, and model the $V$ profile as a single star.

\subsection{BD-16 351}
BD-16 351 (TYC 5856-2070-1) is a poorly studied member of the Columba association \citep{Torres2008-youngNearbyAssoc, daSilva2009-young-associations}.  \citet{Messina2010-RACE-OC-periods} find a photometric rotation period of P = $3.21 \pm 0.01$ for the star.  

Our rotation period search confirms the value of \citet{Messina2010-RACE-OC-periods}.  Our best longitudinal field based rotation period is consistent with this value, although it requires a second order fit to achieve an acceptable $\chi^2$, suggesting a significant quadrupole component to the field.  The best fit radial velocity period is also consistent with this value, although again it requires a second order fit.  The best ZDI period is also consistent with this value, and does indeed have a significant quadrupolar component to the magnetic field ($\sim$25\% of the magnetic energy).  Therefore we adopt the period of P = $3.21 \pm 0.01$ from \citet{Messina2010-RACE-OC-periods}, since it is consistent with our measurements but formally more precise.

\subsection{LO Peg}
LO Peg (HIP 106231, TYC 2188-1136-1, BD+22 4409) is a member of the AB Dor association \citep{Zuckerman2004-ABDor-members, Torres2008-youngNearbyAssoc}. 
\citet{Barnes2005-LOPeg-DI} performed a detailed study of LO Peg using a large dataset of high resolution spectra.  They used Doppler imaging (DI) to produce maps of the surface spot distribution for the star, and derived a rotation period, differential rotation, and inclination for the star.  They found $P = 0.423229 \pm 0.000048$ d ($\Omega_{\rm eq} = 14.86 \pm 0.0027$ rad/day), $\Delta\Omega = 0.0347 \pm 0.0067$ rad/day, and $i = 45.0 \pm 2.5$ degrees.  \citet{Piluso2008-LOPeg-DI} performed DI using a different epoch of data, and found results in agreement with \citet{Barnes2005-LOPeg-DI}, with some difference in the detailed spot distribution.  
\citet{Kiraga2012-photo-periods} measured a photometric period for LO Peg of 0.4231 days, confirming the rotation period of \citet{Barnes2005-LOPeg-DI}.   

Our period searches using longitudinal magnetic fields, radial velocities, and ZDI all produced well defined periods in agreement with \citet{Barnes2005-LOPeg-DI}.  
We performed a differential rotation search following the method of \citet{Petit2002-diff-rot-DI}, assuming their solar-like differential rotation law and searching for the values that maximize the entropy in our ZDI map.  
We find a value of $\Delta\Omega = 0.2 \pm 0.2$ rad/day and $\Omega_{\rm eq} = 14.86 \pm 0.01$ rad/day, which is only marginally significant but in good agreement with \citet{Barnes2005-LOPeg-DI}.  Ultimately we adopt the values of $P = 0.423229 \pm 0.000048$ d and $\Delta\Omega = 0.034714 \pm 0.006692$ rad/day from \citet{Barnes2005-LOPeg-DI}, since their much larger dataset allows for more precision.  

We find a slightly larger \vs\ ($73.1 \pm 1.1$ \kms) than \citet{Barnes2005-LOPeg-DI} ($65.84 \pm 0.06$ \kms).  This leads to a somewhat larger inclination ($66.8 ^{+18.7}_{-8.5}$ compared to $45.0 \pm 2.5$ degrees).  It is possible that our \vs\ is influenced by spots, which have a significant impact on the line profile shape.  
We find the maximum entropy inclination from ZDI is $40 \pm 10$ degrees, which agrees with \citet{Barnes2005-LOPeg-DI}, but is marginally inconsistent with the inclination based on \vs\ period and radius.  We adopt the value from \citet{Barnes2005-LOPeg-DI}, since it is consistent with our value from ZDI, and not impacted as strongly by possible systematic errors in \vs.

\subsection{PW And}
PW And (HD 1405, TYC 2261-1518-1, BD+30 34) is a member of the AB Dor association \citep{Zuckerman2004-ABDor-members, Lopez-Santiago2006-HerLyr-ABDor-assoc, Torres2008-youngNearbyAssoc}.  \citet{Hooten1990-phot-periods} found a photometric period of $1.745$ days, although the value was somewhat uncertain.  
\citet{Strassmeier2006-PWAnd-DI} derived a rotation period for the star of $1.76159 \pm 0.00006$ days from photometry, and performed Doppler imaging of surface spots.  From the DI process, they find \vs\ $= 23.9 \pm 0.2$ \kms\ and $i = 46 \pm 7^\circ$.  Their DI map finds a collection of lower latitude spots, but no polar cap.  

Our period search produces results in good agreement with \citet{Strassmeier2006-PWAnd-DI}, for longitudinal magnetic fields ($1.77 \pm 0.02$ days), radial velocities ($1.76 \pm 0.02$ days), and ZDI ($1.77 \pm 0.2$ days).  Their \vs\ is slightly larger than ours ($22.93 \pm 0.24$ \kms), although the values are close.  Our inclination based on \vs, radius, and period is $i = 57 ^{+33}_{-12}$ degrees, which is consistent with our ZDI maximum entropy value of $i = 45 \pm 15^\circ$.  Both of these are consistent with the value from \citet{Strassmeier2006-PWAnd-DI}, and we adopt their value of $i = 46 \pm 7^\circ$ as it is the formally most precise.

\subsection{HIP 76768}
HIP 76768 (HD 139751, BD-18 4125) is a member of the AB Dor association \citep{Zuckerman2004-ABDor-members, Torres2008-youngNearbyAssoc}.  The star has a photometric rotation period of P = $3.70 \pm 0.02$ days from \citet{Messina2010-RACE-OC-periods}.  Our rotation period search from longitudinal magnetic field values agrees well with this value, as does our period from radial velocity variability, and our period search from ZDI.  Our period is formally less precise, since it was obtained over a shorter time period, therefore we adopt the value of P = $3.70 \pm 0.02$ days from \citet{Messina2010-RACE-OC-periods}.

\subsection{TYC 0486-4943-1}
TYC 0486-4943-1 is a poorly studied star in the AB Dor association \citep{Torres2008-youngNearbyAssoc}.  \citet{Messina2010-RACE-OC-periods} report a photometric period of P = $1.35 \pm 0.02$, however they note that the period was undetected in the periodogram for their complete time series, making this value somewhat uncertain.  We find this period is incompatible with the variability in our longitudinal magnetic field measurements, our radial velocity measurements, and our period search from ZDI.  Therefore we reject this rotation period.  

From our longitudinal field measurements we find a best rotation period of 3.77 days, with a second order fit.  From the ZDI maximum entropy period search, we find a best period of 3.73 days.  These periods phase our LSD profiles in a sensible fashion, and are compatible with the apparent radial velocity variability we find.  Therefore we adopt a rotation period of $3.75 \pm 0.30$ days.

\subsection{TYC 5164-567-1}
TYC 5164-567-1 (BD-03 4778) is a member of the AB Dor association \citep{Torres2008-youngNearbyAssoc}. \citet{Messina2010-RACE-OC-periods} find a photometric rotation period of P = $4.68 \pm 0.06$ days.  Our search for a rotation period finds results that are consistent with this value for the longitudinal magnetic field, radial velocity variability, and the maximum entropy ZDI solution.  Thus we confirm the  P = $4.68 \pm 0.06$ day value of \citet{Messina2010-RACE-OC-periods}, and adopt their value as it is the more precise.

\subsection{HII 739}
HII 739 (Melotte 22 HII 739, HD 23386, V969 Tau, TYC 1803-944-1) is a member of the Pleiades \citep{Hertzsprung1947-Pleiades-old, Stauffer2007-Pleiades-memb}.  It is likely a double star, reported to be a photometric binary by \citet{Soderblom1993-Pleiades-photo-vsini}. It is not an obvious SB2 in our spectra, however, there is a very weak asymmetry to the wings of our LSD line profiles.  Attempting to fit the LSD profile as a combination of two lines produces a range of nearly degenerate solutions, but the velocity separation required is roughly 10 \kms.  There is no clear change in this asymmetry or apparent velocity separation during our observations.  
When treated as a single star and placed on the H-R diagram, the star falls well above the cluster isochrone, and indeed well above the ZAMS.  The mass and radius implied by this H-R diagram position are also inconsistent with our spectroscopic \lgg.  Thus we confirm that the star is a photometric binary, and have tentative spectroscopic evidence for the presence of the secondary.  The mass, radius and turnover times we adopt are based on the association isochrone for this star.

\citet{Magnitskii1987-photo-periods} reports a rotation period, based on photometry, of P = 2.70 days.  This is based on 83 observations over 39 days and, while the data are phased well with this period, the amplitude of variability is not much larger than their error bars, so this period probably has a significant uncertainty.  
\citet{Marilli1997-photo-periods-rough} report a photometric period of P = 0.904 days, however the quality of the data used to make that estimate is not clear.  
\citet{Messina2001-photo-periods} report a photometric period of $0.917 \pm 0.003$ d.  However, this was based on only 14 observations distributed over 9 days (a 7 day run and a 2 day run).  This makes the accuracy of this value questionable, and likely this period is an alias of the true rotation period.  
Considering the \vs\, and radius we find for HII 739, a period near 0.9 days would require an inclination of $\sim 14^\circ$, which is unlikely although not impossible (since the probability distribution of $i$ for randomly oriented rotation axes goes as $\sin i$) but this would imply an extremely large rotational speed ($v_{\rm eq} \sim 60$ \kms).  

The variability in our LSD profiles does not phase coherently with a 0.917 or 0.904 day period.  Considering the uncertain nature of these literature values, we reject these periods.  

The longitudinal magnetic field measurements are consistent with no variability, with the exception of the first two observations, which were obtained 10 days before the rest of the data.  It is possible this change represents an intrinsic evolution of the magnetic field, however the significance of this difference in $B_l$ is not large.  A 2.7 day period does fit this data better than any period near 0.9 days, however the longitudinal field cannot well constrain the rotation period.  

Our period search from ZDI does not provide a unique best period.  This is due to the low amplitude of the signal in the $V$ profiles relative to the noise.  However, the period of $\sim$2.7 days provides a poor fit falling in a local maximum of the periodogram but one of the weakest local maxima.  Periods near 0.9 days provide very bad fits.  By eye, the 2.7 day rotation period does not phase the LSD profiles well, thus we reject this period.  
Instead we adopt a period of 1.577 d that we derived from the 75 day-long continuous Kepler K2 light-curve obtained for HII 739, which was kindly provided to us by J. Stauffer and L. Rebull.  This period phases the longitudinal field and radial velocity data as well as the 2.7 d period, but it corresponds to the global best period from ZDI.  Thus the 1.577 d period is both based on higher quality photometric data, and provide a better fit to our spectropolarimetric data.  

Given the uncertainty in the luminosity and photometric radius of HII 739, we do not use the radius period and \vs\ to derive an inclination.  Instead we search for the inclination which provides the maximum entropy ZDI map, finding the value $i = 51 \pm 20$ degrees.  While this value is somewhat uncertain, it is consistent with the radius, period and \vs\ values derived using the association isochrone.

\subsection{PELS 031}
PELS 031 (Melotte 22 PELS 031, TYC 1247-76-1) is a member of the Pleiades \citep{vanLeeuwen1986-Pleiades-memb-phot, Stauffer2007-Pleiades-memb}. 
A photometric rotation period was reported by \citet{Hartman2010-photo-periods} of P = $2.9190 \pm 0.0003$.  However, this does not phase our $V$ profiles sensibly (and is inconsistent with our ZDI period search), therefore we reject this period.  

The range of periods that are plausible from our \vs\, and radius are roughly 1 to 4.5 days (assuming $90 > i > 10$ degrees).  
From our period search using ZDI, the best period we find is $\sim$2.5 days.  An alternative rotation period at 5.0 days is found, but this is inconsistent with stellar radius and \vs, therefore we reject it as an alias of the real value.  The maximum in entropy at 2.5 days is relatively broad ($\pm 0.1$ days), due to the short time period the observations were collected over.  However, this maximum is unique, and substantially above any other maxima in the periodogram.  This does not provide the optimal phasing of our longitudinal magnetic field measurements, but the phasing is acceptable, and the variability in $B_{\rm l}$ is weak.  A large number of the $V$ profiles of PELS 031 show `crossover' signatures, with small net longitudinal field values, consequently the longitudinal magnetic field is not as well suited to determining the rotation period as a full ZDI fit.  These nearly constant crossover signatures suggest a strong toroidal belt, which is confirmed by the ZDI magnetic map.  The radial velocity periodogram is ambiguous, but the strongest period is $2.7 \pm 0.2$ days.  Therefore we adopt $2.5 \pm 0.1$ days as the rotation period for the star, but note that this is the one case where we do not have a strong confirmation of the ZDI period through other measures.

\subsection{HII 296}
HII 296 (Melotte 22 HII 296, V966 Tau, TYC 1799-963-1) is a member of the Pleiades \citep{Hertzsprung1947-Pleiades-old, Deacon2004-pleiades-aPer-memb}.  
An older rotation period measurement exists for HII 296 from \citet{Magnitskii1987-photo-periods} of P = 2.53 d, based on photometry.  This period appears to phase their observations well, however it is not clear how precise this values is.  A more recent period was measured by \citet{Hartman2010-photo-periods} of P $= 2.60863 \pm 0.00009$ d, based on photometry (Sloan $r$ band).  Our data was not sufficient to derive a reliable unique rotation period for this star, largely due to the weak amplitude of the Stokes $V$ signatures relative to the noise.  However, in our data a period of 2.61 d is one of the two best maximum entropy solutions from ZDI.   This is strongest maximum entropy period that is also consistent with the longitudinal magnetic field variability.  We adopt the value from \citet{Hartman2010-photo-periods}, as it is more precise, and more importantly it phases our observations well for both longitudinal magnetic field and ZDI maps.

\subsection{V447 Lac}
V447 Lac (HD 211472, HIP 109926, TYC 3986-2960-1, BD+53~2831) is a member of the Her-Lyr moving group \citep{Eisenbeiss2013-HerLyr-age}. 
\citet{Strassmeier2000-periods-preDI} find a photometric rotation period of 4.4266 days (based on 74 observations over 88 days). 

Our period search for the star from longitudinal magnetic field measurements yields several ambiguous periods, due to our relatively sparse data set.  However, one of the stronger minima in $\chi^2$ is consistent with the period from \citet{Strassmeier2000-periods-preDI}.  We find no clear radial velocity variability.  The period search from ZDI is similarly ambiguous, however again one of the stronger maxima agrees with \citet{Strassmeier2000-periods-preDI}.  Our observed \vs\ is low ($4.6 \pm 0.3$ \kms), but when combined with the derived radius ($0.81 \pm 0.03$ $R_\odot$), it is consistent with this period (for $i = 29^{+5}_{-4}$). Therefore, our data support the period of \citet{Strassmeier2000-periods-preDI}.  Since they provide no uncertainty estimate we assume a conservative uncertainty of 0.1, and we adopt the value of $P = 4.43 \pm 0.10$ days.

\subsection{DX Leo}
DX Leo (HD 82443, HIP 46843, TYC 1962-469-1, BD+27~1775) is a member of the Her-Lyr moving group \citep{Gaidos1998-Her-Lyr-members-init, Eisenbeiss2013-HerLyr-age}. 
\citet{Messina1999-DXLeo} performed a detailed study of DX Leo, and they find a rotation period of P = $5.377 \pm 0.073$ days, based on photometry.  They also derive an approximate differential rotation of $\Delta\Omega/\Omega_{\rm eq} \sim 0.04$, based on apparently cyclical variations in their measured rotation periods.  They also attempt to map the spot distribution of the star based on this photometry, (using an approximate inclination of $i = 60^\circ$).  
\citet{Strassmeier2000-periods-preDI} measured a photometric rotation period of 5.409 days, based on a smaller dataset (46 observations over 91 days).  This strongly supports the value of \citet{Messina1999-DXLeo},  and we prefer the value of \citet{Messina1999-DXLeo} since it was based on a larger dataset, and being slightly shorter may be closer to the true equatorial period, due to differential rotation.  

Our observations agree with the period of \citet{Messina1999-DXLeo}.  We find a well defined period from longitudinal magnetic field data ($P = 5.18 ^{+0.19}_{-0.16}$ d), and a consistent period from the radial velocity variability.  We also find a consistent period from our ZDI analysis ($P = 5.45 \pm 0.15$ d).  Therefore we adopt the value $P = 5.377 \pm 0.073$ days.  Based on this period and our radius and \vs\ measurements, we derive an inclination of $i = 58.0 ^{+8.0}_{-6.1}$, which is consistent with \citet{Messina1999-DXLeo}.

\subsection{V439 And}
V439 And (HD 166, HIP 544, TYC 1735-927-1, BD+28~4704) is a member of the Her-Lyr moving group \citep{Gaidos1998-Her-Lyr-members-init, Lopez-Santiago2006-HerLyr-ABDor-assoc, Eisenbeiss2013-HerLyr-age}.  
\citet{Gaidos2000-spec-phot-periods} find a photometric rotation period of $6.23 \pm 0.01$ days, based on 33 observations.
\citet{Lopez-Santiago2010-chromosphere-noPeriods} and \citet{Eisenbeiss2013-age-params-noPeriods} quote a rotation period from the literature of 5.69 days, but neither set of authors provides a reference for this value, thus we consider this value unreliable.  Our period search with longitudinal magnetic fields ($P = 6.5 \pm 0.4$ d) and ZDI ($P=6.15 \pm 0.20$ d) produce consistent values, but with significant uncertainties.  The radial velocity variability is very weak, and the period search from radial velocity is somewhat ambiguous, but the dominant period is $6.0 \pm 0.4$ days, with several aliases near 2 days.  These periods are all consistent with our observed \vs\ and derived radius.  Since all our period estimates are in good agreement with \citet{Gaidos2000-spec-phot-periods} we adopt their rotation period.

\section{Spurious signal in low S/N observations}
\label{Spurious signal}

Technical problems were encountered during preparatory runs for this project, for observations of late-type stars (G and K) with low S/N ($<70$ at the peak, per spectral pixel).   In these observations, spurious signal in the diagnostic ``Null'' was found, which appears to also contaminate the Stokes $V$ profile.  
The null spectrum is generated in a similar fashion to the $V$ spectrum, but sub-exposures of different polarization are combined destructively as described by \citet{Donati1997-major}.  If the instrument is functioning properly the null is expected to contain only noise.  
We have encountered this problem with observations from both ESPaDOnS at the CFHT and with Narval at the TBL.  The spurious signal is only clearly visible in LSD profiles of the stars, however in the worst cases its amplitude can approach that of a typical Zeeman signature in these stars.

In light of this, the CFHT engineering run 12BE96 was devoted to investigating the problem.  The conclusions from that run were that the problem stems from imperfect background subtraction during the data reduction phase, and that the problem can be resolved by ensuring a peak S/N above 100. 
For observations of late-type stars with S/N $<70$, the spectral orders far to the blue can contain only a few counts per CCD pixel above an inter-order background of a few hundred.  Thus, a small underestimation of the inter-order background can have a large impact the reduced spectrum, particularly in the blue.  

\begin{figure}
  \centering
  \includegraphics[width=3.3in]{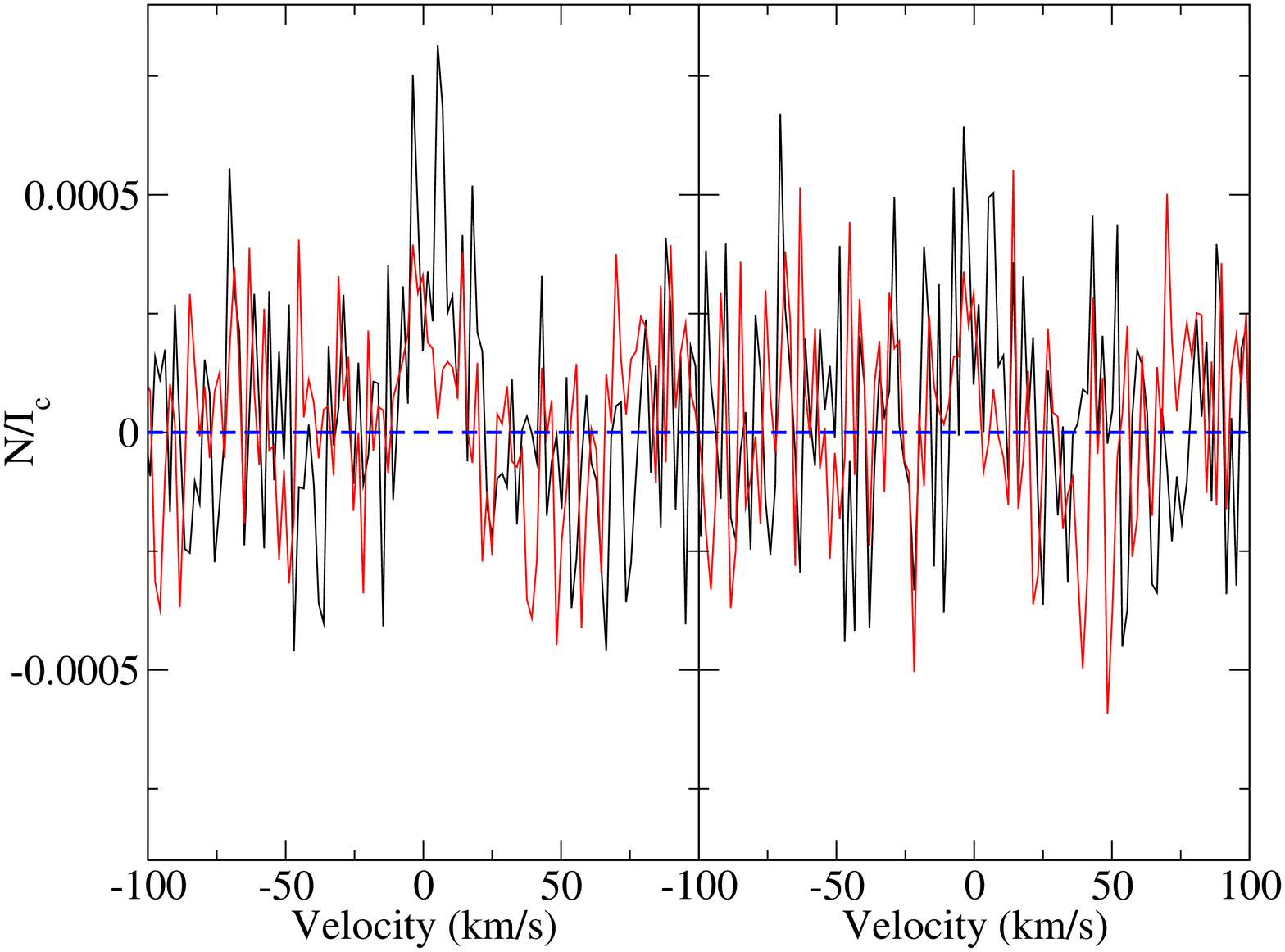}
  \caption{Left: two sample null LSD profiles, with spurious signal (from HII 296).  Right: the same two null LSD profiles calculated using only higher S/N orders (redward of 500 nm).  This example represents the worst case for observations included in our analysis.  The large majority of observations were collected at higher S/N and do not have any detectable spurious signal in the null.   }
  \label{fig-null-profiles}
\end{figure}

On careful inspection of observations badly contaminated with this spurious signal, we find that it is generated in the blue-most spectral orders, where the S/N is lowest.  Specifically, LSD profiles generated using lower S/N spectral orders produce substantially worse spurious contamination, whereas LSD profiles generated using higher S/N spectral orders have little to no spurious contamination.  This is illustrated in Fig.~\ref{fig-null-profiles}.  
We speculate that this problem has not been encountered in low S/N observations of hot stars due to the lower density of spectral lines in the low S/N orders of those observations.  Both hotter and cooler stars have the majority of their spectral lines in the blue, but for hotter stars this is near the peak of the flux distribution.  Thus for hot stars, spurious signal may be present in the low S/N orders in the red, however since very few spectra lines are in the red it would likely not contaminate an LSD profile to any detectable degree.  

This matches our experience with preliminary observations for the HMS project.  For observations made after 2012,  where we maintain a S/N above 100 (typically 150) we do not encounter any significant spurious signal .  However, for some older observations obtained with NARVAL at the TBL in 2009 where the S/N falls below 100, a weak spurious signal in the null profile can be observed.  In these cases, the spurious signal is generated in the blue-most spectral orders, where the S/N is lowest.  

For observations of some stars with peak S/N between 100 and 150, while no statically significant spurious signal is found in individual observations, averaging over all observations of a star can produce a weak but significant spurious signal in the mean null LSD profile.  This suggests that there may still be a weak spurious signal in the observations at this S/N level.  Restricting the LSD analysis to higher S/N orders of the star (e.g.\ $>500$ nm) eliminates the spurious signal in these mean LSD profiles.  

Because of this spurious S/N problem, in the observations for this program we always aimed for a peak S/N above 100, and preferably above 150.  Furthermore, for all our observations we restrict our analysis to the red part of the spectrum, $>500$ nm.  This ensures we avoid contamination by any potential spurious signal.  Since there is very little real signal in those blue-most orders, discarding them has virtually no impact on the sensitivity of our observations.  Thus the restriction to $>500$ nm can be applied to all observations without any loss of data quality.

\section{Emission indices}
\label{Emission indices}

For all the observations, we calculated indices characterizing the emission in a few lines, following the procedure of \citet{Marsden2014-Bcool-survey1}.  For the calcium H and K lines, we calculated a Mount Wilson S-index following the method of \citet{Wright2004-SIndex-calib}.  We used the calibration of the S-index for ESPaDOnS and Narval from \citet{Marsden2014-Bcool-survey1}, thus these values are directly comparable to the Mt. Wilson S-index measurements.  The S-index was calculated using the flux in two triangular filters centered on the H and K lines, divided by the flux in two rectangular filters on either side of the H and K lines.  These fluxes were scaled by calibration coefficients for ESPaDOnS and Narval from  \citet{Marsden2014-Bcool-survey1}.  We calculated similar indices for the calcium infrared triplet (Ca IRT) and H$\alpha$.  The Ca IRT index consists of three rectangular filters centered on the lines in the triplet, and a pair of rectangular filters on either side of the triplet defining the continuum level.  We also calculated an index for H$\alpha$ emission, consisting of a rectangular filter centered on H$\alpha$, and a pair of rectangular filters on either side defining the continuum level.  

\begin{figure*}
  \centering
  \includegraphics[width=3.3in]{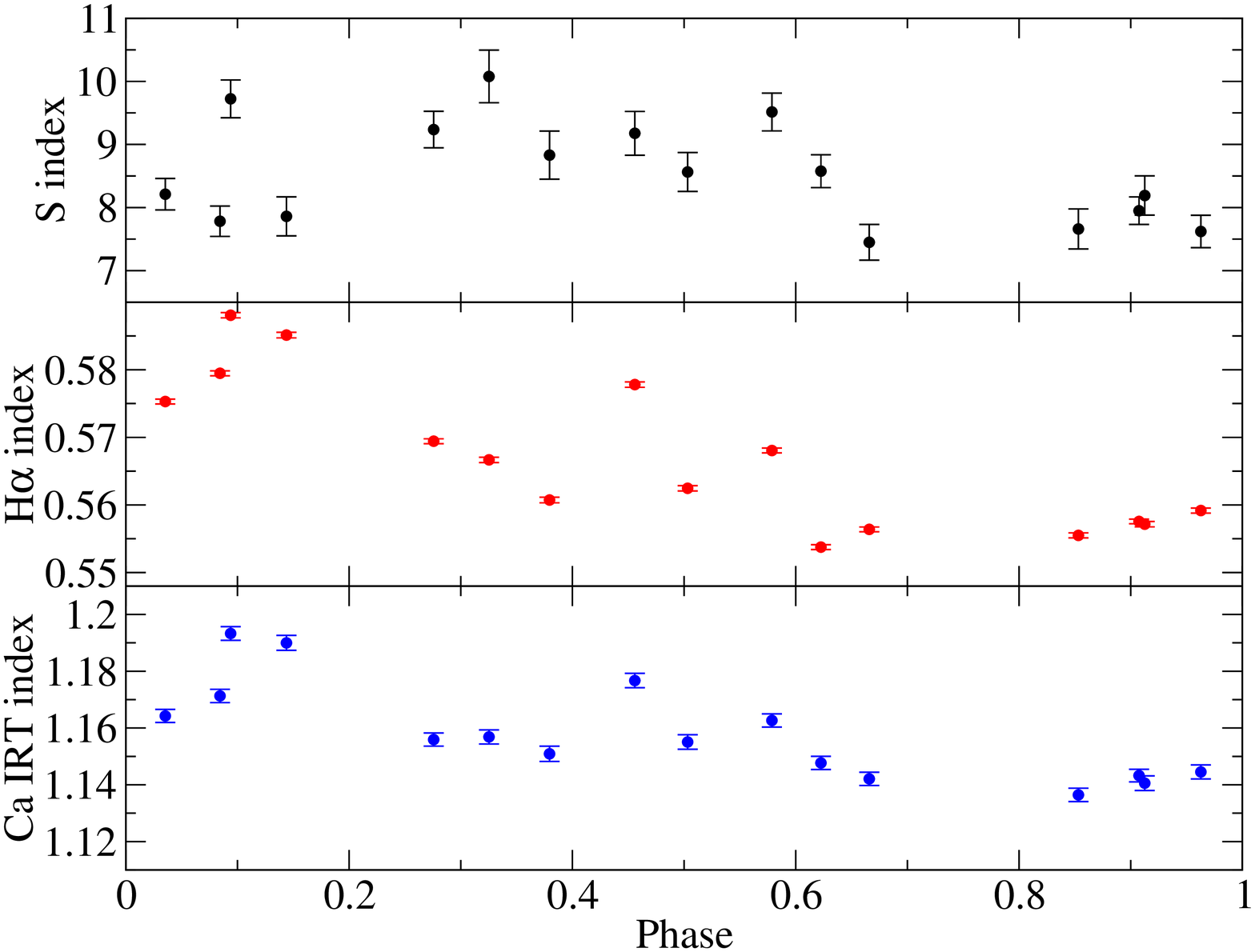}
  \caption{Emission indices measured for TYC 6349-200-1, phased with the rotation period derived in Sect.~\ref{rotational-period}.  Plotted are the S index (for Ca {\sc ii} H and K emission), and similar indices constructed for H$\alpha$ and the Ca infrared triplet.  }
  \label{fig-sindex}
\end{figure*}

Chromospheric emission should be modulated with stellar rotation.  However due to its complex structure, and large amount of intrinsic variability, it provides a poor means of measuring stellar rotation.  Consequently, we phase the emission indices with the stellar rotation period, and generally they show coherent variability, but we do not attempt to measure a stellar rotation period from them.  An example of emission indices phased with their stellar rotation period is shown in Fig.~\ref{fig-sindex}.

Since the chromospheric structure is more complex than the large-scale photospheric magnetic structure, the chromospheric variability is more complicated than the disk integrated longitudinal magnetic field.  Furthermore, the chromospheric structure may change more rapidly than the photospheric magnetic field, so variability between rotation cycles may be much larger for these emission indices than the observed magnetic fields.

\begin{table*}
\centering
\caption{Emission line indices for the stars in this study.  Presented are the means and standard deviations (characterizing variability) for the full observed datasets. }
\begin{tabular}{lccccccc}
\hline\hline
Star & Assoc. & \multicolumn{2}{c}{S index} & \multicolumn{2}{c}{H$\alpha$ index} & \multicolumn{2}{c}{Ca IRT index} \\ 
     &        & mean   & stdev.               & mean     & stdev.                     & mean & stdev. \\ 
\hline
TYC 6349-0200-1 &  $\beta$ Pic & 8.53 & 0.79 & 0.567 & 0.011 & 1.158 & 0.017 \\ 
      HIP 12545 &  $\beta$ Pic & 6.80 & 0.53 & 0.581 & 0.014 & 1.180 & 0.019 \\ 
TYC 6878-0195-1 &  $\beta$ Pic & 3.96 & 0.33 & 0.494 & 0.009 & 1.109 & 0.013 \\ 
       BD-16351 &      Columba & 1.58 & 0.29 & 0.464 & 0.009 & 1.092 & 0.015 \\ 
         LO Peg &       AB Dor & 2.52 & 0.29 & 0.593 & 0.015 & 1.186 & 0.016 \\ 
         PW And &       AB Dor & 1.45 & 0.08 & 0.515 & 0.012 & 1.188 & 0.014 \\ 
      HIP 76768 &       AB Dor & 4.78 & 1.16 & 0.563 & 0.017 & 1.140 & 0.021 \\ 
TYC 0486-4943-1 &       AB Dor & 2.55 & 0.37 & 0.452 & 0.006 & 1.024 & 0.006 \\ 
 TYC 5164-567-1 &       AB Dor & 1.53 & 0.15 & 0.413 & 0.005 & 1.040 & 0.007 \\ 
        HII 739 &     Pleiades & 0.53 & 0.01 & 0.368 & 0.003 & 1.048 & 0.007 \\ 
       PELS 031 &     Pleiades & 1.93 & 0.26 & 0.454 & 0.006 & 1.086 & 0.011 \\ 
        HII 296 &     Pleiades & 1.05 & 0.05 & 0.425 & 0.006 & 1.098 & 0.013 \\ 
       V447 Lac &      Her-Lyr & 0.52 & 0.01 & 0.334 & 0.002 & 0.892 & 0.005 \\ 
         DX Leo &      Her-Lyr & 0.61 & 0.01 & 0.363 & 0.004 & 0.968 & 0.004 \\ 
       V439 And &      Her-Lyr & 0.46 & 0.02 & 0.321 & 0.002 & 0.885 & 0.006 \\ 
\hline\hline
\end{tabular} 
\label{table-var-param} 
\end{table*}

\section{Convective turnover times}
\label{Convective turnover times}

In order to calculate accurate Rossby numbers, realistic convective turnover times are necessary.  
In this work we used convective turnover times computed from the evolutionary models in Sect \ref{HRD_evolutionary_tracks}, and take the value at one pressure scale height above the base of the convection zone.  The location of one pressure scale height was chosen as this yields values the most consistent with \citet{Noyes1984-CaHK-Rossby} and \citet{Cranmer2011-massloss-with-Rossby}.  This makes our Rossby numbers the most directly comparable to many literature values.   Changing the location in the model where the convective turnover time is taken significantly changes the resulting turnover time.  However, this does so in a uniform fashion for all our stars, thus it does not change the quality of the correlation between Rossby number and mean large-scale magnetic field strength, discussed in Sect.~\ref{Magnetic trends in young stars}.  The exponent of the power law describing this correlation is not significantly affected by the choice of location, however the coefficient in front of the power law does change.  This is important in that it impacts specific value of the Rossby number at which magnetic saturation may occur.

For the comparison sample of older field stars, discussed in Sect.~\ref{Comparison with older field stars}, we calculated convective turnover times in the same way as for our younger sample, using the same theoretical evolutionary tracks.  \teff\ was taken from the references for individual stars in that section. We re-derived luminosities for the stars in a homogeneous fashion.  Luminosity was computed using 2MASS $J$-band photometry, Hipparcos parallaxes, and the bolometric correction of \citet{Pecaut2013-PMS-BC-withJ}, and assuming negligible extinction.  The exception to this is HD 131156A \& B, which is unresolved in 2MASS, thus we use $V$-band photometry and the relevant bolometric correction from \citet{Pecaut2013-PMS-BC-withJ}.  The \teff\ and luminosity were compared to evolutionary tracks and used to derive convective turnover times.  

For the comparison sample of T Tauri stars, discussed in Sect.~\ref{Comparison with T Tauri stars}, convective turnover times were calculated in the same fashion as for our older star samples.  Luminosity and \teff\ were taken from the literature sources in Sect.~\ref{Comparison with T Tauri stars}, and compared to the evolutionary tracks to derive convective turnover times.  We did not re-derive luminosities for these stars, given the complexities involved due to large amounts of variability and extinction.  Note that there is potentially a large systematic uncertainty in these convective turnover times.  As these stars become fully convective, the location of the dynamo may change, thus it is not clear at what point in the star to take the convective turnover time.  We have continued to use the value at one pressure scale height above the base of the convection zone, however this approaches one pressure scale height above the center of the star.

% Don't change these lines
\bsp	% typesetting comment
\label{lastpage}
\end{document}